\documentclass[]{interact}

\usepackage{epstopdf}% To incorporate .pdf illustrations using PDFLaTeX, etc.
\usepackage[caption=false]{subfig}% Support for small, `sub' figures and tables

\usepackage[backend=bibtex,style=nature,doi=true,url=false]{biblatex}
\usepackage{physics}
\usepackage{commath}

\usepackage{comment}
\usepackage{color}
\usepackage[normalem]{ulem}
\usepackage{upgreek}
\usepackage{siunitx}
\usepackage{mathtools}
\usepackage{textcomp}

\usepackage[colorlinks,citecolor=blue]{hyperref}

\usepackage{zebra-goodies} % for \todo

\theoremstyle{plain}% Theorem-like structures provided by amsthm.sty

\theoremstyle{definition}

\theoremstyle{remark}

\def\v{\vb}

\newcommand{\etal}{\textit{et al}.\@ }

\begin{document}

\title{Interatomic potentials: Achievements and challenges} 

\author{
\name{Martin~H. M\"user\textsuperscript{a}\thanks{CONTACT M.~H. M\"user. Email: martin.mueser@mx.uni-saarland.de}, Sergey~V. Sukhomlinov\textsuperscript{a}, and Lars Pastewka\textsuperscript{b,c}}
\affil{\textsuperscript{a}Dept. of Materials Science and Engineering, Saarland University, Saarbr\"ucken, Germany\\\textsuperscript{b}Dept. of Microsystems Engineering, University of Freiburg, Freiburg, Germany\\\textsuperscript{c}Cluster of Excellence \emph{liv}MatS, Freiburg Center for Interactive Materials and Bioinspired Technologies,
University of Freiburg, Freiburg, Germany}
}

\maketitle

\begin{abstract}
Interatomic potentials approximate the potential energy of atoms as a function of their coordinates. Their main application is the effective simulation of many-atom systems. Here, we review empirical interatomic potentials designed to reproduce elastic properties, defect energies, bond breaking, bond formation, and even redox reactions. We discuss popular two-body potentials, embedded-atom models for metals, bond-order potentials for covalently bonded systems, polarizable potentials including charge-transfer approaches for ionic systems and quantum-Drude oscillator models mimicking higher-order and many-body dispersion. Particular emphasis is laid on the question what constraints ensue from the functional form of a potential, e.g., in what way Cauchy relations for elastic tensor elements can be violated and what this entails for the ratio of defect and cohesive energies, or why the ratio of boiling to melting temperature tends to be large for potentials describing metals but small for short-ranged pair potentials. The review is meant to be pedagogical rather than encyclopedic. This is why we highlight potentials with functional forms sufficiently simple to remain amenable to analytical treatments. Our main objective is to provide a stimulus for how existing approaches can be advanced or meaningfully combined to extent the scope of simulations based on empirical potentials.
\end{abstract}


\tableofcontents

\section{Introduction}

Interatomic potentials are functions of nuclear coordinates approximating the electronic ground state energy, or for metals, the electronic free energy of a system.
Forces on atoms, as needed in molecular-dynamics simulations, can be obtained by calculating the gradient of these  functions with respect to the nuclear coordinates.
Frequently, the terms {\it interatomic potential} and {\it force field} are used synonymously.
There is, however, a subtle and sometimes important difference between the two.
Many-body force fields are not necessarily gradients of a scalar function, in contrast to ``real'' interatomic forces when electrons are given enough time to equilibrate.
In this review, we focus on force fields that can be represented as gradients of scalar functions approximating the exact interatomic potentials. 
Before getting started on details,
let us take a step back.

In his {\it Lectures on Physics}, Richard Feynman wondered 
what statement would contain the most information in the fewest words, if
all of scientific knowledge were to be destroyed in some cataclysm,
and only one sentence could be passed to the next generations of
scientists~\cite{Feynman1964Book}:
{\it I believe it is the atomic hypothesis
that all things are made of atoms --
little particles that move around in perpetual motion,
attracting each other when they are a little distance apart,
but repelling upon being squeezed into one another.
}
Two-body potentials reflect this generic behavior, except those
describing two ions carrying a charge of identical sign, since they repel each other even at large separation.

In fact,  quite a bit can be learned from studying two-body potentials.
For example, the crystalline structure of many elemental or binary crystals can
be rationalized, as it is done in any better text book on solid state physics.
%~\cite{born31} 
%
Moreover, the functional form of {\it constitutive laws}
is often independent of the
precise nature of the potentials, such as Hooke's law, which states  restoring
forces of solids in equilibrium to be linear in small deformations.  % "in" should be correct
Even many non-linear constitutive equations do not depend on the details
of the potential.
The exponents with which compressibility or specific heat diverge as the 
temperature or pressure approaches the fluid-vapour critical point are identical 
for all substances irrespective of their specific interactions~\cite{Guggenheim1945JCP,Kadanoff1967RMP,Fisher1967RPP}. 
The way how the viscosity of polymers increases with molecular
weight can also be described with the help of two-body potentials,
as long as they prevent two polymers from crossing each
other~\cite{Kremer1990JCP}.
These are but a few examples for the success of two-body potentials.
The only thing that does depend on chemical detail,
it almost seems, are  boring prefactors.
Unfortunately, or, depending on your viewpoint, fortunately,
this is not quite right.

Realistic parameterizations of two-body potentials, inluding Morse~\cite{Morse1929PR},
Buckingham~\cite{Buckingham1938PRSL}, or Lennard-Jones (LJ)~\cite{LennardJones1931PPS}, favor close-packed assemblies
of atoms, e.g., face centered cubic or hexagonal closed packed lattices.
Thus, neither molecular crystals as those formed by oxygen or nitrogen at 
small temperature, nor layered crystals like graphite at ambient conditions 
could be thermodynamically stable. 
Even those elements that do like to close pack may not be describable by 
two-body potentials.
For example, the proper description of the elastic properties
of metals and their ductility hinges on many-body terms~\cite{daw1983PRL}.
The importance of directed interactions and thus of many-body terms 
is particularly apparent for carbon, where depending on the hybridization
of individual carbon atoms, different interatomic forces ensue. 
In the worst case, history dependence of interatomic, or rather, interionic 
forces can arise.
To illustrate this point, assume a NaCl molecule dissociates into ions 
in water and the water is evaporated later so that two isolated ions 
emerge. 
If, however, the NaCl molecule had been separated slowly in an inert gas 
environment before the nuclear degrees of freedom had been brought to their
final destinations, two neutral atoms would have formed.
This is because the 
ionization energy of sodium exceeds the electron affinity of chlorine. 
Real powerful potentials should mimic such a process correctly and
account for charge transfer in the appropriate way, though the notion
that forces arise as an unambigous function of nuclear coordinates would 
have to be abondened. 
From this discussion, we can see that no chemistry and none of its implication 
(batteries, laptops, soccer, or life in a more general sense) could arise 
if atomic systems could be accurately described with two-body potentials.
In brief, without many-body interactions, the world in general,
and science in particular, would be much less exciting than it actually is.

The root of many-body potentials is that two atoms change their interaction when additional atoms are present, because their electronic structure changes.
For instance, two hydrogen atoms prefer to form a strong 
covalent bond between them rather than to remain lonely.
However, as soon as an oxygen
%lp2all: lady
enters the scene, the hydrogens stop being homo and happily form a heteronuclear water molecule with the oxygen.

Many-body potentials were much advanced over the last few decades.
Realistic, large-scale simulations of several thousand
atoms can nowadays be conducted even on commodity computers, at least
for selected compounds such as simple metal alloys or hydrocarbons.
Force-field based simulations of increasingly complex systems
and processes become possible, including those during which
atoms rehybridize in the course of a chemical reaction~\cite{brenner_empirical_1990,brenner_second-generation_2002,VanDuin2001-iv}.
However, the need for further improvement remains, in particular for systems in which two or even more different bonding types (ionic, metallic, covalent, dispersive) combine.
This text is meant to provide a rudimentary understanding for why specific functional forms for potentials were chosen and what a given (class of) potential may or may not achieve. 
With this, we hope to stimulate insight on how to combine or to generalize  different  potential classes so that systems held together by different bonding types can be better simulated in the future.
Since an incredible number of papers on potentials has been published, we are certain to have missed important contributions, even if we did our best to identify the original key literature. 

Focusing on concepts leaves us little room to provide readers with the best
possible parameterization for a given substance and application.
Performance evaluations of potentials and their parameterization
containing information like \textit{the specific parameterization by author A using
potential B reproduces properties C and D of the material E but does a
poor job on properties F and G} are probably better looked up in one of
several excellent and important databases~\cite{Tadmor2011JOM,becker_considerations_2013,trautt_facilitating_2015,de_tomas_graphitization_2016,hale_evaluating_2018,de_tomas_transferability_2019}.
In addition, we wish to refer to many excellent 
reviews~\cite{carlsson_beyond_1990,brenner_art_2000,Ackland2012Inbook,sinnott_three_2012,finnis_concepts_2012,foiles_contributions_2012,pastewka_bond_2012,shin_variable_2012,plimpton_computational_2012,liang_reactive_2013,behler_perspective_2016,harrison_review_2018,deringer_machine_2019,deringer_gaussian_2021,mishin_machine-learning_2021}
or even text books~\cite{Finnis2005Book,tadmor_modeling_2011}
on interatomic potentials emphasizing specific parameterization of potentials 
more than this review.
Here, we discuss materials- or element-specific numbers only when we explore the transferability of a potential, which is its ability to accurately predictit properties, structures, and/or stoichiometries, which it was not adjusted for. 
In addition, one system is highlighted for each class of bonding type. 
To this end, we chose argon as the representative of systems bonding through van-der-Waals, because its dispersive interactions are half way between those of water and CH$_2$ or CH$_3$ units in hydrocarbon chains.
Carbon, copper, and rocksalt (NaCl) complement the list to represent covalently bonded materials, metals, and ionic systems.

\section{Classification, construction, and consequences of (many-body) potentials}

\subsection{Brief classification of interaction potentials}

The way in which interactions are incorporated into potentials is so diverse that they can be classified according to several main criteria:
the chemical nature of the bond, whether a bonding topology is prescribed or allowed to change, two-body versus many-body interactions, and the degree of empiricism or level of theory used to construct the functional form of the potential and to gauge adjustable parameters.

Atoms with open valence shells form covalent or metallic bonds, while atoms with closed valence shell interact through van der Waals forces. 
The latter include repulsion at short separation and dispersive forces, which result from the mutual induction of dipoles and other multipoles owing to quantum-mechanical fluctuations of the valence shell.
Each bonding type has its own characteristics how interatomic forces change with interatomic separation and as a function of their environment, which motivates the distinction  between open-shell and closed-shell potentials.
Van der Waals or ``non-bonded'' interactions have energies on the order of thermal energy at room temperature and are thus weaker than covalent and metallic bonds.
In addition, there can be Coulombic interactions due to partial or close-to-integer charges on atoms or ions. 
Interactions between atoms can also be a combination of all of the above, the most prominent example involving the hydrogen bond. 

Potentials with a fixed bonding topology range from simple bead-spring models~\cite{Kremer1990JCP} to highly sophisticated, chemically realistic valence force fields containing explicit bond-angle or torsional terms~\cite{weiner_amber:_1981,mackerell_all-atom_1995,pearlman_amber_1995,jorgensen_development_1996,mackerell_all-atom_1998,mackerell_extending_2004,case_amber_2005,salomon-ferrer2013overview}.
Two atoms interact differently with each other, depending on whether they are considered bonded or non-bonded to each other, even if all involved atoms carry the same chemical symbol, or, if they correspond to the same coarse-grained entities, such as a so-called united atom reflecting, for example, a CH$_2$ or CH$_3$ group~\cite{jorgensen_development_1996}.
The design of valence force fields is in a rather mature state~\cite{weiner_amber:_1981,mackerell_all-atom_1995,pearlman_amber_1995,jorgensen_development_1996,mackerell_all-atom_1998,mackerell_extending_2004,case_amber_2005,salomon-ferrer2013overview}, which is why we will discuss them only peripherally.
This does not prevent us from highly recommending their use for any situation, in which bond breaking and bond formation can be ruled out, simply because they are computationally lean while being sufficiently accurate for many purposes. 
Nonetheless, weaknesses remain, in particular the frequently poor treatment of higher-order and many-body dispersion as well as of charge transfer, which, however, are all best explained assuming free atoms or ions  rather than bonded entities as reference.
In a biophysical context, a proper treatment of dispersive interactions can be particularly important when molecules transfer between aqueous and hydrophobic environments, since the dispersive interactions with carbon-hydrogen chains are  stronger than with water molecules~\cite{Walters2018JPCB}. 
Adsorbing higher-order or many-body effects by tweaking pertinent  Lennard-Jones interaction coefficients is then even more hazardous than for homogeneous phases. 
Thus, readers being interested in improving bonded potentials for biomolecular simulations have no excuse to stop reading here. 

The distinction between two-body and many-body potentials should be self-explanatory.
The latter can be further subdivided into explicit or implicit.
An explicit many-body potential does not require on-the-fly adjustments of parameters like atomic charges or dipoles. 
In contrast, implicit many-body potentials necessitate the determination of such quantities.
This is commonly achieved by minimizing expressions for the potential energy with respect to degrees of freedom representing a coarse-grained version of the full electronic response. 
Also explicit many-body potentials can be categorized into further subgroups.
They can be cast in terms of analytical functions, or numerical tables, or they can be extrapolated from a large %any set is "finite" 
set of high-dimensional data as is done in machine-learned potentials~\cite{behler_perspective_2016,deringer_machine_2019,deringer_gaussian_2021,mishin_machine-learning_2021}.

Potential classes are sometimes also named differently depending on the quantum-mechanical framework from which they are motivated.
For example, embedded-atom potentials~\cite{daw1983PRL,daw1984PRB} are best motivated from density-functional theory, while second-moment tight-binding potentials~\cite{Ducastelle1970JDP,finnis_simple_1984} arise, as their name says, quite naturally from the tight-binding approximation.
Potentials lacking a direct theoretical justification are also called empirical.
While we could elaborate much more on how to further classify potentials, we feel compelled to cut to the chase. 

\subsection{Constructing (many-body) potentials}

Formally, it is possible to expand the potential energy
$U(\{ {\bf r}\})$ of a (classical) many-atom
system into two-body, three-body, and higher-order contributions,
where $\{{\bf r}\}$ represents the positions of atoms, i.e.,
\begin{eqnarray}
U(\{{\bf r}\}) & =  &
\sum_{i} U_1({\bf r}_i) + 
\sum_{i<j} U_{2}({\bf r}_i, {\bf r}_j) 
% \nonumber\\ & + &
+ \sum_{i<j<k} U_{3}({\bf r}_i, {\bf r}_j, {\bf r}_k)  + \cdots.
\label{eq:vExpand}
\end{eqnarray}
Here, the index in the $U(...)$'s on the r.h.s. of the equation gives the order of the many-body expression.
For non-elemental systems, the $U_n(...)$ also depend explicitly on the atomic indices, or more precisely on the nature of the atoms interacting with each other. 
The underlying assumption of the expansion is that electrons are in
a well-defined state, e.g., in their ground state or in
thermal equilibrium.

\begin{figure}[htb]
\begin{center}
\includegraphics[width=0.5\textwidth]{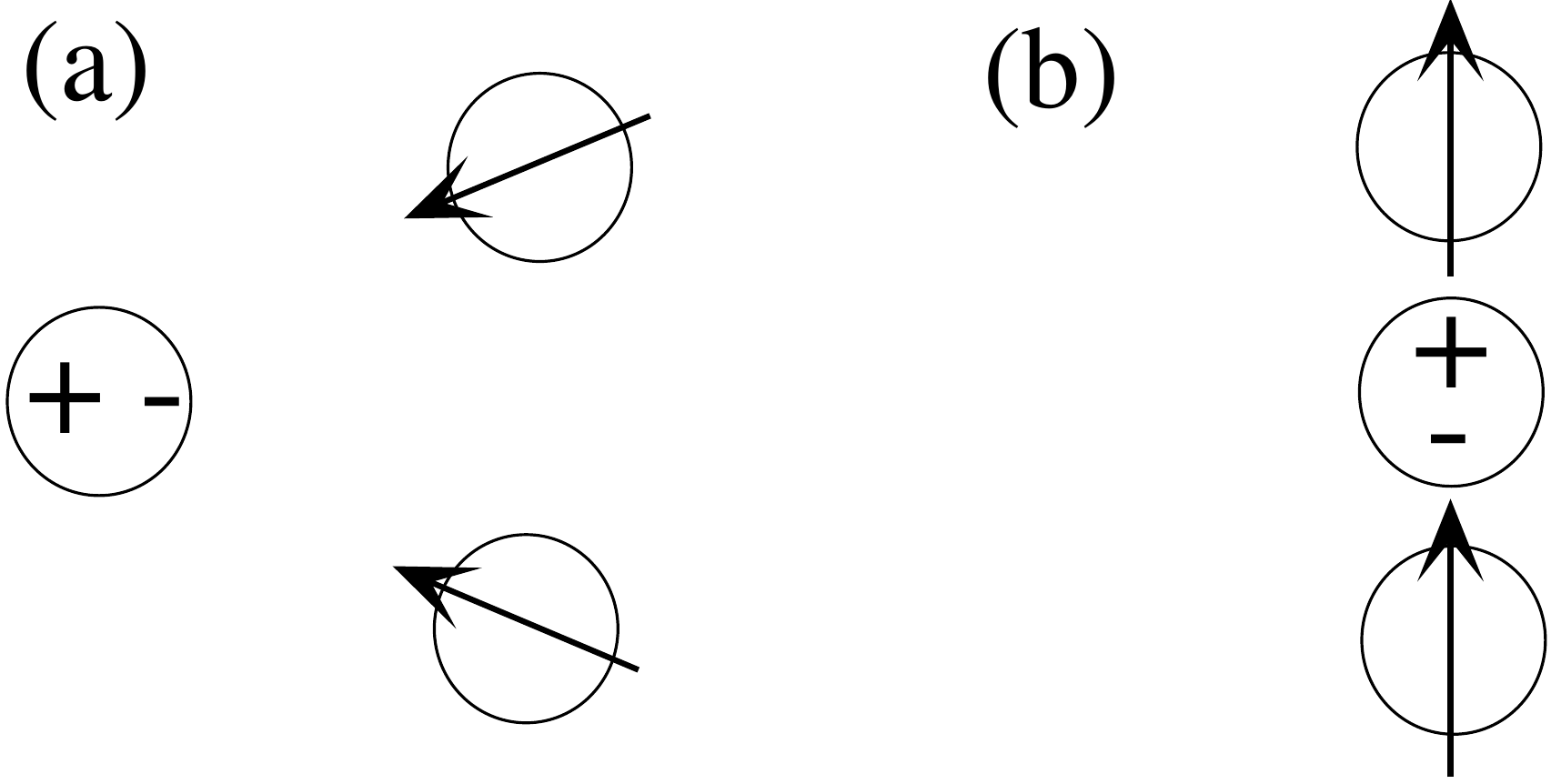}
\end{center}
\caption{\label{fig:atm}
Effect of an external dipole, marked by a positive and negative
sign, on secondary, induced dipoles indicated by arrows.
Depending on the location and orientation of the central dipole,
be it permanent or dynamic,
the  induced dipoles repel (a) or attract (b) each other.
}
\end{figure}

As long as external fields are present, two-body and higher-order interactions generally do not obey Galilean or rotational invariance, which is illustrated in Fig.~\ref{fig:atm}.
It depicts two atoms with full valence shells, which are polarized by the electrostatic field of an external dipole. 
Depending on the relative position of the external dipole to the two atoms, the atoms will experience an  electrostatic repulsion, as in panel (a), or attraction, as in panel (b), in addition to their previous interaction.  

In the absence of external fields, the two-body potential can be reduced
to depend only on the interatomic distances $r_{ij}$, in which case Eq.~\eqref{eq:vExpand} reduces to
\begin{equation}
    U(\{{\bf r}\})  = \sum_{i<j} U_2(r_{ij}) +  \sum_{i<j<k} U_3(r_{ij}, r_{jk}, r_{ki}) + \dots
    \label{eq:vExpandGalilei}
\end{equation}
Expansions of the form given by Eq.~\eqref{eq:vExpandGalilei} are sometimes called \emph{cluster potentials}~\cite{carlsson_beyond_1990,tadmor_modeling_2011}.
Unfortunately, even this Galilean invariant expansion Eq.~\eqref{eq:vExpand} is of limited (direct) use for systems of practical interest, because its convergence can be extremely slow.
This is best seen when considering an ion in close proximity to a metallic surface.
The ion induces a charge distribution in the metal that depends on the shape of the metal surface turning a formal expansion into explicit many-body terms useless. 
Likewise, comparing local structural motives to an existing database, as is done with machine-learned potentials, does not appear to be promising in this example, although they frequently allow to effectively truncate an expansion similar to that outlined in Eq.~\eqref{eq:vExpandGalilei}, however, with atoms not being in vacuum but surrounded by other atoms. 

Equation~\eqref{eq:vExpandGalilei} converges relatively quickly only for closed-shell atoms, meaning atoms whose valence shell is filled, like noble gas atoms and singly charged alkali cations or halogen anions~\cite{Sangster1976AP}. 
Even then convergence might not be truly impressive.
Two-body interactions commonly used to simulate condensed phases of noble gas atoms are \textit{effective} pair potentials rather than \textit{true} pair potentials accurately reproducing the second virial coefficient~\cite{Barker1968AJC}.
The difference between effective and true potentials compensates to some extent for omitted many-body effects.
Thus, pair and three-body potentials trained exclusively in the bulk risk to be inaccurate when applied to surfaces and gases. 
Despite its frequent ineffectiveness, Eq.~\eqref{eq:vExpandGalilei} allows the difference between two opposite philosophies to the construction of potential function to be discussed: the bottom-up and the top-down design, which would lead to the construction of non-empirical and empirical potentials, respectively. 

In the bottom-up approach, pen-on-paper quantum chemistry, density-functional theory (DFT), or any related electronic structure calculations provides analytical results and/or small-scale data, such as forces on individual atoms as a function of the atomic configurations $\{\mathbf{r}\}$ or energies of diatomic molecules as a function of bond length.
Such information can motivate the functional form of a two-body potential or be used to construct soulless look-up tables. 
From a philosophical point of view, it could be argued that inputting experimental information on dimers, trimers, etc., falls into the bottom-up approach, as the electronic structure problem is solved by nature rather than by computers.  
Of course, the downside of letting nature do that job is that it is quite difficult to get enough experimental data on small molecules and clusters that would allow the higher-order terms in the series of Eq.~\eqref{eq:vExpandGalilei} to be determined to any significant accuracy.
Only pair potentials can probably be deduced to high precision from measurements of the vibrational and rotational spectra without having to make serious restrictions on their functional form.

In the top-down design, no atomic-scale information is provided, but instead collective, typically macroscopic properties, such as elastic properties, equations of state (EOS), surface energies, or the temperature dependence of the specific heat including heats of melting and evaporation.
In the sense of an inverse problem, or reverse coarse-graining, potential energy functions are constructed such that the available information  is reproduced. 
Historically, the first attempts to design potentials followed this approach.
The arguably most systematic top-down design makes use of the virial expansion~\cite{Thiesen1885AP,Masters2008JPCM}, in which the pressure $p$ of a many-particle system in thermal equilibrium and in the absence of external fields is expanded  into a power series of the number density $\rho$,
\begin{equation}
p = \rho k_\textrm{B} T \left\{
 1 + B_2(T)\rho + B_3(T) \rho^2 + \cdots \right\},
\end{equation}
where $k_\textrm{B}T$ is the thermal energy, while
$B_2(T)$ and $B_3(T)$ denote the second and third virial coefficient,
respectively.
The so-called cluster expansion~\cite{Mayer1941JCP,Chandler1972JCP} allows the various virial coefficients
to be related to two-body, three-body, and higher-order
interactions.
For example, knowledge of the second virial coefficient
\begin{eqnarray}
B_2(T) & = & - \frac{1}{2} \int\! \mathrm{d}^3r\, 
\left[\exp\{-\beta U_2(r) \}-1\right]
\label{eq:secondVirial} 
% \\ B_3(T) & = & ...,
\end{eqnarray}
allows the pair potential to be reconstructed by an inverse procedure, though a numerically stable inversion works best if $U_2(r)$ is represented as  a function with few adjustable parameters. 
The third virial coefficient, $B_3(T)$, depends on two- and three-body interactions, and so on.
Thus, in principle, two substances may have similar second but different third virial coefficients so that knowledge of $B_3(T)$ allows deviations from pair additivity to be quantified and adjustable parameters of a three-body potential $U_3(r_{12},r_{23}, r_{31})$ to be gauged. 

Unfortunately, the virial expansion does not converge for every state point, due to the existence of thermodynamic discontinuities a.k.a. phase transformations~\cite{Lebowitz1964JMP}.
This is one reason why the expansion of Eq.~\eqref{eq:vExpandGalilei} cannot be fully 
parametrized from a cluster expansion after all. 
Yet, early calculations performed by van der Waals in the spirit of a cluster expansion, 
revealed that the leading-order corrections to the ideal-gas
EOS requires the two-particle interaction in simple
gases to obey~\cite{vanDerWaals1873PhD}
\begin{equation}
U_2(r) =- \frac{C_{6}}{r^6},
\label{eq:dispersInter}
\end{equation}
at large distances.
Today, $C_6$ is called the dipole-dipole dispersion coefficient. 

About 50 years after van der Waals, London~\cite{London1930ZFP} rationalized why atoms and molecules with closed valence shells attract each other through pairwise additive $1/r^6$ interactions by coupling the quantum-mechanical ground state fluctuations of their dipoles in the far-field approximation. 
Extending London's treatment to higher-order electrostatic multipoles and beyond second-order perturbation theory leads to refinements of the theory, which are outlined in Sect.~\ref{sec:closed-shell}. 
Despite the existence of corrections, the bottom-up approach reveals quite clearly that the exponent in Eq.~\eqref{eq:dispersInter} is essentially exact,
at least as long as electrostatic interactions can be treated as instantaneous (or the speed of light as infinitely large)~\cite{Israelachvili1972-ju,Israelachvili1972-qc}.
Tweaking the exponent to better match reference data would quickly result in overfitting of the potential:
a better match of the reference data would deteriorate the description of interatomic forces between two isolated particles at large distances. 
Thus, bottom-up and the top-down design of interatomic potentials are complementary to each other
but should converge to similar results assuming sufficient and sufficiently accurate input.

\subsection{Consequences of many-body interactions}
\label{sec:consequences}

A common way to assess the relevance of many-body interactions is to determine what percentage of the cohesive energy is related to the exact or to an effective two-body interaction. 
Such an analysis may be misleading, because  defect energies or elastic properties may not be described well even if the total energy of a crystal or the radial distribution function of a liquid reproduces \textit{ab-initio} or experimental data.
We can always adjust a pair potential that reproduces the cohesive energy of a specific crystalline phase (discussed extensively below) or that fits a specific liquid pair distribution function (see e.g. Refs.~\cite{Reith2003-av,Lyubartsev1995-na}).
These properties often benefit from the annihilation of or from the insensitivity to many-body contributions, but the resulting potentials are then not \emph{transferable} between crystal structures, temperatures, or other states of the system.
Even three-body correlation functions may be poor properties on which to gauge potentials.
Liquid copper and liquid argon just above their crystallization temperature both assemble in structures similar to that of random-sphere packing~\cite{Sukhomlinov2020JCP}, despite their potentials being utterly different. 

Induced dipoles or charge transfer decrease the total energy (and hence make the ``bonds'' stronger), or they would not occur. 
In contrast, most other many-body interactions weaken bonds, i.e., the energy per bond in metals decreases with increasing coordination number $Z$, typically with $1/\sqrt{Z}$ for metals and even more quickly for covalently bonded systems (see Sect.~\ref{sec:open-shell} or Refs.~\cite{Ducastelle1970JDP,abell1985PRB,Heine1991-mw}).
This scaling has important consequences not only on what crystal structure is assumed but also on the relative increase of boiling relative to the melting temperature, which is roughly 4\% for argon but 100\% for copper. 

\begin{figure}[htb]
\begin{center}
\includegraphics[width=0.75\textwidth]{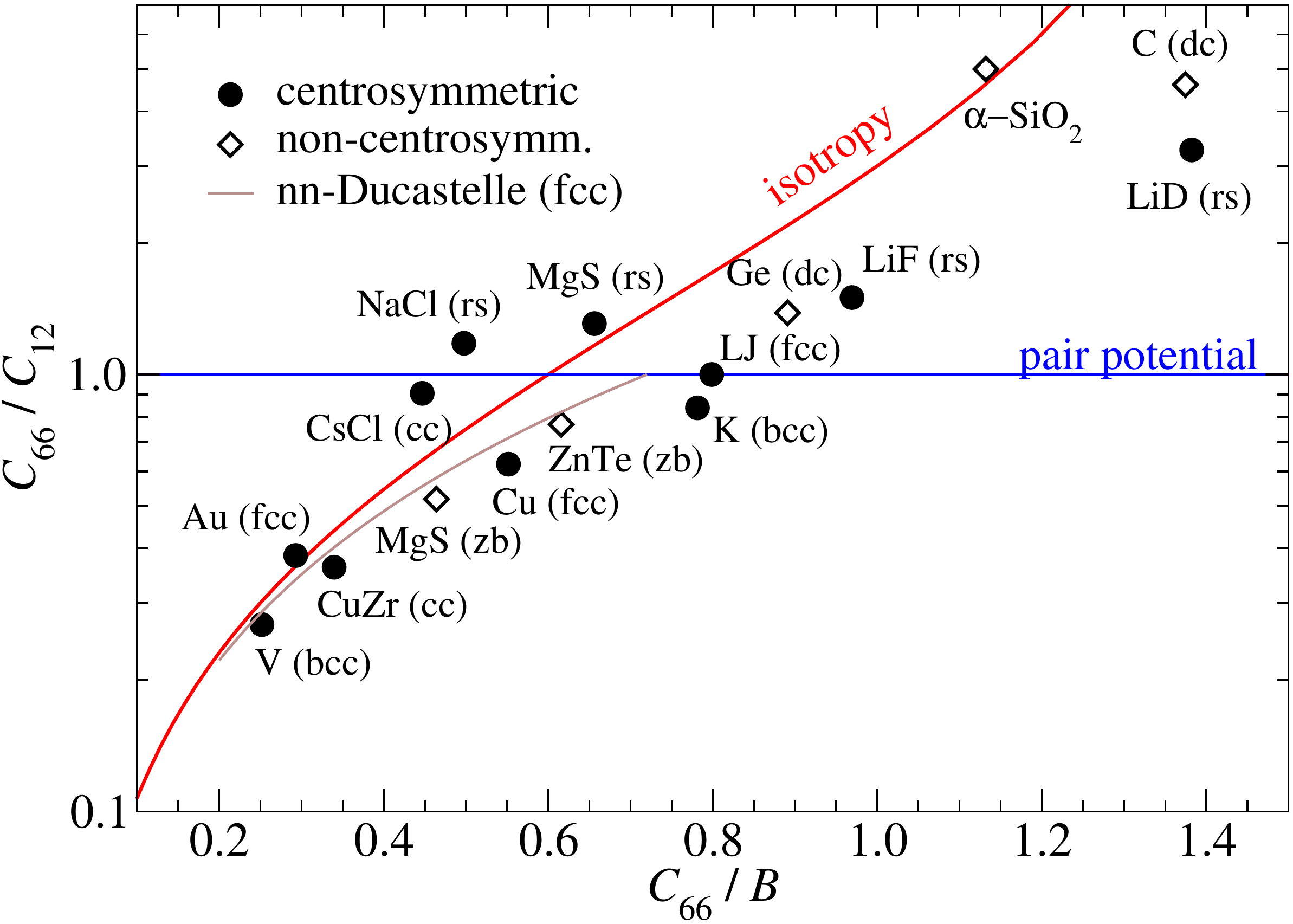}
\end{center}
\caption{\label{fig:cauchyRelation}
Ratio of elastic tensor elements $C_{66}$ and $C_{12}$ as a function of $C_{66}$ in units of the bulk modulus $B$ for a variety of cubic systems. ($C_{66} \equiv C_{44}$ in cubic systems.)
The blue line holds for pair potentials, while the red line reflects elastically isotropic materials.  
The thin gray line assumes the generic model proposed by Ducastelle, introduced in Sect.~\ref{sec:eam}, which has one dimensionless parameter affecting the $C_{66}/C_{12}$ and $C_{66}/B$ ratios in the nearest-neighbor approximation. 
Centrosymmetric lattices (closed circles) include face-centered cubic (fcc), body-centered cubic (bcc), rocksalt (rs), and caesium chloride (cc), while diamond cubic (dc) and zinc blende (zb) lack inversion symmetry (open diamonds). 
Experimental and simulated data are included for selected crystals (Cu~\cite{Simmons1971}, Au~\cite{Simmons1971}, K~\cite{Simmons1971}, V~\cite{Simmons1971}, C~\cite{Simmons1971}, Ge~\cite{Simmons1971}, NaCl~\cite{Simmons1971}, LiD~\cite{James1982JPC}, LiF~\cite{Simmons1971}, MgS(rs)~\cite{Fu2010PSS}, MgS(zb)~\cite{Fu2010PSS}, ZnTe~\cite{Lee1970JAP}, CsCl~\cite{Slagle1967JAP}, $\alpha-$SiO$_2$~\cite{Carpenter1998AM}) as well as calculated elastic constants for CuZr~\cite{Du2014}.
}
\end{figure}

Another frequent consequence of many-body terms is that bond lengths become shorter with decreasing coordination or increasing bond order. 
One of the best known example may be the bond length between two carbon atoms:
$a_0 \approx 1.54$~\AA 
($Z_\textrm{H} = 3$, ethane, C$_2$H$_6$),	
$1.34$~\AA ($Z_\textrm{H} = 2$, ethylene, C$_2$H$_4$)	
$1.20$~\AA ($Z_\textrm{H} = 1$, acetylene, C$_2$H$_2$), where $Z_\textrm{H}$ states how many hydrogen atoms each carbon is bonded to. 
Similarly, the spacing between layers near a free metal surface contract~\cite{Gupta1981PRB}, while common short-range pair potentials would predict them to expand. 
Historically, the non-additivity of atomic potentials lead to the dismissal of the atomic hypothesis among many prominent scientists in the 19'th century, see Ref.~\cite{Capecchi2010AHES} for a well written and very enlightening account of the debate.
The assumption was that atoms, so they exist, should only interact through central potentials and be pairwise additive.
Cauchy and Poisson~\cite{Capecchi2010AHES} and also potentially St. Vernan~\cite{Love1927book} had shown that this reduced the number of independent elastic tensor elements, i.e., from 21 to 15 for a triclinic crystal and from three to two for a cubic crystal.
Specifically, they found, see  Sect.~\ref{sec:Cauchy} for a derivation, that any permutation of the four indices of an elastic tensor element should leave it unchanged so that relations, now known as Cauchy relations, like, for example, $C_{1212} = C_{1122}$, or, $C_{12}=C_{66}$ (in cubic systems $C_{44}=C_{66}$) in Voigt or Nye notation, would hold.
It then came as a blow to the atomic hypothesis when
Voigt noticed that none of the crystals he had investigated satisfied the Cauchy relation within experimental errors~\cite{Voigt1910Book}.
Figure~\ref{fig:cauchyRelation} epitomizes results on the violation of the pair-potential assumption.
The latter is also frequently stated in terms of the Cauchy pressure $P_\textrm{C} = (C_{12}-C_{44})/2$, which vanishes for athermal, inversion-symmetric, classical crystals interacting with pair potentials.  

Rather than abandoning the assumption of the pairwise additivity of potentials, many scientists rejected the atomic hypothesis alltogether.
However, Rutherford's scattering experiments in the early 20'th century removed all legitimate doubt about it, even if quite a few muddleheads keep taking issue with it up to this day in happy concert with deniers of human-made climate change. 
Finally, Born~\cite{Born1954book} reconciled the properties of the elastic tensor of real materials with the atomic hypothesis by showing that one (of several) reasons for the breakdown of Cauchy relations is the existence of many-body interactions.

It should certainly not be concluded that solids obeying the Cauchy relations automatically satisfy the pair-potential assumption, since there can be fortuitous or symmetry-induced cancellation of many-body effects.
For example, the dipole polarizability of anions in the rocksalt structure cannot reveal itself in the elastic tensor for symmetry reasons.

\section{Two-body potentials}
\label{sec:pop2BPs}
In this section, we introduce the most generic two-body potentials for pairs of atoms forming covalent and metallic bonds in the condensed phase as well as potentials for ionic and van der Waals interactions.
We discuss these two classes separately, because atoms with open electron shells bond primarily through the sharing of electrons while closed-shell atoms interact predominantly through van der Waals or ionic forces. 
Both classes reflect Feynman's mantra that \textit{[atoms attract] each other when they are a little distance apart, but repel upon being squeezed into one another}~\cite{Feynman1964Book}.
Interatomic potentials with well-motivated functional forms can reproduce the equation of state, reasonably well if their small number of adjustable parameters are gauged on a few accurate reference values. 
However, parameters are only transferable to other structures or properties if the pair-potential assumption is justified. 

\begin{figure}[hbtp]
    \centering
    \includegraphics[width=0.475\textwidth]{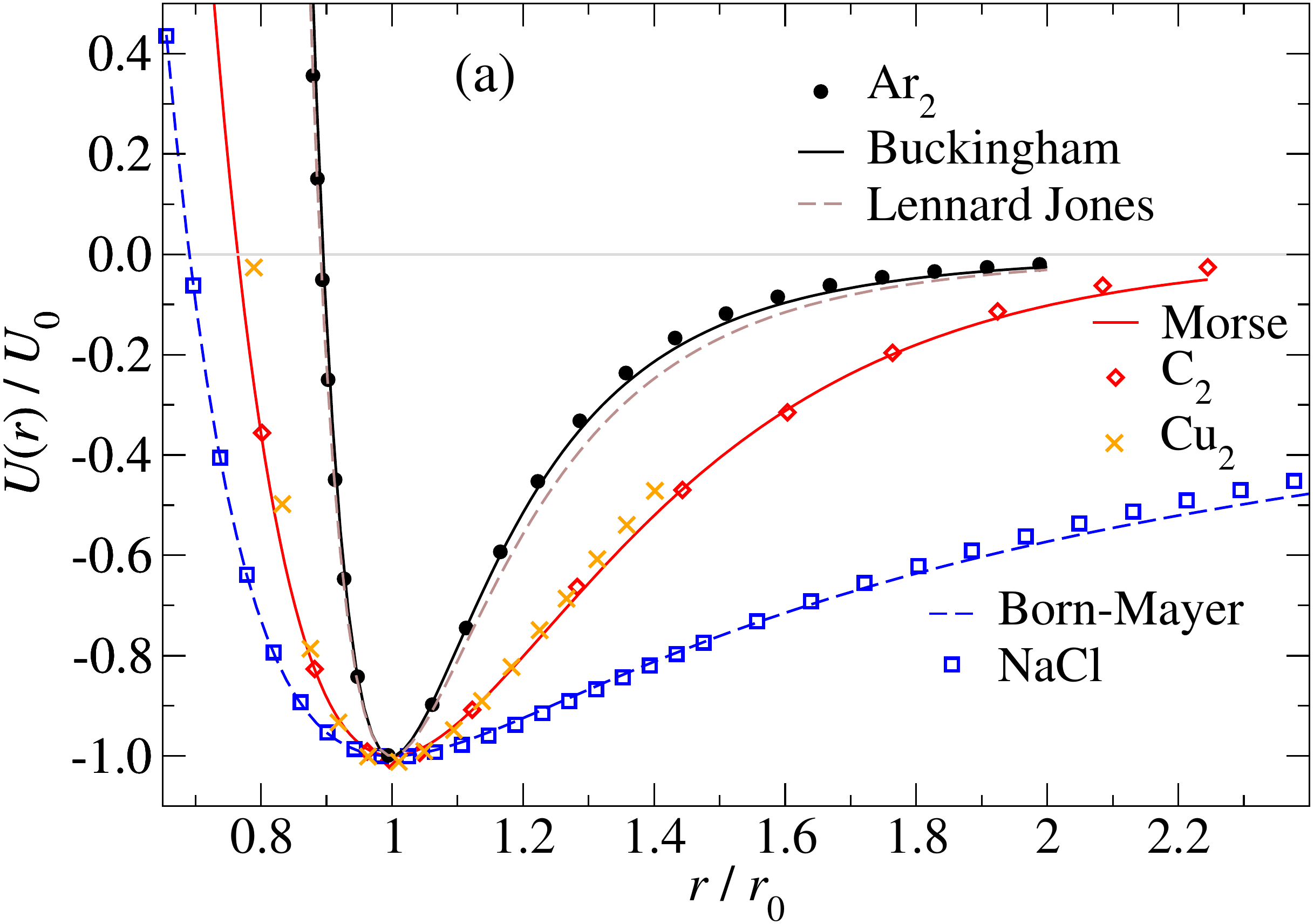}\hspace*{3mm}
    \includegraphics[width=0.475\textwidth]{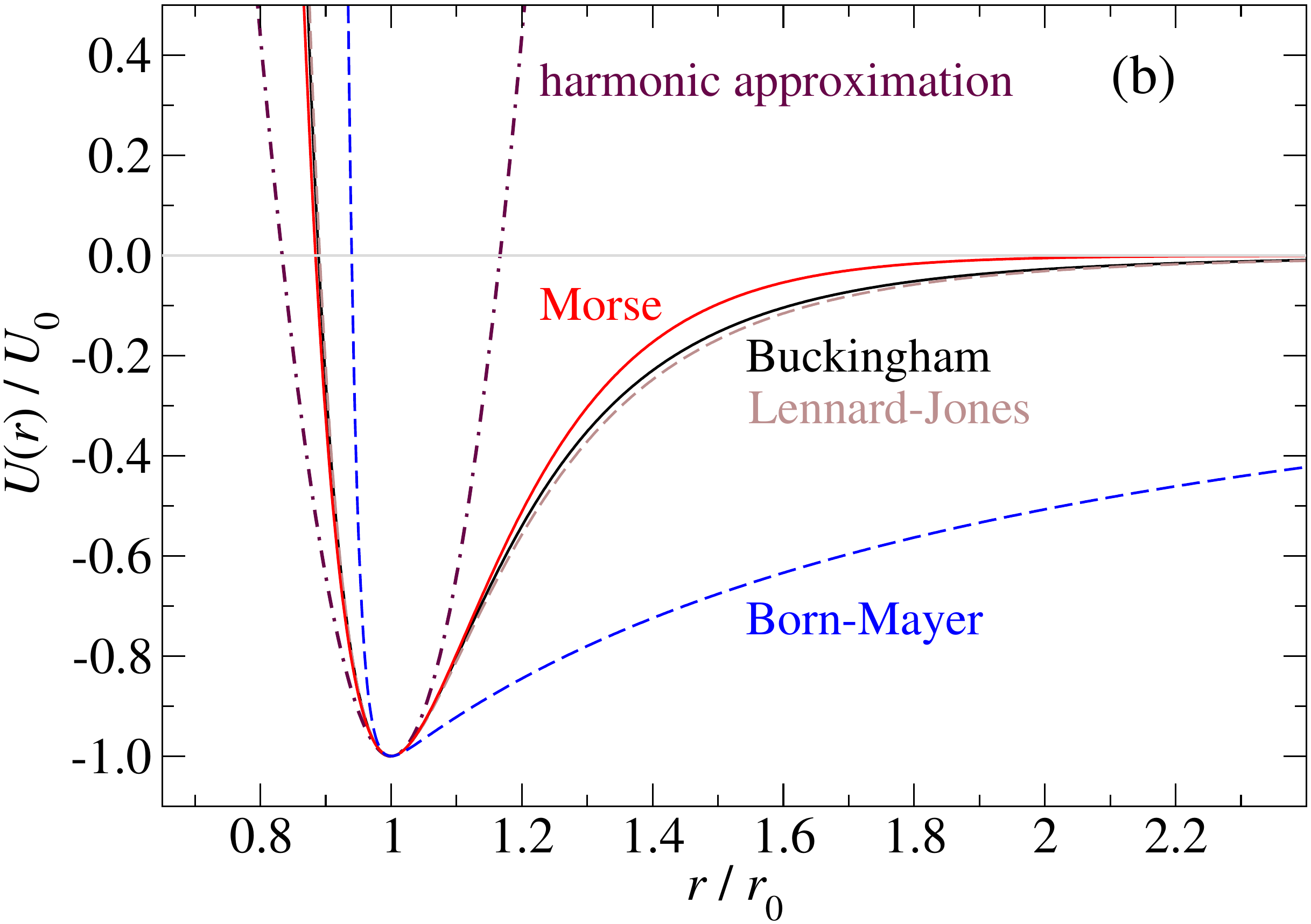}
    \caption{\textbf{(a)} Two-body potentials of different diatomic molecules, $U(r)$, normalized to the binding energy $U_0$ as a function of the bond length $r$ in units of the equilibrium bond length $r_0$. 
    Symbols reflect high-accuracy reference data, while lines show fits of pair potentials introduced in Sect.~\ref{sec:pop2BPs} to the energy minimum. 
    Reference data include the Aziz potential for Ar$_2$~\cite{Aziz1986MP}, and quantum chemical  calculations for C$_2$~\cite{Booth2011JCP}, Cu$_2$~\cite{Mishin2001PRB}, and NaCl~\cite{Dapp2013EPJB}.
    Numbers used for the fits are listed in Table~\ref{tab:compMolSolParam}.
    \textbf{(b)} Two-body potentials with a curvature in the minimum corresponding to that of the Lennard-Jones potential. 
    \label{fig:allPot}
    }
\end{figure}

An element specific analysis of the potentials will be made for carbon (C), copper (Cu), sodium chloride (NaCl), and argon (Ar), which are representative of covalent, metallic, ionic, and van-der-Waals bonding, respectively.
To set the stage for further discussion, their pair-potential energies are shown in Fig.~\ref{fig:allPot} as a function of the molecular bond length $r$. 
The reference data shown therein is not necessarily the most accurate on the market, but their overall errors should be minor compared to those of lean pair potentials, i.e., the data should be sufficiently accurate to test if a given functional form of the two-body potential is justified.

\subsection{Morse potential}
\label{sec:Morse}
\begin{figure}
    \centering
    \includegraphics[width=0.75\textwidth]{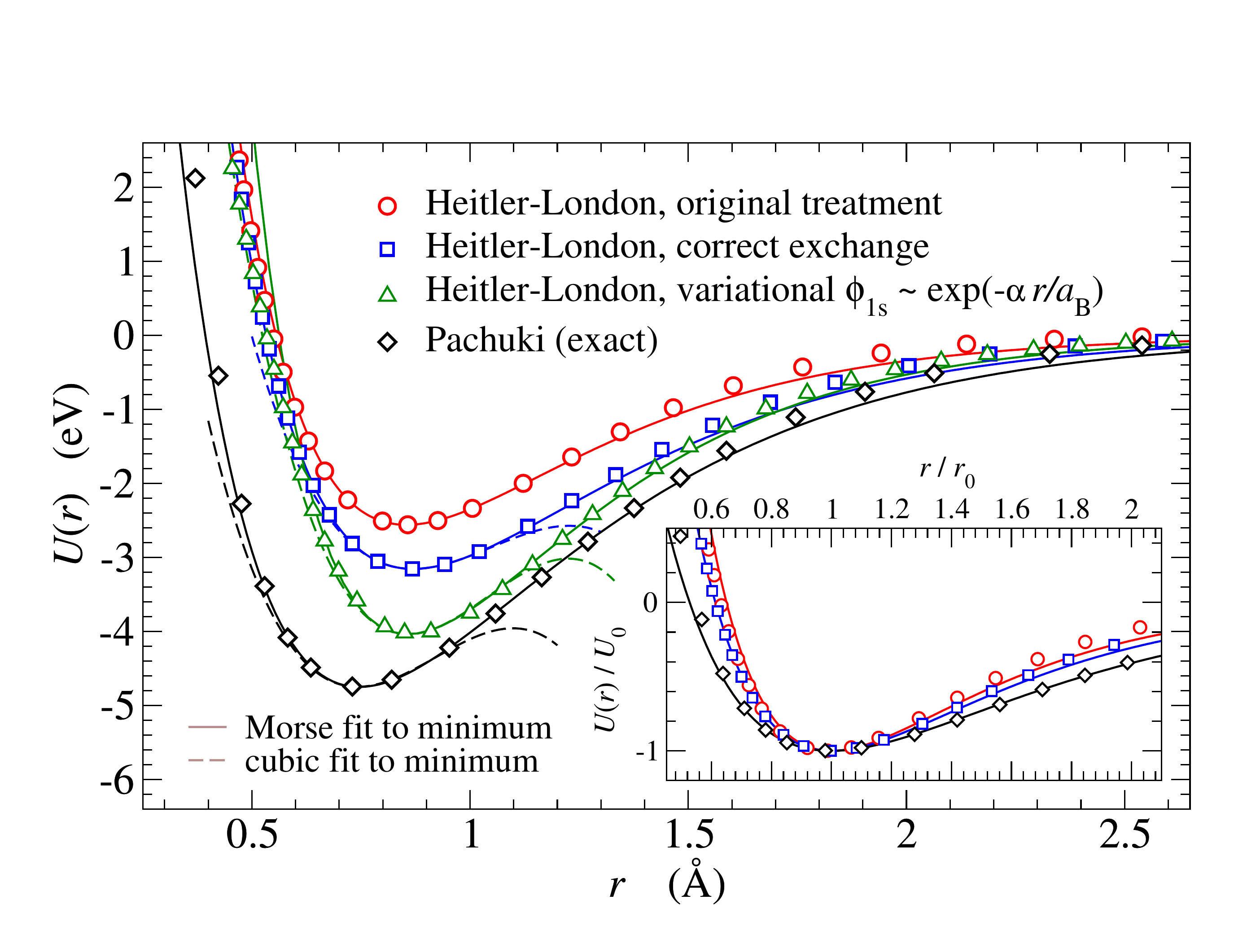}
    \caption{Binding energy assuming different quantum chemical approaches to the binding in H$_2$. Full lines are fits to that data near the minimum assuming the functional form proposed by Morse. Dashed lines represent cubic fits to the minimum. The inset shows the same data but normalized to the respective equilibrium bond lengths and binding energies. 
    }
    \label{fig:VBvsMOvsMorse}
\end{figure}

One of the early, great successes of quantum mechanics was Heitler and London's quantitative description of the chemical bond in a hydrogen molecule in 1927~\cite{heitler_wechselwirkung_1927}.
Their theory is text-book material and will only be sketched here.
Heitler and London linearly combined the atomic wave functions of the two hydrogen atoms forming a H$_2$ molecule, antisymmetrized the spin-wave function to reflect the Fermi principle, and evaluated the binding energy using first-order perturbation theory, thereby providing a lower bound for the molecular binding energy $U_0$. 
Sugiura~\cite{sugiura_uber_1927} succeeded in identifying the correct solution for the exchange-energy integrals. Errors of order  40\% for $U_0$ and bond length $a_0$ remain, see Fig.~\ref{fig:VBvsMOvsMorse}.
Treating $a_\textrm{B}$ in the electronic ground-state wave functions $\Phi\propto\exp(-r/a_\textrm{B})$ as a variational parameter, errors can be reduced by
a factor close to two.

Morse~\cite{Morse1929PR} approximated the analytical expressions occurring in the Heitler-London treatment with
\begin{equation}
\label{eq:Morse}
    U(r) = U_0 \left( e^{n\,(1-r/r_0)} - 2\, e^{n\,(1-r/r_0)/2} \right),
\end{equation}
where $U_0$ is the molecular binding energy, also called dissociation energy, $r_0$ the equilibrium bond length,
and $n$ a dimensionless number. 
Historically, Morse did not introduce the parameter $n$ but rather $a = n/(2\,r_0)$, which is still commonly used today.
Moreover, he expressed the energy relative to the ground state, $\Delta U(r) \coloneqq U(r)-U_0$, 
\begin{equation}
\label{eq:Morse2}
    \Delta U(r) = U_0 \left(1 - e^{-a(r-r_0)}\right)^2.
\end{equation}
To ease comparisons between different potentials and to yield $U(r\to \infty) = 0$, we choose forms similar to Eq.~\eqref{eq:Morse} throughout this article. 

Morse~\cite{Morse1929PR} recognized that the vibrational levels of many diatomic molecules, both homo- and hetero-nuclear, could be described quite accurately when treating $U_0$, $r_0$, and $n$ as adjustable parameters.
The parameter $n$ allows the curvature of $U(r)$ in the minimum to be adjusted independently from the ratio $U_0/r_0^2$, however, no additional fine tuning of higher-order derivatives is possible.
Given its simplicity, the Morse potential approximates reference data impressively well, as can be appreciated in Fig.~\ref{fig:allPot} for Cu$_2$ and C$_2$.
It even represents the curves obtained for H$_2$ at different accuracy levels at which the hydrogen molecule is described and substantially better so than cubic fits to the energy minimum, see Fig.~\ref{fig:VBvsMOvsMorse}.. 
Of course, there are also exceptions, where the Morse potential fails, e.g., Cr$_2$, which is described in Sect.~\ref{sec:funky}.

An important result of the Heitler-London treatment is that short-range repulsion 
between atoms increases (approximately) exponentially with decreasing distance as
two atoms approach each other.
All highly-accurate potentials use repulsion that does not stray too far away
from such a dependence as the pressure of (simple) bulk systems increases approximately exponentially at high compression with pressure for all bonding classes, at least as long as external pressures do not substantially exceed the bulk modulus $B$, which we will get back to in Sect.~\ref{sec:lattice-props}.

The simplest generalization of the Morse potential is a double exponential~\cite{abell1985PRB} of the form 
\begin{equation}
\label{eq:genMorse}
    U(r) = \frac{U_0}{n-m} 
    \left\{
     m \,e^{n\left( 1-{r}/{r_0}\right)} - 
     n \, e^{m\left( 1-{r}/{r_0}\right)}
    \right\},
\end{equation}
which reduces to Morse if the additional parameter $m$ is set to $n/2$.
As Abell~\cite{abell1985PRB} noted, ``\textit{this choice is
based in part on analytical convenience, but also on the
physical grounds that atomic orbitals decay exponentially
with $r$}''.
However, these long-range asymptotics are irrelevant at very small separations between nuclei, as briefly discussed in Sect.~\ref{sec:shortRangeCorr}.

\subsection{Mie/Lennard-Jones and Born-Mayer/Buckingham potentials}
\label{sec:dispersionIonic}
The so-called Lennard-Jones potential~\cite{LennardJones1931PPS} is the arguably most used potential in molecular simulations.
It describes the interactions between entities with closed-electron shells quite well. 
This includes not only interactions between noble gas atoms, but also the intra- and intermolecular forces between, say, two CH$_2$ repeat units in polyethylene as long as the two units are not directly connected by a covalent bond. 
The functional form of the regular LJ potential is 
\begin{equation}
\label{eq:LJ1}
    U(r) = U_0 \left\{
    \left(\frac{r_0}{r}\right)^{12} -
    2\,  \left(\frac{r_0}{r}\right)^{6}
    \right\}.
\end{equation}
Here $U_0$ is the binding energy and $r_0$ the equilibrium bond length.
The more common way to write the LJ potential is
\begin{equation}
\label{eq:LJ2}
    U(r) = 4 \epsilon \left\{
    \left(\frac{\sigma}{r}\right)^{12} -
     \left(\frac{\sigma}{r}\right)^{6}
    \right\},
\end{equation}
where $\epsilon \coloneqq U_0$ and $\sigma := r_0 /\sqrt[6]{2}$.
However, comparison to other potentials and analytical manipulations are more easily done when starting from Eq.~(\ref{eq:LJ1}). 

The attractive $1/r^6$ term is the leading-order dispersive interaction.
The exponent can be motivated rigorously from perturbation theory, 
as sketched in Sect.~\ref{sec:QDO}.
In contrast, the exponent from the repulsive $1/r^{12}$ term is a mixture of convenience and good luck. 
It results from the squaring of the $1/r^6$ term, which was particularly beneficial at times when numerics was not yet done by machines but by humans.
This facilitated the work of those who were dismissively called \textit{Rechenknechte} by theoretical physics professors in Germany in the beginning of the 20'th century. The term translates to \textit{computing or arithmetic servants} and is nowadays used for computers.
However, when treating the exponent as a fit parameter in a generalized Lennard-Jones potential
\begin{equation}
\label{eq:LJ-n6}
    U(r) = \frac{U_0}{n-6} \left\{
    6 \left(\,\frac{r_0}{r}\right)^{n} -
    n\,  \left(\frac{r_0}{r}\right)^{6}
    \right\} \mbox{ with } n>6,
\end{equation}
the exponent $n$ generally turns out close to its generic value of $n = 12$. 

Replacing the exponent $6$ in Eq.~\eqref{eq:LJ-n6} with a number $0<m<n$ yields the Mie potential~\cite{Mie1903ADP}.
The regular LJ potential can therefore be regarded as a limiting case of Mie's functional form, a so-called 12-6 Mie potential.
In contrast, it may not be entirely appropriate to downgrade a Mie potential with arbitrary exponents as a generalized LJ potential,
since Mie's work~\cite{Mie1903ADP} predated those of Lennard-Jones~\cite{Jones1924PRSL,LennardJones1931PPS} by more than two decades and in fact, it is not clear to us who was the first to use $n = 12$ as exponent in the repulsion.
Jones, later known as Lennard-Jones, assumed it to be $n = 15$~\cite{Jones1924PRSL} for argon. 

Rather than tweaking the exponent of a Mie potential, it is probably more meaningful to replace it with an expression that somehow accounts for the Pauli repulsion between closed electron shells, whose density decreases as an exponential function of the interatomic distance. 
Slater~\cite{Slater1928PR} was the first to identify such a dependence for helium dimers by using a similar approach as Heitler and London~\cite{heitler_wechselwirkung_1927}.
The functional form he identified as being most appropriate for a limited range of interatomic distances was a single exponential, as in the  Morse potential.
Later, Born and Mayer~\cite{Born1932ZFP} used the same functional form for the description of repulsion in simple ionic crystals without further theoretical justification but merely by arguing that it fits experimental results better than an inverse power as in Mie repulsion~\cite{Born1932ZFP}.
Buckingham~\cite{Buckingham1938PRSL} made a similar observation for noble-gas solids. 
Using exponential repulsion and $1/r^m$ attraction then leads to the two-body potential
\begin{equation}
    U(r) = \frac{U_0}{n-m}\,
    \left\{ m \exp\left[n\left(1-\frac{r}{r_0}\right)\right]
    - n\,  \left(\frac{r_0}{r}\right)^{m}
    \right\},
\end{equation}
which is often called Buckingham potential ($m = 6$) but also Born-Mayer potential ($m=1$) for two ions of dislike charges, in which case the potential parameters must be constrained to satisfy
\begin{equation}
    \frac{n\,U_0 r_0}{n-1} = \frac{Q^2}{4\pi\varepsilon_0},
    \label{eq:MieConstraint}
\end{equation}
where $Q$ is the charge of one ion and $\varepsilon_0$ the vacuum permittivity. 
Born-Mayer potentials may also contain dispersive interactions in addition to exponential repulsion and Coulomb potentials. 
For two like ions, there is no bound state and repulsion is dominated by Coulomb interactions at typical interionic distances. 

Later, Buckingham~\cite{Buckingham1958TFS} found that the repulsion between hydrogen atoms in the lowest triple state and that between helium atoms are better described when replacing the prefactor of the exponential term with an appropriate polynomial.
He motivated this with theoretical results obtained using Heitler-London type approaches, which is in line with more recent comparisons~\cite{OConnor2015JCP}.
Thus, the term Buckingham potential may also refer to a potential, in which the constant prefactor to the exponential repulsion is replaced  with a polynomial in $r$.
Van Vleet \textit{et al}.~\cite{VanVleet2016JCTC} provide a contemporary discussion on how to construct the polynomials.

\subsubsection{Higher-order dispersion and induction}
\label{sec:dispCorr}

Dispersive interactions are not limited to the leading-order dipole-dipole interactions, yielding the $1/r^6$  attraction. 
For example, dipole-quadrupole interactions cause a correction of
\begin{equation}
    \Delta U(r) = -\frac{C_{8}}{r^8},
\end{equation}
while the quadrupole-quadrupole and the dipole-octopole interactions lead to corrections that asymptotically scale as $1/r^{10}$.
Cipcigan \etal~\cite{Cipcigan2016JCP} recently summarized the hierarchy of dispersive interactions together with those resulting from electrostatic induction.

The most important correction to Coulomb interactions in ionic systems results from the large electrostatic, dipolar polarizability of anions, which was first considered by Rittner~\cite{Rittner1951JCP}.
The energy gained by a dipole placed in an external electrostatic field $\mathbf{E}$ is $\Delta U_1 = -\mathbf{p}\cdot\mathbf{E}$, while the on-site energy required to create it is $\Delta U_1 = p^2/(2\alpha)$, where $\alpha$ is the polarizability of the anion.
The dipole adjusts itself so that $\Delta U_1 + \Delta U_2$ is minimized. 
This yields a correction of 
\begin{equation}
\label{eq:Rittner}
    \Delta U(r) = -\frac{\alpha'\,Q^2}{8\pi\varepsilon_0\, r^4}
\end{equation}
to the Born-Mayer potential applied to a heteronuclear, diatomic molecule, where $\alpha' \coloneqq \alpha/(4\pi\varepsilon_0)$.
Assuming bare Coulomb interactions, the polarizability of the cation is assumed to be negligible compared to that of the anion.
As already recognized by Rittner~\cite{Rittner1951JCP}, corrections to the correction in Eq.~\eqref{eq:Rittner}, arise due to the induction of the cation and the mutual induction of cation and anion.
In addition, electrostatic field gradients induce quadrupoles and higher-order derivatives higher-order multipoles.
Their incorporation may necessitate damping of the interactions~\cite{Tang1984JCP,Cipcigan2019RMP} or further fine tuning of the on-site interaction between different multipoles beyond linear response to ensure a systematic improvements of the potential. 

Rittner~\cite{Rittner1951JCP} found that adding this type of correction allowed him to reproduce important molecular properties of alkali halide molecules while assuming integer charges on the ions. 
He deemed Pauling's criterion for the fraction of ionic character of a bond (dipole moment divided by bond length times elementary charge) false, because it fails to include the far-from-negligible polarization deformation on the ions. 
Madden and Wilson~\cite{Madden1996CSR} provided much ammunition in favor of Rittner's conclusion by also considering crystalline structures of many ionic binaries. 

It is important to note that the Rittner and related higher-order corrections are \emph{not} pairwise additive.
This is because an additional charge or charge distribution would change the induced dipole and thereby its interaction with the first charge. 
As a consequence, polarizability in many-atom systems is generally not solved in closed form, as described in more detail in Sect.~\ref{sec:polarizable}.

\subsubsection{Damping interactions at short distances}
\label{sec:shortRangeCorr} 

Obviously, higher-order dispersion and induction only matters significantly at small interatomic distances at which point the charge distributions of the interacting atoms start to overlap, which leads to a reduction or damping of the interactions.
To prevent artificially large attractions from occurring at small separations, pertinent terms of the potentials are multiplied with damping functions $f_n(r)$. 
They are generally constructed such that the overall potential goes linearly to zero with $r$ at small $r$, while the damping function quickly approaches unity with increasing $r$.
To this end, Tang and Toennies~\cite{Tang1984JCP} proposed to replace a dispersive interaction scaling with $1/r^n$ according to
\begin{equation}
\label{eq:dampingFctn}
    \frac{1}{r^n} \to \frac{1}{r^n} f_n(r/b),
\end{equation}
where $b$ is a parameter which depends on the nature of the two interacting atoms and potentially also on the index $n$, while
\begin{equation}
    f_n(x) = 1 - \frac{\Gamma(n+1,x)}{n!},
\end{equation}
$\Gamma(n+1,x)$ being the incomplete Gamma function, which, in the case of a non-negative integer $n$, can be expressed as
\begin{equation}
    \Gamma(n+1,x) = n!\, e^{-x} \sum_{k=0}^n \frac{x^k}{k!}.
\end{equation}
The Tang and Toennies damping function is supposedly the most widely used one. 

Not only dispersive but also regular Coulomb interactions between charges or dipoles can be damped, rewind, \emph{should} be damped with functions like that defined in Eq.~(\ref{eq:dampingFctn}) to reflect the finite extent of electronic shells and to avoid grossly exaggerated attraction at small separation. 
In fact, as early as 1962, Dalgarno~\cite{Dalgarno1962AIP} discussed damping,  or, ``shielding'' as it was called at the time, in the framework of shell potentials accounting for atomic polarizability. 

\subsubsection{Electrostatic screening at large distances}
\label{sec:electrostaticScreening}

When atoms approach each other much more closely than their typical nearest-neighbor spacing, as it can happen during high-energy ion bombardment, it is more appropriate to assume the bare Coulomb repulsion, $U_\textrm{C}(r)= Z_1 Z_2 e^2/(4\varepsilon_0 r)$ between the nuclei as asymptotic reference rather than the exponential repulsion reflecting the large-$r$ asymptotics.
Here, $Z_{1,2}$ are the nuclear charges of the interacting atoms or ions. 
Due to the propensity of matter to be locally charge neutral, electrostatic interactions are screened at ``large'' distances, i.e., at distances approaching equilibrium bond lengths, as is also the case, for example, between charged colloids in an electrolyte. 
To lowest order, the bare Coulomb interaction in a charge-neutral system is screened with an exponential $f(r)=\exp(-r/\lambda)$, where $\lambda$ is called the screening length.
The resulting potential $U(r)=f(r)U_\text{C}(r)$ is known as the Yukawa potential~\cite{Yukawa1935-bd}.

Ziegler, Biersack, and Littmark (ZBL)~\cite{ziegler_stopping_1985} proposed 
an empirical screening function $f(r) = \phi(r/\lambda)$ with
\begin{equation}
    \phi(x) =
    0.1818 e^{-3.2x}
    +
    0.5099 e^{-0.9423x}
    +
    0.2802 e^{-0.4029x}
    +
    0.02817 e^{-0.2016x},
\end{equation}
where only the screening length $\lambda=0.46850~\text{\AA}/(Z_1^{0.23} + Z_2^{0.23})$ depends explicitly on the nuclear charges.
ZBL may still be the most used short-range repulsion, despite reported improvements~\cite{Nordlund1997NIMP}, which, however, lack the mentioning of specific functional forms. 

In practice, it is necessary to switch between short and long-distance asymptotics to predict repulsion at intermediate $r$.
This is commonly done through switching functions as discussed in the appendix of Ref.~\cite{Juslin2005JAP} .

\subsection{Combining rules for two-body potentials}
\label{sec:combiningRules}
% W. L. Bragg, Phil. Mag. 40, 169 (1920).
Bragg~\cite{Bragg1920PMJS} observed that the lattice constants of many crystals can be reproduced quite accurately when atomic radii are assigned to individual elements and the assumption is made that two nearest-neighbors touch in ideal crystalline structures. 
This observation implies no rigorous but an approximate constraint for how (two-body) potentials parameterized for individual elements can be combined to mixed interactions. 
An arsenal of propositions was made in that regard. 
The simplest and most wide spread for the Lennard-Jones potential are the Lorentz-Berthelot~\cite{Lorentz1881AP,Berthelot1898} rules.
They take the arithmetic mean of the length scale parameter $\sigma$, whereby Lorentz predated Bragg's observations by 40 years, and, the geometric mean of the binding energies $\epsilon$~\cite{Lorentz1881AP,Berthelot1898}, expressed here using $\epsilon$ rather than $U_0$,
\begin{eqnarray}
\label{eq:LJcombiningRadius}
\sigma_{AB} & = & \frac{\sigma_A + \sigma_B}{2} \\ 
\epsilon_{AB} & = & \sqrt{\epsilon_A \epsilon_B}.
\end{eqnarray}
The major flaw of Lorentz-Berthelot is that it generally violates a rigorous result for the mixed dispersion coefficient,  $C_{6,AB} \coloneqq 4\epsilon_{AB}/\sigma_{AB}^6$, which is given further below in Eq.~\eqref{eq:sumRuleDisp}.
An apparently reasonable approximation to Eq.~\eqref{eq:sumRuleDisp} for combined dispersion coefficient reads~\cite{Zeiss1977MP}
\begin{equation} \label{eq:MeathDispersionCombo}
    C_{6,AB} \approx \frac{2\,\alpha_A\,\alpha_B\,C_{6,AA}\,C_{6,BB}}
    {{\alpha_B^2}C_{6,AA} + {\alpha_A^2} C_{6,BB}},
\end{equation}
where $\alpha_X$ is the polarizability of atom or ion $X$,
while the geometric mean $\sqrt{C_{6,AA} \,C_{6,BB}}$ merely provides an upper bound for $C_{6,AB}$~\cite{Zeiss1977MP}. 
We abstain from reviewing combining rules not reproducing  Eq.~\eqref{eq:MeathDispersionCombo} or better motivated combining rules, other than Lorentz-Berthelot.
Instead, we content ourselves noting that Tang and Toennis~\cite{Tang2003JCP} found the potential depths and locations of hetero nuclear noble-gas molecules to be well reproduced when using their equations (4) and (5), which satisfy  Eq.~\eqref{eq:MeathDispersionCombo} from this work, 
as combining rules. 
We would expect that reasonable answers can also be obtained for closed-shell (united) atoms when using Eq.~\eqref{eq:MeathDispersionCombo} in conjunction with the Lorentz rule, from where the prefactor for the $1/r^{12}$ repulsion can be deduced. 

A better model for repulsion than that used by Mie~\cite{Mie1903ADP} or Lennard-Jones~\cite{LennardJones1931PPS} is the exponential repulsion from Born and Mayer~\cite{Born1932ZFP}, originally Slater~\cite{Slater1931PR}, which we write here as
\begin{equation}
    U_{ij}(r_{ij}) = A_{ij}\,e^{-r_{ij}/\sigma_{ij}},
\end{equation}
where $A_{ij}$ typically lies in the range of a few to several dozen keV~\cite{Abrahamson1969PR}.
Values for $A$ and $\sigma$ pertaining to our selected reference structure can be read off from Tab.~\ref{tab:compMolSolParam} by associating $A$ with $U_0\exp(n)$ and $\sigma$ with $r_0/n$.
Assuming that repulsion originates from he overlap of exponentially decaying charge densities yields the harmonic means for the length scale,
\begin{equation}
    \sigma_{ij}^{-1} = \frac{\sigma_{ii}^{-1}+\sigma_{jj}^{-1}}{2}.
\end{equation}
A standard assumption, similar to the Lorentz-Berthelot rule, is to apply a geometric mean for the energy prefactors, $A_{ij} = \sqrt{A_{ii}\,A_{jj}}$.
Abrahamson~\cite{Abrahamson1969PR} compiled a list for $1/\sigma_{ii}$ and $A_{ii}$ for all neutral elements up to the atomic number 105.

\begin{table}
    \centering
    \begin{tabular}{|cc|c|c|c|c| }\hline
         & &  $U_0$ (eV) & $r_0$ (\AA)& $n$ & $r_0/n$ (\AA) \\ \hline
         Ar & (mol) & 0.01234 & 3.757 & 14.6\phantom{1} &  0.26$^*$ \\
         Ar & (fcc) & 0.01155 & 3.811 & 14.1\phantom{6} & 0.27$^*$ \\ \hline
         C & (mol) & 6.318\phantom{34} & 1.250 & 5.82 & 0.21$^*$  \\
         C & (dc)  & 2.868\phantom{34} & 1.804 & 5.34 & 0.34$^*$  \\ \hline
         Cu & (mol) & 1.823\phantom{12} & 2.275 & 6.80 & 0.33$^*$  \\
         Cu & (fcc) & 0.3964\phantom{5} & 2.717 & 7.81 & 0.35$^*$  \\ \hline
         NaCl & (mol)  & 5.152$^*$\phantom{3} & 2.442 & \phantom{1}7.91\;& 0.31$^*$\\
         NaCl & (rs)  & 7.199$^*$\phantom{4} & 2.309 & 7.47 & 0.31$^*$\\ \hline 
    \end{tabular}\vspace*{2mm}
    \caption{Comparison of pair-potential parameters obtained after fitting adjustable coefficients to accurate reference data on diatomic molecules (mol) and on crystals, where dc and rs are the diamond cubic and rock salt structure, respectively. Numbers indexed with a star are functions of the adjustable parameters. Used potentials are, Born-Mayer/Buckingham for argon and sodium chloride, and the Morse potential for carbon and copper. Plots of the respective potentials are shown in Fig.~\ref{fig:allPot}. The first seven neighbor shells are included into the fit.}
    \label{tab:compMolSolParam}
\end{table}

An alternative combining rule for the prefactors $A_{ij}$ was proposed by Smith~\cite{Smith1972PRA} having in mind atoms or ions with noble-gas electronic configuration. 
He argued that repulsion is an energy excess related to the deformation of the electron density caused by the Fermi principle.
He treated that deformation in a way as if each of the two interacting atoms or ions were in contact with an infinitesimally thin but impenetrable wall, which is positioned a distance $r_{ij}\sigma_{ii}/(\sigma_{ii}+\sigma_{jj})$ from atom $i$.
The excess energy associated with each side of the wall can then be calculated from the pair repulsion.
This nonrigorous, yet plausible model leads to the combining rule 
\begin{equation} \label{eq:SmithRule}
    A_{ij} = 
    \left( A_{ii} \frac{\sigma_{ij}}{\sigma_{ii}}\right)^{\sigma_{ii}/(2\sigma_{ij})} 
    \left( A_{jj} \frac{\sigma_{ij}}{\sigma_{jj}}\right)^{\sigma_{jj}/(2\sigma_{ij})}. 
\end{equation}
An identical rule had already been found empirically by Gilbert~\cite{Gilbert1968JCP} for alkali halide monomers a few years earlier.
Besides proposing his semi-empirical explanation, 
Smith~\cite{Smith1972PRA} added to that list various pairs consisting of noble gas atoms, metal ions, and halogen anions, each with a complete shell. 
Böhm and Ahlrichs~\cite{Bohm1982JCP} found the combining rule for $A_{ij}$ in Eq.~\eqref{eq:SmithRule} to be more accurate than others and further complemented the list of investigated pair repulsions to diatomic molecules beyond those consisting of atoms or ions with closed electron shell. 
Recent work~\cite{VanVleet2016JCTC}, which is motivated by the analysis of electronic-charge density overlap between two atoms, proposes altered combining rules, in which $\sigma_{ij}$ is the geometric mean of $\sigma_i$ and $\sigma_j$.
In addition, more complicated mixing rules apply than those ``derived'' by Smith and the prefactors have a certain $r_{ij}$
 dependence.
 
Given its success, it might be in place to further comment on the arguments leading to Eq.~\eqref{eq:SmithRule}.
First, it explains quite naturally that repulsion is not necessarily a pair-wise additive quantity.
The electron distribution of a first atom is more easily squished by a second atom if there is no third atom already squishing the maltreated electron cloud of the first atom.
The poor electrons simply have no more volume of refuge. 
Thus, the more atoms squeeze against a central atom, the larger the central atom should appear to be.
Table~\ref{tab:compMolSolParam} reflects this trend for single-component systems not only for the equilibrium lengths but also for $\sigma = r_0/n$.
It is particularly strong for the covalently bonded carbon atoms and still quite noticeable for copper. 
In fact, from own unpublished work on alkali metals, our impression is that the assumption of pairwise additive repulsion is particularly poor for small coordination numbers.
Second, potentials adjusting atomic charges on the fly might want to include the effect that this charge has on the atomic size and thus on its repulsive interaction. 
Neither of these two points appears to have attracted the attention it deserves.

\subsection{Funky two-body potentials}
\label{sec:funky}

Simple, computationally lean potentials generally lead to simple behavior without producing the intricate dynamics of highly viscous liquids or the complex structure of amorphous solids.
The phase diagram of Lennard-Jonesium, like that of noble gas atoms other than helium consists of 
a closed-packed structure at low temperature, a low-viscosity fluid phase in a narrow temperature range at (what would be characteristic for) ambient pressure, and the gas phase at large temperature.
While nature manages to produce elements with extended liquid phases at moderate pressure and temperature, like mercury or gallium, they do not become highly viscous.
Even binaries often do not form highly viscous fluids and instead either phase separate or crystallize upon cooling from the liquid phase. 
Notable exceptions to this rule at elevated temperature are SiO$_2$ and CuZr, which both resist crystallization upon cooling to a significant degree while being describable with relatively simple potentials, e.g., by the 
silica potential proposed by van Beest, Kramer, and van Santen (BKS)~\cite{vanBeest1990PRL} or by the embedded-atom model (EAM) for CuZr~\cite{mendelev_using_2007,mendelev_development_2009,cheng_atomic-level_2011}.
Unfortunately, BKS necessitates long-range electrostatic interactions to be evaluated while EAM lacks pairwise additivity, even if it can be computed at the cost of order ${\cal{O}}(Z_\textrm{loc}N)$, like pairwise additive potentials, where $Z_\textrm{loc}$ is the number of atoms within the cutoff radius and $N$ the number of atoms.  

To study processes taking place on times exceeding vibrational periods by several decades and to address \emph{generic} questions typical for disordered solids or highly viscous fluids, it can be advantageous to work with potentials producing the observed macroscopic behavior while being computationally as lean as possible.
Such potentials may not be representative of any real material, but, so the hope, produce the correct qualitative behavior for the right reason.
In the worst case, they provide a model for what nature could be and thereby allow theories for the fracture of disordered solids~\cite{Shi2006PRB,He2019PRL}, the supercooling of liquids~\cite{Wahnstrom1991PRA,Kob1995PRE} or the rigidity of glasses~\cite{lerner_mechanical_2019,rainone_pinching_2020} to be tested. 
Moreover, for many of the frequently qualitative questions to be answered, it can be advisable to sacrifice accuracy in the interactions rather than to make compromises in system size or cooling rate.  
For example, the anomaly of the specific heat in a bulk-metallic glass former can have serious artifacts when the linear system size is not at least twice the density correlation length~\cite{Sukhomlinov2019CMS}.
Likewise, determining the proper scaling for the vibrational density of states with frequency in amorphous solids requires the use of large system sizes and long simulation times ~\cite{laird1991_localized_low-frequency_modes, schober2004size, lerner2016_statistics_low-frequency_modes, Mizuno2017-ag} and thereby the use of lean potentials.

\begin{figure}[hbtp]
    \centering
    \includegraphics[width=0.6\textwidth]{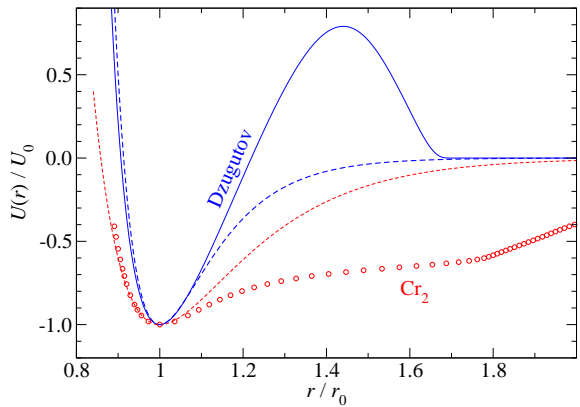}
    \caption{Two examples of funky pair potentials, $U(r)$, normalized to their binding energy $U_0$ and their bond distance $r_0$. The solid blue line represents a Dazugutov potential, which was constructed to make a mono-atomic system avoid crystallization. Circles depict experimental results on the pair potential of the chromium dimer, Cr$_2$. Dashed lines are Morse potential fits to the energy minimum.}
    \label{fig:funky}
\end{figure}

The common recipe to construct simple potentials keeping systems from crystallizing quickly is to build frustration into them.
Dzugutov~\cite{Dzugutov1992PRA} achieved this by introducing a hump in the pair potential. 
Its functional form is given by 
\begin{eqnarray}
    U(r)  =  A\,\left(r^{-m} - B \right)\,e^{c/(r-a)} \Theta(r-a) 
     + B e^{d/(r-b)} \Theta(r-d),
\end{eqnarray}
where the parameters $A$, $B$, $a$, $b$, and $m$ used for Fig.~\ref{fig:funky} are those from the original work.
The most important feature of the Dzugatov potential certainly is that the hump is located near next-nearest neighbors spacings in a closed-packed structure if nearest neighbors settle in the vicinity of the energy minimum.
As a consequence, atoms like to adopt local icosahedral structures, which can arrange similar to the order found in quasi crystals lacking long-range order, though large system sizes and small cooling rates may be required to prevent the crystallization into a bcc structure~\cite{Dzugutov1993PRL}. 

For mixtures of particles, frustration can be encoded to some degree through artificial combining rules in Lennard-Jones binaries, as for example, by setting all $\epsilon_{ij} = 1$ while choosing $\sigma_{AA} = 1.2$, $\sigma_{AB} = 1.1$, $\sigma_{BB} = 1$~\cite{Wahnstrom1991PRA}.
This first attempt by Wahnström is quite prone to crystallization as revealed by Kob and Anderson~\cite{Kob1995PRE}.
In an attempt to mimic a Stillinger-Weber potential designed to mimic amorphous Ni$_{80}$P$_{20}$~\cite{Weber1985PRB}, 
they suggested a binary Lennard-Jones potential in which dislike atoms have a large binding energy while being incompatible in size: $\epsilon_{AA} = 1$, $\epsilon_{BB} = 0.5$, $\epsilon_{AB} = 1.5$, and $\sigma_{AA} = 1.0$, $\sigma_{AB} = 0.8$, and $\sigma_{BB} = 0.88$~\cite{Kob1995PRE}. 
To further reduce the risk of having unnoticed or even worse noticeable crystalline reference phases, continuous distributions of Lennard-Jones radii can be used~\cite{lerner_mechanical_2019,rainone_pinching_2020}.
For a comparison of different glass-forming models, the reader is referred to a recent work by Ninarello, Berthier, and Coslovich~\cite{Ninarello2017PRX}.

We note in passing that pair potentials with humps similar to the one present in the Dzugutov potential are occasionally used to model mono-atomic bcc phases, in particular that of iron.
This is because the third neighbor shell in bcc sits at $\sqrt{11/3} \approx 1.91$ times the radius of the nearest-neighbor shell and still $\sqrt{11/4}\approx 1.65$ times that of the second nearest-neighbor shell. 
Thus the first two shells comfortably fit into the potential well while the third shell has to pay no penalty.
In contrast to bcc, the next-nearest neighbor shell in fcc or hcp would be utmost unhappy. 

Interestingly, real pair potentials of bcc-forming metals can deviate substantially from the Morse potential, as is the case, for example for the chromium dimer, Cr$_2$, whose pair potential is included in Fig.~\ref{fig:funky} and which appears to be challenging to compute accurately even with the most advanced post-Hartree Fock methods~\cite{Vancoillie2016JCTC}.
Despite the funky shape of ``exact'' pair potentials of some transition metal dimers, they still do not allow quantitative predictions to be made for elemental bcc metals, because real interactions simply happen not to be pairwise additive.
It would lead to a sometimes substantial overestimation of the shear modulus, whereby dislocations are artificially suppressed in crystalline materials (their energy depends predominantly on the shear elastic constants) and thus brittle fracture be enhanced. 
Pair potentials also lack the important effect that a (metallic) bond between two atoms is strengthened when at least one of the two reduces its coordination as it would happen in a propagating crack.
This lack biases material behavior toward brittleness. 

Finally, pair potentials with a hump can also be designed to mimic chemical reactions as they occur, for example, during polymeric chemical reactions of step growth~\cite{Deng2016P}  and also radical-reaction based chain-growth~\cite{Deng2017P}.
It remains a philosophical question if such potentials should be deemed reactive, since the notion of a chemical reaction traditionally requires electron transfer or hybridization changes resulting in altered pair interaction, as is induced, for example between two hydrogen atoms due to the presence of an oxygen atom. 
Pair potentials can mimic neither one by definition so that they do not classify as reactive, in our opinion.
Nonetheless, a well-designed bump potential can reproduce steric effects while reproducing the large energy barrier that needs to be overcome for two monomers to bond, whereby the complex interplay of boundary condition, chain relaxation, and chain growth can be studied. 

\subsection{Relating potentials and elastic properties}
\label{sec:Cauchy}

Elastic properties are among the most important properties of solids.
Any potential should therefore be tested for its ability to reproduce the elastic tensor of crystalline references.
Ideally, they include not only thermodynamically stable but also hypothetical structures or finite stress to enhance transferability.
Non-existing reference data can be obtained from DFT or related methods.
One of the problems surfacing quickly is that elastic constants are not uniquely defined, except at zero stress and temperature.
Different definitions lead to deviations of similar order as the external stress~\cite{Birch1947PR,Birch1952-br,WALLACE1965RMP,Barron1965-gj,Wallace1967-fo,Wallace1970-ks,Wang1993PRL,Wang1995PRB}.
This can matter, for example, under geophysical conditions, so that properly converting between different elastic tensors may be critical, which we will come back to briefly at the end of this section.
To set the stage for many of the calculations throughout this article, we briefly review central aspects of the theory of elasticity, while deriving the Cauchy relations, whose violation has been and remains central to the development of potentials. 
We attempted our summary to be more condensed than other texts, while being palatable for readers not familiar with the topic.

We assume a % homogeneous or affine 
deformation of a crystal, in which atomic positions after deformation are $r_{n\alpha} = (\delta_{\alpha\beta}+ u_{\alpha\beta})\, r_{n\beta}^0 + \chi_{n\alpha}$.
Here, atoms are enumerated by $n$, Cartesian components are represented by Greek letters, and the Einstein summation convention applies. 
The upper index 0 in  $r^0_{n\alpha}$  denotes an (equilibrium) position in a reference structure, the tensor $u$ is the macroscopic displacement gradient, while $\chi_{n\alpha}$ is a local atomic displacement.
The deformation is affine if all $\chi_{n\alpha}$ vanish. 
The Eulerian strain tensor $\varepsilon$ is the symmetrized displacement gradient,  $\varepsilon_{\alpha\beta} = (u_{\alpha\beta}+u_{\beta\alpha})/2$.
From this definition, the squared distance between atoms $i$ and $j$, $S_{ij} := r_{ij\alpha} r_{ij\alpha}$, after an affine deformation can be written as
\begin{equation}
    S_{ij} = \left( \delta_{\alpha\beta} + 2\eta_{\alpha\beta}\right) r_{ij\alpha}^0 r_{ij\beta}^0,
\end{equation}
where $\eta_{\alpha\beta} := (u_{\alpha\beta} + u_{\beta\alpha} + u_{\gamma\alpha} u_{\gamma\beta})/2=\varepsilon_{\alpha\beta} + \varepsilon_{\alpha\gamma}\varepsilon_{\gamma\beta}/2$ is called the Lagrangian strain tensor~\cite{Landau1986-ua}.
Other strain tensors exist but are not relevant here. 

Stress tensor elements are generally defined as the first derivative of a thermodynamic potential w.r.t. a strain tensor element and normalized to the volume of the reference, e.g., 
\begin{equation}
     \sigma_{\alpha\beta} := \frac{1}{v^0_\text{pa}}
    \left.
    \frac{\partial U_\text{pa}}{\partial \eta_{\alpha\beta}}\right|_{\eta=0},
    \label{eq:stressgeneral}
\end{equation}
is called a Cauchy stress.
Here, we have normalized energies and volumes per atom (pa). 
Note that the evaluation of the derivative at $\eta=0$ only means that the derivative is taken at the reference but not that the stress disappears. 
Other stress tensor definitions exist, but they only differ when evaluated at nonzero or finite strain.
For example, Eq.~\eqref{eq:stressgeneral} defines the second Piola-Kirchhoff stress, if evaluated at nonzero $\eta$.

For systems interacting through (central) pair potentials, $\sigma_{\alpha\beta}$ can be easily computed.
To this end, we first reexpress $U_2(r_{ij})$ as $\tilde{U}_2(S_{ij})$ so that
\begin{eqnarray} \label{eq:stressDefinition}
    \sigma_{\alpha\beta} & = & \frac{1}{v^0_\text{pa}} \sum_{j} \tilde{U}_2'(S_{ij}) r_{ij\alpha}^0 r_{ij\beta}^0,
\end{eqnarray}
where the prime in $\tilde{U}'$ indicates the first derivative.
For simple lattices under isotropic stress, the summation can be sorted according to neighbor shells, where $s = 0$ denotes the nearest-neighbor shell, $s = 1$ the next nearest-neighbor shell, and so on. Thus, 
\begin{equation}
\label{eq:pairStress}
    \sigma_{\alpha\beta} = \frac{1}{2v^0_\text{pa}} \sum_s  a_s 
    % \left. \frac{\partial \tilde{U}}{\partial S}\right\vert_{S=a_s^2}  
    U_2'(a_s) \nu_s^{\alpha\beta}.
    % r_{s}\,U_\text{pa}'(r_{s}) \nu_s^{\alpha\beta}.
\end{equation}
Here $a_s$ is the distance between a ``central'' atom $i$ and an atom $j$ in shell $s$ and $\nu_s^{\alpha\beta}$ is the second rank shell tensor, whose elements are defined as
\begin{equation}
    \nu_s^{\alpha\beta} = \sum_{j\in \textrm{shell }s}  \frac{r_{ij\alpha}}{a_s}  \frac{r_{ij\beta}}{a_s}.
\end{equation}
For sufficiently symmetric shell structures, the second-rank shell tensor simply turns out to be $Z_s/D$, where $Z_s$ is the number of atoms in neighbor shell $s$ and $D$ the spatial dimension~\cite{Sukhomlinov2016JPCM}. 
In static equilibrium, the external hydrostatic pressure is nothing but $p = -\sigma_{\alpha\alpha}/D$. 

Elastic constants are defined as the change of stress with strain, i.e., as second-order derivative of a thermodynamic potential w.r.t. to strain.
This implies that results differ at non-zero stress for Eularian and Lagrangian strain tensors, see also Eq.~\eqref{eq:diffRuleEulerLag}. 
Moreover, it matters if atomic reference positions, i.e., the Wykhoff positions of atoms in a unit cell are kept fixed or change with strain, as could be expressed by assuming the $\chi_{n\alpha}$ to depend on strain whenever applicable.
Such structural non-affine relaxation can happen when atomic sites lack inversion symmetry, as when shearing a diamond structure, or, as an extreme example, amorphous materials~\cite{Lemaitre2006-ta}.

\textit{In-silico}, it is an easy matter to constrain all $\chi_{n\alpha}$ to zero, experimentally less so. 
Elastic constants defined for $\chi_{n\alpha}\equiv 0$ are marked with an upper zero.
In that case, taking the derivative of the two sides in Eq.~\eqref{eq:stressDefinition} simply yields
\begin{equation}
    C^0_{\alpha\beta\gamma\delta} = \frac{2}{v^0_\text{pa}} \sum_{j} \tilde{U}_2''(S_{ij}) r_{ij\alpha}^0 r_{ij\beta}^0 r_{ij\gamma}^0 r_{ij\delta}^0. 
\end{equation}
By introducing the fourth-rank shell tensor,
\begin{equation}
\label{eq:shellTensor}
    \nu_{s}^{\alpha\beta\gamma\delta} \coloneqq  \sum_{j\in s}
    \frac{r^0_{ij,\alpha}}{r^0_{ij}}\,
    \frac{r^0_{ij,\beta}}{r^0_{ij}}\,
    \frac{r^0_{ij,\gamma}}{r^0_{ij}}\,
    \frac{r^0_{ij,\delta}}{r^0_{ij}}.
\end{equation}
and by using $4\,S^2\,\tilde{U}_2''(S) = \{r^2 U_2''(r)- rU_2'(r)\}$, the elastic tensor reads
\begin{equation} \label{eq:CauchyFinal}
    C^0_{\alpha\beta\gamma\delta} = \frac{1}{2 v^0_\text{pa}}\sum_s
    \{a_s^2 U_2''(a_s)- a_s U_2'(a_s)\} \nu_{s}^{\alpha\beta\gamma\delta}
    =
    \frac{1}{2 v^0_\text{pa}}\sum_s a_s^2 k(a_s) \nu_{s}^{\alpha\beta\gamma\delta}.
\end{equation}
Structural relaxation always reduces the energy so that $C^0_{44}$ of diamond would violate the Cauchy relation even more than already evidenced in Fig.~\ref{fig:cauchyRelation}.

The elastic tensor is invariant w.r.t. to any permutation of its indices, including for example, $C_{1122} = C_{1212}$, or, in Voigt notation $C_{12} = C_{66}$.
It also includes implicitly the recipe for what elastic tensor obeys the Cauchy relations at finite hydrostatic pressure $p$, namely the second-order derivative of the internal energy w.r.t. the Lagrangian strain.
However, most simple crystals violate it as is obvious from Fig.~\ref{fig:cauchyRelation}.

Equation~\eqref{eq:shellTensor} allows the elastic tensor of simple crystals obeying pair potentials to be computed in a straightforward fashion, in particular when knowing the fourth-rank shell tensor.
Its independent components are compiled in Tab.~\ref{tab:ntensor} for selected lattices along with other information useful to quickly compute properties of simple crystals. 
\begin{table}[htb]
\centering
\begin{tabular}{|c|c|c|c|c|c|c|c|}
\hline
 & $Z_0$ & $Z_1$ & $a_1/a_0$ & $v_\textrm{pa}/a_0^3$& $\nu^{11}_0/Z_0$ & $\nu^{1111}_0/Z_0$ & $\nu^{1122}_0/Z_0$\\ \hline
sc  &  6 & 12 & $\sqrt{2}$ & 1 &1/3 & 1/3 & 0\\ \hline
fcc & 12 & 6 & $\sqrt{2}$ & $1/\sqrt{2}$ &1/3 & 1/6 & 1/12\\ \hline
bcc &  8 & 6 & $2/\sqrt{3}$ & $4/\sqrt{27}$ & 1/3 & 1/9& 1/9 \\ \hline
dc  &  4 &12 & $\sqrt{8/3}$ & $8/\sqrt{27}$ & 1/3 & 1/9& 1/9  \\ \hline
hcp & 12 & 6 & $\sqrt{2}$ & $1/\sqrt{2}$ & 1/3 & 5/24 & 5/72 \\ \hline
\end{tabular} \vspace*{2mm}
\caption{\label{tab:ntensor}
Useful dimensionless numbers for selected crystal systems using standard orientation of the lattices: simple cubic (sc), face-centered cubic (fcc), body-centered cubic (bcc), diamond cubic (dc), and hexagonal close packed (hcp).
Additional elements for hcp are $\nu^{1133}_0/Z_0 = 1/18$, $\nu^{3333}_0/Z_0 = 2/9$, and $\nu^{112}_0/Z_0=\mp 1/(24\sqrt{3})$, the negative sign applies to A-layer atoms and the negative for B-layer atoms if B is shifted by $(1/2,1/\sqrt{12},\sqrt{2/3})$ w.r.t. A.  
In diamond, each atom on (0,0,0) and (1/4,1/4,1/4) contribute
$\nu^{123}_0/Z_0 = 1/(3\sqrt{3})$ and $\nu^{123}_0/Z_0 = -1/(3\sqrt{3})$, respectively.}
\end{table}

For short-range potentials only the first two shells contribute substantially to the elastic tensors.
Using the information collected in this section so far, we find for fcc, bcc, and sc, using Voigt instead of tensor notation,
\begin{subequations}
\label{eq:cubicPairSecShell}
\begin{eqnarray}
  C_{11} & \approx & \frac{1}{a_0}
  \begin{cases}
  \sqrt{2}\, k_0 + 2\sqrt{2}\, k_1 & \textrm{ (fcc)} \\
  k_0 /\sqrt{3} + \sqrt{3}\, k_1 & \textrm{ (bcc)} \\
  k_0 + 2\, k_1 & \textrm{ (sc)} 
  \end{cases} \\
% \end{eqnarray}
% and
% \begin{eqnarray}
  C_{12} & \approx & \frac{1}{a_0}
  \begin{cases}
  k_0/\sqrt{2} & \textrm{ (fcc)} \\
  k_0 /\sqrt{3} \phantom{a b c} & \textrm{ (bcc)} \\
  k_1 & \textrm{ (sc).}
  \end{cases}
\end{eqnarray}
\end{subequations}
with the effective shell spring constant
\begin{equation}
\label{eq:effectiveSpringConstant}
k_s  \coloneqq k(a_s) =  U''(a_s) - U'(a_s)/a_s.
\end{equation}
Note that for fcc, bcc and sc, $C_{ij}\equiv C_{ij}^0$ since their primitive unit cell contains only a single atom.

To obtain the relation between different elastic tensor definitions, it is useful to deduce from the definition of the Lagrangian strain that
\begin{equation}
    \frac{\partial \eta_{\alpha\beta}}{\partial \varepsilon_{\gamma\delta}} = 
    \delta_{\alpha\gamma} \delta_{\beta\delta} + \frac{1}{2}\left(\delta_{\alpha\gamma} \varepsilon_{\delta\beta} + \varepsilon_{\alpha\gamma} \delta_{\beta\delta}
    \right).
\end{equation}
Thus, for a second-order derivative of an arbitrary function $f$, it follows that
\begin{equation} \label{eq:diffRuleEulerLag}
    \frac{\partial^2 f}{\partial\varepsilon_{\alpha\beta} \partial\varepsilon_{\gamma\delta}} =
    \frac{\partial^2 f}{\partial\eta_{\alpha\beta} \partial\eta_{\gamma\delta}} +
    \frac{\partial f}{\partial \eta_{\gamma\alpha}}\,\delta_{\beta\delta}.
\end{equation}
The differentiation rule of Eq.~\eqref{eq:diffRuleEulerLag} also allows one to determine the second-order derivative of the term $pv/v_0$, where $p$ is a constant (external) pressure. 
Given the relative change of a volume element,
$ {\Delta v}/{v_0} = \varepsilon_{\alpha\alpha} + \left( \varepsilon_{\alpha\alpha} \varepsilon_{\beta\beta}  - \varepsilon_{\alpha\beta} \varepsilon_{\beta\alpha} \right)/2 + O(\varepsilon^3)$ and using Voigt notation, $\Delta C_{ij} := p (\partial^2 v/\partial\eta_i\partial \eta_j)/v_0$ can be deduced to be
\begin{equation}
\label{eq:pressEffectOnCij}
    \Delta C_{ij} = p \Delta_{ij}
\end{equation}
with $\Delta_{ii} = -1$ for all $i$,  $\Delta_{ij\ne i} = 1$ for $i$ and $j$ both $\le 3$, and
else $\Delta_{ij} = 0$.

Finally, we note that for a solid to be stable, the elastic tensor must be positive definite, which leads to the Born stability criteria~\cite{Born1940MPCPS,Mouhat2014PRB}, e.g., $C_{11}>C_{12}$, $C_{12}>-C_{11}/2$, and $C_{44}>0$ in cubic materials. 
At constant pressure, $C_{ij}+\Delta C_{ij}$ must be positive definite. 
For a more general discussion on how imposed boundary conditions affect mechanical stability, we refer to an instructive work by Wang \textit{et al.}~\cite{Wang1993PRL}, who studied a simple model for gold.  

\section{Many-body potentials for closed-shell systems}
\label{sec:closed-shell}

This section focuses on the description of many-body effects in systems composed of atoms and ions in which the constituents can be said to have a closed electron shell.
Paradigms are noble gases and alkali metal halides, however,
many of the methods and insights conveyed here pertain to a broader context such as valence force fields. 
The central goal of this section is to highlight the role that atomic dipoles play in the interaction between atoms.
These dipoles may originate from quantum mechanical ground-state fluctuations or be induced by an  electrostatic field, which arises, for example, on a chlorine atom in rock salt due to a lattice distortion breaking the local inversion symmetry.

Incorporating many-body effects of atomic dipoles---as well as higher-order electrostatic multipoles, which are somewhat meant to be referred to implicitly whenever the term dipole is mentioned---can be achieved in many different ways.
For neutral atoms, their effect can be incorporated into potentials in the spirit of  the expansion of Eq.~(\ref{eq:vExpand}).
However, as soon as ions linger around, it is more efficient to model the dipoles explicitly. 
This can be done either by placing a formal dipole on the atom, or, by introducing a Drude model, that is, by coupling the center of mass of the displaced electron shell with a spring to a nucleus, or, to a given interaction site in a molecule. 
Finally, the dipole can be treated either classically, or, quantum mechanically. 
The latter appears to be a promising route to model many-body dispersion quite accurately, albeit at a large computational cost. 
For reasons of completeness, we state that Drude models would  better be named after Lorentz, since Drude~\cite{Drude1900AP} considered free electrons in a metal, while Lorentz~\cite{Lorentz1909book} attached (dissipative) springs between electrons and nuclei to model the optical response of bound charges in insulating matter. 

\subsection{Explicit many-body dispersion}
\label{sec:atm}

Axilrod and Teller~\cite{Axilrod1943JCP} and independently
Muto~\cite{Muto43JPMSJ} (ATM) extended the second-order perturbative treatment of
the quantum dipole fluctuations leading to the leading-order $1/r^6$ London forces to third-order perturbation theory.
The resulting three-body ATM potential reads (for three like atoms)
\begin{equation}
U_{{\rm ATM}}(\{{\bf r}\}) =
C_9 \sum_{i<j<k}
\frac{1+3\cos\vartheta_{ijk} \cos\vartheta_{jki} \cos\vartheta_{kij}}{r^3_{ij}\,r^3_{jk}\,r^3_{ki}},
\label{eq:atm}
\end{equation}
where the sum runs over all unique triangles formed by three atoms $i$-$j$-$k$.
Interior angles of the triangle formed by the three atoms $i$-$j$-$k$ are denoted by the Greek letter $\vartheta_{ijk}$, where the first index denotes the atom at the apex of the angle.
The cosines can be computed straightforwardly from the individual bond lengths,
$\cos\vartheta_{ijk}=(r^2_{ij}+r^2_{ik}-r_{jk}^2)/(2 r_{ij}r_{ik})$.
A frequently used approximation for (mixed) dispersion coefficients is obtained from geometric averages in atomic units, i.e.,  $C_{9,ABC} = \sqrt{C_{6,A} C_{6,B} C_{6,C}}/\sqrt{\mathrm{H}}$,  where $A$, $B$, $C$ are element specific and $\mathrm{H}$ is 1~Hartree. 
More rigorous results can be deduced from the Casimir-Ponder integral as described in Refs.~\cite{vonLilienfeld2010JCP,Tang2012JCP}.

The usual course of action when using ATM in conjunction with LJ or Buckingham is to explore improvements on predicted properties.
Traditionally, ATM corrections to the binding energy for noble gas crystals like argon are estimated to be of order 10\%~\cite{Barker1968AJC}.
For very polarizable systems, they can be much larger.
Von Lilienfeld and Tkatchenko~\cite{vonLilienfeld2010JCP} found them to account for 50\% of the binding between two graphene sheets.
In addition, ATM interactions can correct for systematic deficiencies in the elastic and vibrational properties that cannot be overcome in the realm of pair potentials~\cite{Barker1970PRB}.
We see it as beneficial to analyze $U_\textrm{ATM}$ as an independent quantity in terms of a lattice ATM constant, which can be defined in analogy to the Madelung constant as 
\begin{equation}
\label{eq:ATMconstant}
\alpha_\textrm{ATM} \coloneqq a_0^9 \sum_{1<j<k} \frac{1+3\cos\gamma_1\cos\gamma_j\cos\gamma_k}{r_{1j}^3 r_{jk}^3 r_{k1}^3}.
\end{equation}
We obtain $\alpha_\textrm{ATM} \approx 57.548$ for an ideal fcc crystal from which one sixth can be assigned to each atom. 
For hcp, we obtain $\alpha_\textrm{ATM} \approx 57.563$, which can topple the balance in favor of fcc. 
The differences between these two numbers is small, yet, larger than between the results for $C_6$, i.e., $C_6 \approx 14.4538$, $C_{12} \approx 12.1319$ for fcc, versus, $C_6 \approx 14.4545$, $C_{12} \approx 12.1323$ for hcp.
In order for $\alpha_\textrm{ATM}$ to be larger for hcp than for fcc, interactions beyond next-nearest neighbors must be included.
Unfortunately, we did not manage to find recent literature results on the simple-to-compute ATM lattice sums beyond relatively rough, initial estimates by Axilrod~\cite{Axilrod1951JCP}.
Nonetheless, different authors~\cite{Jansen1966AQC,Lotrich1997PRL} have previously concluded that three-body dispersion favors fcc over hcp, although the corrections due to nuclear quantum fluctuations appears to be substantially more important in that regard~\cite{Rociszewski2000PRB}. 

\subsection{Classical polarizable potentials}
\label{sec:polarizable}

The electrostatic field on a central atom or molecule is produced by other charges, dipoles, or higher-order multipoles.
This leads to a hierarchy of inductive interactions~\cite{Cipcigan2019RMP}, which we have already touched upon in Sect.~\ref{sec:dispCorr}.
It contains the $1/r^4$ attraction between a charge and an induced dipole. 
Next in line is the (asymptotic) $1/r^6$ attraction between a permanent dipole and an induced dipole or that between a charge and an induced quadrupole. 
Unfortunately, the interactions related to induced dipoles or induced higher-order multipoles, which will be ignored for the most part in the following, are not pairwise additive.
Two like charges placed at $\pm \Delta \mathbf{r}$ from an atom will not induce a dipole on that atom and thus not lead to twice the energy gained if only one of the two charges were present. 

Since adjacent existing (static, molecular) dipoles try to align themselves to an external electrostatic field in an attempt to minimize the potential energy, induced dipoles tend to be parallel to pre-existing static dipoles.
This effect is well known to increase the mean dipole moment of water molecules from 1.85~D in the gas phase to approximately 2.7~D in the liquid, whereby dipole-dipole interactions essentially double.  

The polarizability of a homogeneous medium having cubic or higher symmetry is stated in terms of its dielectric constant $\varepsilon_\textrm{r}$, which implicitly reflects the feedback that dipoles have on each other. 
In atomic or molecular systems, $\varepsilon_\textrm{r}$ can be well estimated through the Clausius-Mossotti relation~\cite{AshcroftMermin76} 
\begin{equation}
\label{eq:ClausiusMossotti}
    \frac{\varepsilon_\textrm{r}-1}{\varepsilon_\textrm{r}+2} = \sum_i \frac{\rho_i \alpha_i}{3\varepsilon_0},
\end{equation}
where $\rho_i$ is the number density of species $i$ and $\alpha_i$ its (orientationally averaged) polarizability.
Once the right-hand side of Eq.~\eqref{eq:ClausiusMossotti} is greater or equal one, $\varepsilon_\textrm{r}$ is no longer a finite positive number so that the model has reached its physically meaningful limit.
The dipoles ``want'' to grow \textit{ad infinitum}, at which point the system becomes metallic. 
The underlying polarizable potential needs to be extended to either suppress this so-called polarization catastrophe, e.g., by shielding the Coulomb interaction at small distances~\cite{Tang1984JCP}, or, to actually allow the system to become conducting.
In advanced shell potentials being a compromise between polarizable and charge-transfer models, this can be achieved by not constraining an electron cloud to one particular atom~\cite{Leven2019JPCL}.

When modeling charged or polar systems by atomistic means, one would certainly want to reproduce the dielectric response function of a medium correctly, e.g., when simulating the condensation of water on a surface~\cite{Ranathunga2020L}, the Helmholtz double layer on an electrode~\cite{Reed2007JCP,Dapp2013JCP}, or the damped Coulomb interaction of charged colloids in water. 
For the simulation of simple geometries and homogeneous media, it may be possible to achieve this with effective potentials~\cite{Robbins1988JCP} or with the concept of induced mirror charges~\cite{Arnold2013E}.
However, many important problems lack the symmetry or isotropy required to pursue such approaches so that the dielectric response has to be solved on the fly.
This can be achieved with molecular approaches encoding the polarizability into the potential energy surface. 

One possibility is to induce \emph{ideal} dipoles or higher-order multipoles on atoms or specific interaction sites in a molecule~\cite{Buckingham1967ACP,Wilson1996JPC}.
This is done by adding an energy contribution for each inducible multipole $\mathbf{p}_i$. 
The usual interatomic potential $U(\{\mathbf{r}\})$ is the one that minimizes $U(\{\mathbf{r}\},\{\mathbf{p}\})$, e.g.,
\begin{equation}
    U(\{\mathbf{r}\}) = \min_{\{ \mathbf{p}\}} U(\{\mathbf{r}\},\{ \mathbf{p}\}),
\end{equation}
where 
\begin{equation}
\label{eq:fullDipolePot}
U(\{\mathbf{r}\},\{ \mathbf{p}\}) = U_\textrm{sr}(\{\mathbf{r}\}) +  U_\textrm{C}(\{\mathbf{r}\},\{ \mathbf{p}\})+ \sum_i \frac{ p_i^2}{2\alpha_i}
\end{equation}
with respect to the dipoles. In Eq.~\eqref{eq:fullDipolePot}, contributions to the short-range (sr) interaction are separated from those due to Coulomb (C) interactions , which may contain contributions from static dipoles (for molecular rather than atomic simulation) in addition to those from point charges.
Since, Eq.~\eqref{eq:fullDipolePot} is a second-order polynomial in the induced dipoles, a well-defined minimum exists as long as the Hessian related to the  dipoles is positive definite. 
A brute-force inversion of the Hessian to yield the exact minimum is generally inadvisable.
Alternatives are conventional minimization techniques, such as those based on conjugate gradients~\cite{nocedal_numerical_2006} or extended Lagrangians~\cite{Parrinello1980PRL,Car1985PRL}.
In the latter case, the dipoles, or other adjustable degrees of freedom such as those describing the shape of the periodically repeated system~\cite{Parrinello1980PRL} or prefactors to electronic wavefunctions~\cite{Car1985PRL}, are assigned an inertia and propagated and relaxed along with the atomic coordinates.
It is beyond the scope of this review to discuss the pros and cons of extended Lagrangians in detail. 
It suffices to say that their implementation is relatively simple.
However, they introduce an effective delay on the molecular dynamics~\cite{Tangney2002JCP}.
In addition, relaxation to the energy minimum can be slow when the Hessian has strongly differing eigenvalues, which automatically happens for large $\varepsilon_\textrm{r}$.
Then, the dipolar response functions are ``stiff'' at  small length scales but ``soft'' at the continuum scale. 

One disadvantage of placing \textit{ideal} dipoles on atoms or ions is that the evaluation of Coulomb interactions is substantially complicated, even if solutions exist to include their effect into the (fast) Ewald summation~\cite{Hummer1998JPCA}.
Another deficiency is that higher-order multipoles are ignored that are generated when an electron cloud displaces with respect to a nucleus. 
These drawbacks can be remedied with Drude particles, or ``Drudes'', in which a fixed Drude charge $q_\textrm{D}$ is coupled harmonically with a spring constant $k_\textrm{D}$ to an atom or ion, to which a charge $-q_\textrm{D}$ is added~\cite{Dick1958PR,deLeeuw1998PRB,Lamoureux2003JCP}. 
The last summand on the r.h.s. of Eq.~\eqref{eq:fullDipolePot} must then be replaced with $\sum_i k_{\textrm{D},i} u_i^2/2$, where $\mathbf{u}_i$ is the displacement of one Drude charge w.r.t. the atom or interaction site that it is bonded to.
On-site Coulomb interactions between all charges on the Drude, which despite similar spelling is not to be confused with \emph{The Dude} from \emph{The Big Lebowski}, must be switched off. 
Minimization of the total energy w.r.t. to the Drude displacements can be done in a similar fashion as for ideal dipoles. 

In order for the Drude to reproduce the correct polarizability, the relation
\begin{equation}
    \alpha = q^2_\textrm{D}/k_\textrm{D}
\end{equation}
must be obeyed. 
The limiting case of ideal dipoles is obtained for $q_\textrm{D}, k_\textrm{D} \to \infty$ while keeping $\alpha$ constant. 
With an appropriate choice of $k_\textrm{D}$ and sign for $q_\textrm{D}$, the correct quadrupolar induction of a spherically symmetric Drude can be matched to a homogeneous field.
Traditionally, the three independent parameters of Drude oscillators, which in addition to $q_\textrm{D}$ and $k_\textrm{D}$ are also assigned an inertia or mass $m_\textrm{D}$, were chosen to best match the frequency dependence of the dielectric constant at high frequencies~\cite{Dick1958PR}.
Ideally, those values would be close to results obtained from a fit to accurate reference data on forces and energies, for which $k_\textrm{D}$ and possibly even $q_\textrm{D}$ are treated as adjustable parameters. 
Large discrepancies between different parameterization procedures should probably be seen as a sign that something is missing or wrong with the potential. 

\begin{figure}
    \centering
    \includegraphics[width=\textwidth]{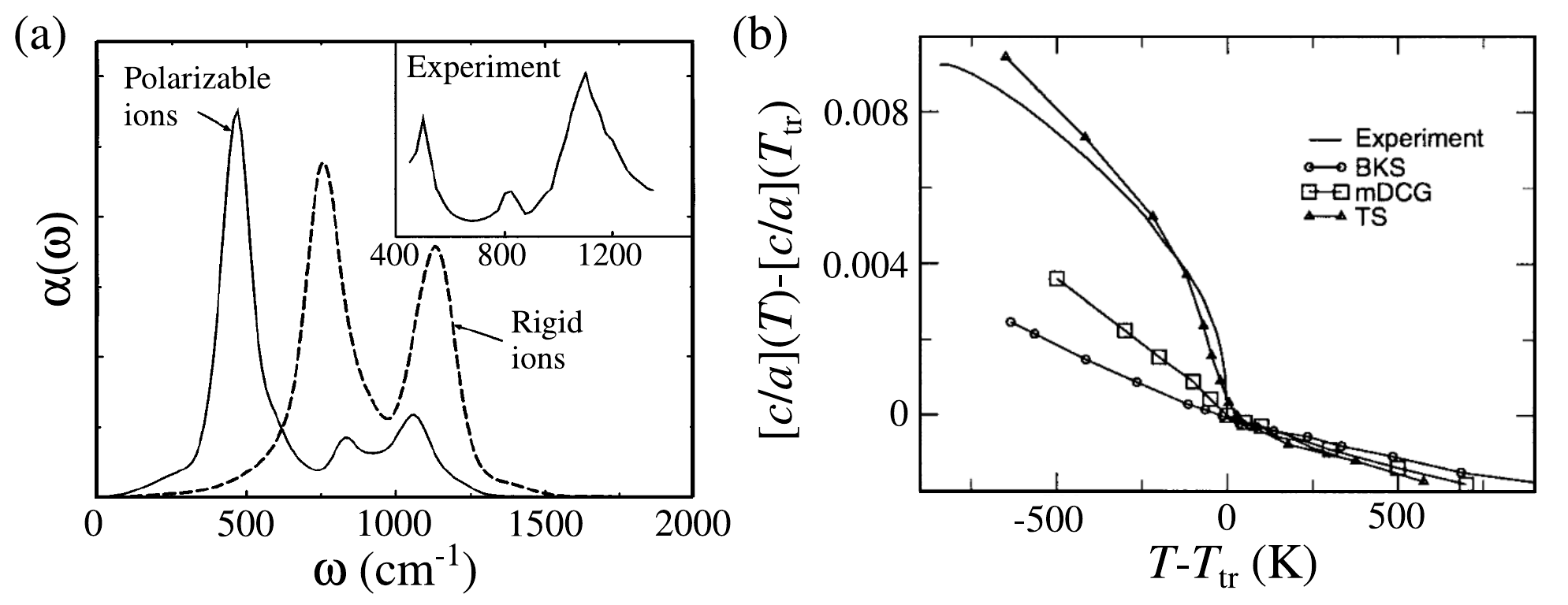}
    \caption{Influence of polarizability on properties of SiO$_2$. (a) Infrared absorption spectrum of amorphous SiO$_2$ (silica) computed one time with a rigid, fixed-charge ion model (dashed line) and one time with a dipol-polarizable anion (solid line). The polarizable model matches qualitatively the experimental results shown in the inset. (b) $c/a$ ratio of crystalline SiO$_2$ (quartz) computed with the rigid ion potentials of van Beest, Kramer and van Santen (BKS)~\cite{vanBeest1990PRL}, the fluctuating-charge potential of Demiralp, {\c C}a{\u g}in and Goddard after modifying parameters (mDCG)~\cite{Demiralp1999-ft}, and the polarizable force field of Tangney and Scandolo (TS)~\cite{Tangney2002JCP-2}. Only the polarizable force field reproduces the $c/a$ anomaly at the $\alpha$-$\beta$-quartz transition with temperature $T_\text{tr}$.
    (a) is reprinted with permission from Wilson, M., Madden, P. A., Hemmati, M., and Angell, C. A. \textit{Phys. Rev. Lett.} \textbf{77}, 4023 (1996) (Ref.~\cite{Wilson1996PRL}). Copyright (1996) by the American Physical Society.
    (b) is reprinted from Herzbach, D., Binder, K. and M. H. M\"{u}ser, \textit{J. Chem. Phys.} \textbf{123}, 124711 (2005) (Ref.~\cite{Herzbach2005JCP}), with the permission of AIP Publishing.
    }
    \label{fig:silica}
\end{figure}

A stringent test for the correctness but also for the relevance of dipolar polarizability can be obtained when comparing force-field-based infrared absorption spectra to reliable reference data. 
For example, in the case of amorphous silica, obtaining the correct number, positions, and intensities of peaks requires the polarizability of the oxygen atoms to be included~\cite{Wilson1996PRL} (Fig.~\ref{fig:silica}a).
Accounting for electrostatic induction also appears necessary to accurately reproduce bond-angle distributions in silica and related systems~\cite{Tangney2002JCP-2}.
While global bond-angle histograms may seem fine for pair potentials, deviations between symmetry-specific angles in crystalline structures from the ideal tetrahedral bond angle have the wrong sign for the most commonly used SiO$_2$ pair potential~\cite{vanBeest1990PRL} but the correct sign and magnitude for the polarizable Tangney-Scandolo potential~\cite{Tangney2002JCP-2,Herzbach2005JCP}.

The incorporation of dipolar polarizability is particularly crucial for structures in which highly polarizable (united) atoms or ions, most notably anions, are located in sites deviating strongly from inversion symmetry. 
This concerns in particular oxygen whenever it is two-coordinated, as it is in water, but also in low-temperature tetrahedral network formers like silica. 
If competing local orders exist, neglecting the polarizability of anions will artificially favor the order with the smaller deviation from inversion symmetry,  because the anion is shorn of its ability to reduce its energy in the larger electrostatic field of the symmetry-broken phase.

Including dipolar polarizability can also be required to describe macroscopic structural changes occurring during the transition between high- and low-symmetry phases in a \emph{qualitatively} correct fashion, as is the case for the $\alpha-\beta$ transition in quartz.
Using pair potentials, the $c/a$ ratio shows no anomaly as a function of temperature at the transition temperature.
For charge-transfer potentials, the slope of the $c/a$ ratio changes at the transition temperature, while it is discontinuous, as in experiment, when dipolar polarizability on oxygen is added to the pair potential~\cite{Herzbach2005JCP} (Fig.~\ref{fig:silica}b).
This discussion may show that the analysis of phase transformations can benefit the validation of potentials, most notably polarizable potentials, even if applications will evolve primarily around liquid or amorphous phases.  

An affine deformation of a highly symmetric ionic crystal, as for example,  rocksalt (B1) or cesium chloride (B2), does not lead to an electrostatic field on a central lattice position created by the atoms residing on other positions.
Thus, the central atom does not develop a dipole.
However, a non-isotropic deformation, e.g., a uniaxial strain, can reduce the cubic symmetry, whereby the charge density on the anion may develop a quadrupole moment.
Consequently, unlike dipolar polarizability, quadrupolar polarizability can induce a Cauchy violation in the said crystals, which is why it can be said to be more important than dipolar polarizability, at least from a continuum-mechanics point of view. 
Madden, Wilson, and coworkers demonstrated that
inclusion of quadrupolar polarizability can substantially increase the agreement between experimental and simulation results, at least in the case of simple ionic systems like AgCl~\cite{Wilson1996JPC} or MgO~\cite{Rowley1998JCP}.
An important aspect of their work~\cite{Rowley1998JCP} is that they managed to gauge adjustable parameters independently from one another, even those describing a charge-density change on an anion due to electrostatic polarization from those being caused by short-range repulsion. 

\subsection{Quantum-Drude oscillators}
\label{sec:QDO}

The  starting point for the description of dispersive interactions are two, or more, isolated atoms, described by their free-atom Hamiltonians $\hat{H}_0^{(1,2)}$.
The atoms interact through their dipoles $\mathbf{p}_{1,2}$ via $\{\mathbf{p}_1\cdot \mathbf{p}_2 r_{12}^2-3(\mathbf{p}_{1}\cdot\mathbf{r}_{12})(\mathbf{p}_{2}\cdot\mathbf{r}_{12})\}/(4\pi\varepsilon_0r_{12}^5)$ plus potentially through higher-order multipoles. 
Second-order perturbation theory then leads to a pairwise-additive $1/r_{12}^6$ interactions between the atoms.

Asymptotic $1/r^6$ interactions are also obtained when replacing the atomic Hamiltonians with quantized Drude oscillators~\cite{London1937TFS,Whitfield2006CPL,Cipcigan2019RMP}, as will be shown here below.
Their three, free parameters per atom, $q_\textrm{D}$, $k_\textrm{D}$, and $m_\textrm{D}$ can be chosen to match the correct polarizability $\alpha_\textrm{D} = q_\textrm{D}^2/k_\textrm{D}$ and leading-order dispersive coefficient $C_6$, which allows dispersive and inductive interactions to be treated on a common footing.
The justified hope is that higher-order and many-body dispersive interactions can be mimicked reasonably well simultaneously by properly selecting the third remaining parameter~\cite{Cipcigan2016JCP}. 
If the nucleus is also quantized, $m_\textrm{D}$ must be replaced with a reduced mass $\mu_\textrm{D}$.

The Hamiltonians of two identical, three-dimensional quantum Drudes can be decoupled into three pairs with  
\begin{equation}
    \hat{H}_k = \sum_{i=1}^2 \left(\frac{\hat{p}_{ki}^2}{2\mu_\textrm{D}} + \frac{k_\textrm{D}}{2} u_{ki}^2\right) + \frac{g_k}{4\pi\varepsilon_0}\,\frac{q_\textrm{D}^2 u_{k1} u_{k2}}{r_{k12}^3}.
    \end{equation}
The parameters $g_k$ ($k = 1,...,3$, no summation convention) take the values $1$ or $-2$ depending on whether the displacements $u_{1,2}$ are orthogonal or parallel to $\mathbf{r}_{12}$.
The oscillators can be decoupled further through the transformation $u_{k\pm} = (u_{k1} \pm u_{k2})/\sqrt{2}$ leading to three isolated oscillators pairs described by Hamiltonians of the form
\begin{equation}
    \hat{H}_{k\pm} = \frac{\hat{p}_\pm^2}{2\mu_\textrm{D}} + \frac{1}{2}\left(k_\textrm{D} \pm  \frac{g_k}{4\pi\varepsilon_0}\,\frac{q_\textrm{D}^2}{r_{k12}^3} \right) u_{k\pm}^2.
\end{equation}
In leading order, their combined excess ground-state energy w.r.t. that of two uncoupled oscillators, $U_0 = \hbar\omega_\textrm{D}/2$ with $\omega_\textrm{D} = \sqrt{k_\textrm{D}/\mu_\textrm{D}}$ per free quantum Drude pair is
\begin{equation}
    \Delta U_{k0}(r_{12}) = - \frac{\hbar\omega_\textrm{D}}{8}\,\left( \frac{g_k\,q_\textrm{D}^2\,}{4\pi\varepsilon_0 k_\textrm{D}} \right)^2 \,\frac{1}{r_{12}^6}.
\end{equation}
Adding up the three Drude pairs, using $\alpha' = q^2_\textrm{D}/(4\pi\varepsilon_0 k_\textrm{D})$, and repeating the entire exercise for potentially dislike Drudes, $A$ and $B$, then yields,
\begin{equation}
    C_{6,AB} = \frac{3\hbar}{2}\, \frac{\alpha'_A\alpha'_B\omega_A\omega_B}{\omega_A+\omega_B},
\end{equation}
which is the same combination rule as Eq.~\eqref{eq:MeathDispersionCombo}.
Using the true atomic reference Hamiltonians and placing the dimer on the $z$-axis, the correct mixed dispersion coefficient reads~\cite{Eisenschitz1930ZP}
\begin{eqnarray} \label{eq:sumRuleDisp}
    C_{6,AB} & = & {6}\,
    \left( \frac{\textrm{e}}{4\pi\varepsilon_0} \right)^2\,
    {\sum_{n_{A},n_{B}}^{}}'
    \frac{ 
    \vert z_{0\,n_{A}}\vert^2 \,
    \vert z_{0\,n_{B}}\vert^2
    }{E_{n_{A}}^{(A)}-E_0^{(A)} + E_{n_{B}}^{(B)}-E_0^{(B)} } ,
\end{eqnarray}
where $z_{0\, n}$ is the matrix element $\langle 0 \vert z \vert n\rangle$ of an atom
and where the primed sum indicates that at least one of the two quantum numbers $n_A$ and $n_B$ must differ from zero. 

\begin{figure}
\begin{center}
\includegraphics[width=0.75\textwidth]{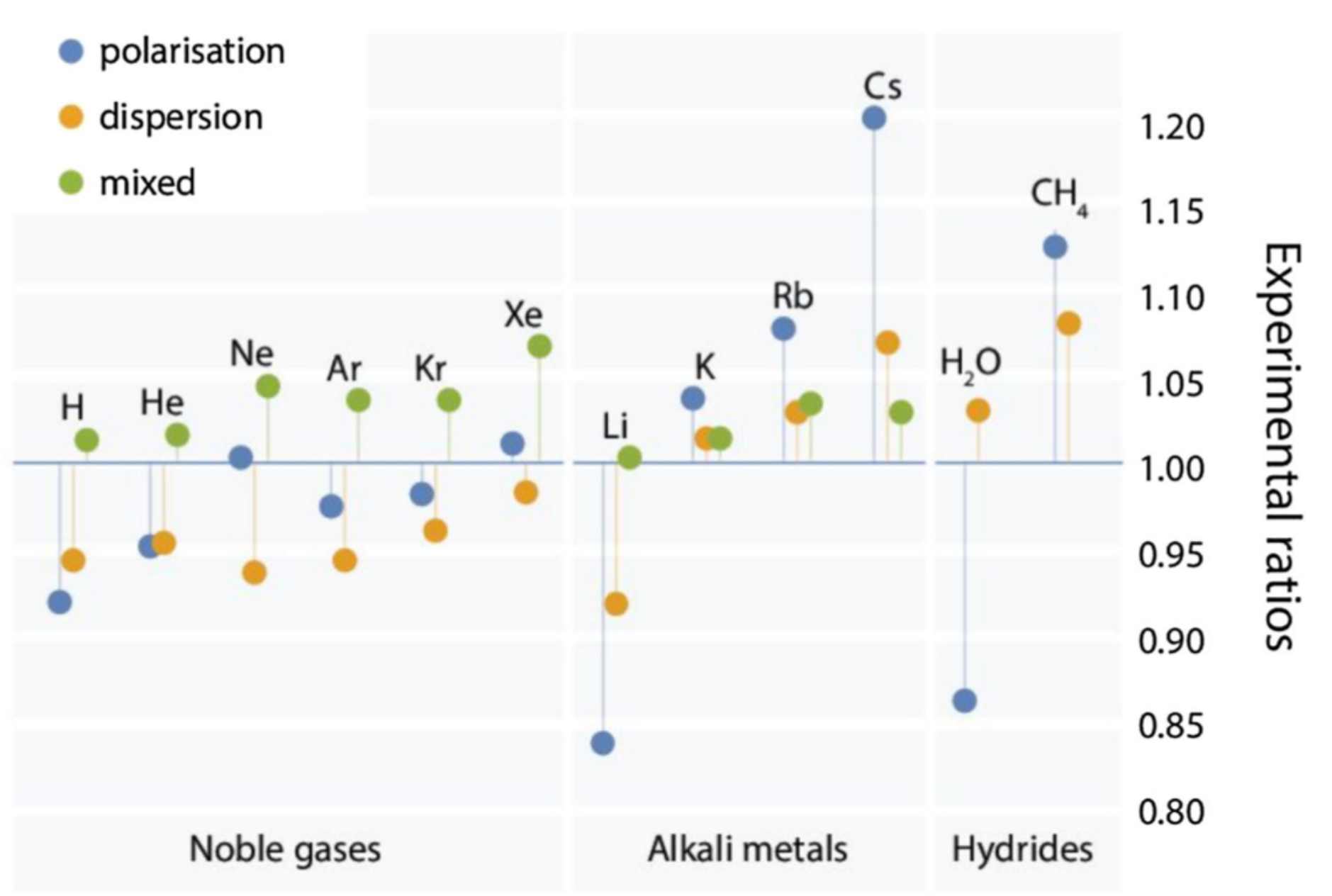}
\end{center}
\caption{\label{fig:martyna} Experimental values for the left-hand sides of Eq.~\eqref{eq:martyna}.
Reprinted from \textit{J. Comput. Phys.}, 326, Cipcigan, F. S., Sokhan V. P., Crain, J., and Martyna, G. J., Electronic coarse graining enhances the predictive power of molecular simulation allowing challenges in water physics to be addressed, 222-233, Copyright (2016), with permission from Elsevier.
%From Ref.~\cite{Cipcigan2016JCP}.
}
\end{figure}

Typical parameterizations of quantum Drudes~\cite{Cipcigan2019RMP} yield  $\hbar \omega_\textrm{D}$ of order \SI{0.5}{H} to \SI{1}{H} for hydrogen and noble gas atoms as well as for small, closed-shell molecules, $q_\textrm{D}/\textrm{e}$ in between 0.7 and 1.4, and $m_\textrm{D}/m_e$ from 0.1 to 0.6. 
Since a regular quantum Drude is fully defined by three parameters, many terms related to higher-order polarizability $\alpha_l$ ($l = 1$, dipole, 2 = quadrupole, 3 = octupole) are constrained to take fixed ratios.
Some of them are given by~\cite{Cipcigan2016JCP}
\begin{subequations}
\label{eq:martyna}
\begin{eqnarray}
\sqrt{\frac{20}{9}} \, \frac{\alpha_2}{\sqrt{\alpha_1\alpha_3}} & = & 1 \textrm{ (polarization)}\\
\sqrt{\frac{49}{40}} \, \frac{C_8}{\sqrt{C_6 C_{10}}} & = & 1  \textrm{ (dispersion)}\\
\frac{C_6 \alpha_1}{4 C_9} & = & 1  \textrm{ (mixed)}.
\end{eqnarray}
\end{subequations}
Fig.~\ref{fig:martyna} reveals that these ratios tend to reproduce experimental results within 10\% error.

Once a quantum Drude model is defined, various strategies exist to deduce the energy and interatomic forces following from it. 
Brute-force diagonalization is not advisable for reasons of computational efficiency but also because the model is no longer quadratic in the Drude displacements, as soon as they seize to be negligible compared to interatomic distances, i.e., at typical condensed-phase, nearest-neighbor spacings. 
Moreover, any solution strategy should allow for the possibility to replace the harmonic coupling with one in the spirit of finitely-extensible non-linear coupling, as to potentially match simultaneously more than just one inductive or dispersive coefficient in addition to $\alpha$ and $C_6$.

A common strategy to handle quantum Drudes is the use of path-integral techniques~\cite{Tuckerman1993JCP,Ceperley1995PRM,Habershon2013ARPC,Herrero2014JPCM}.
The trouble with quantum Drudes is that the number of system replicas to be simulated is of order one Rydberg divided by thermal energy, which is a few hundred at room temperature. 
One can get away with slightly fewer replicas with diffusion Monte Carlo~\cite{Anderson1976JCP}, which is closely related to path integrals, in conjunction with a diagrammatic expansion of interacting Drudes~\cite{Jones2009PRB}.
Unfortunately, improvements do not seem such that numerical costs fall below an approach, in which induction is handled classically and ATM interactions are included explicitly. 
This turns the quantum Drude oscillator approach into something like a sleeping beauty, which could be awaken if a handsome prince managed to cut down the computational overhead w.r.t. classical polarizable potentials plus two-body like short-range potentials to a factor of order ten or less.

\subsection{Shell models and many-body repulsion}
\label{sec:shellModels}

The interaction of atoms results in the deformation of (closed) electronic shells beyond the induction of electric multipoles through electrostatic fields, reflected, for example, by an approximately rigid translation of the valence shell with respect to the remaining ion. 
A reduction of the Coulomb interaction between charges and dipoles due to the overlap of electronic shells can be described with damping functions, which were introduced in Sect.~\ref{sec:shortRangeCorr}.
The electrostatic field, its gradients but also the Fermi principle and hence repulsion deforms the electronic density compared to the superposition of the free-ion or free-atom references. 
These deformations were first approximated as being spherical in the so-called breathing-shell model~\cite{Schroder1966SSC} and described on a common footing with dipolar polarizability~\cite{Dick1958PR}. 
Later, non-spherical shell deformations were also considered with deformable-shell~\cite{Basu1968PSS}, deformable-ion~\cite{Wilson1996JPC}, or distortable-ion~\cite{Rowley1998JCP} potentials. 
Variables describing the state of deformation are the ionic radius, or rather, the deviation $\delta_i$ from the radius of the free ion, plus variables describing the shape of the deformation. 
In atomistic simulations, the shell variables, including the induced dipoles, are assumed to minimize the potential energy.

The starting point of much of the original literature~\cite{Schroder1966SSC,Basu1968PSS,Sangster1976AP} assumes a harmonic expansion of the potential energy in terms of small variables, i.e., displacements, dipoles, and breathing modes.
However, for general situations, the potentials are better cast in terms of implicit many-body potentials, the ingredients of which are the repulsive interactions and on-site energy penalties for the deformation of the shell with respect to free ions.
In such models, the short-range repulsion can be expressed, for example, in a breathing-shell potential, through an expression of the type~\cite{Rowley1998JCP,Matsui1998JCP,Tangney2003JCP}
\begin{equation}
\label{eq:breathingShell}
    U_\textrm{sr} = \sum_{i} U_{\textrm{bre},i} \cosh(\delta_i/\sigma_{i}) +
\sum_{i,j>i} U_{\textrm{rep},ij}e^{-(r+\delta_i+\delta_j)/(b_i+b_j)} .
\end{equation}
Here, $U_{\textrm{bre},i}$, $\sigma_{i}$, $U_{\textrm{rep},ij}$, and $b_i$ are constant coefficients, while the $\delta_i$ minimize $U_\textrm{sr}$.
The first summand on the r.h.s. of Eq.~\eqref{eq:breathingShell} reflects the on-site coupling in a heuristic fashion, i.e., it can be parameterized to yield the correct shell stiffness at small $\delta_i$ while suppressing embarrassing ion shrinkage at large compression. 

Accounting for anisotropic shell deformation requires tensors of rank one and higher to be included as arguments in the functions appearing on the r.h.s. of Eq.~\eqref{eq:breathingShell} such that they reduce to a scalar in a meaningful way, i.e., without violating the isotropy of space.
Anisotropic shell distortions seem to be required to simultaneously improve phonon spectra and the binding energy differences of competing ionic structures compared to rigid-shell potentials merely reflecting dipolar polarizability~\cite{Schroder1966SSC,Wilson1996JCP}. 
We note in passing that aspherical atoms modeled with potentials assuming fixed bonding topography has been pioneered by Price, Stone, and coworkers~\cite{Price1994JACS}.
Deviations from spherical symmetry then stems predominantly from intramolecular interactions so that shape parameters can be treated as fixed relative to a molecular coordinate system. 

A central motivation for introducing flexible shell models was to reproduce the experimentally observed violation of the Cauchy relation in simple salts and their phonon dispersion at high symmetry points~\cite{Schroder1966SSC,Sangster1976AP}.
Here, it might be worth observing that the Cauchy pressures tend to be minor for alkali halides but quite substantial for alkaline earth oxides like MgO. 
One central difference between these two classes of simple salts is that a free, singly-charged halogen anion is stable while a free, doubly charged chalcogen anion is not. 
Thus, assuming doubly charged ionic references without accounting for charge transfer upon a change in bond length could be difficult to justify. 

\begin{figure}
    \centering
    \includegraphics{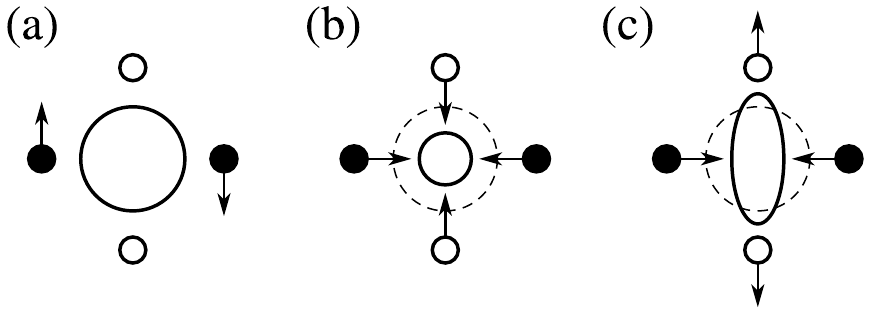}
    \caption{Deformation of an electron shell in response to a deformation of the macroscopic body. The electron shell is schematically depicted by the central circle. (a) During shear deformation, the displacement does not lead to a deformation of the shell. There is hence no influence of shell deformation on the shear modulus $C_{44}$. (b) If the shell deforms isotropically, then an inwards motion of the black atoms will lead to an inwards force on the white atoms, implying $C_{12}< C_{44}$. (c) For anisotropic shell deformation, an inwards motion of the black atoms will lead to an outwards force on the white atoms, leading to $C_{12}>C_{44}$.
    Adapted from \textit{J. Phys. Chem. Solid.}, \textbf{34}, Sangster, M., Interionic potentials and force constant models for rocksalt structure crystals, 355–363, Copyright (1973), with permission from Elsevier (Ref.~\cite{Sangster1973JPCS}).
    }
    \label{fig:shell-deformation}
\end{figure}

Interestingly, charge-transfer and breathing-shell models have a similar effect on the Cauchy pressure.
When the application of a stress in $x$-direction induces ions to shrink in all spatial directions, the external pressure required to maintain the crystal shape in the $y$ or $z$ direction is reduced, which implies that  $C_{12}$ is reduced compared to a rigid-shell model with fixed or zero $\delta_i$. 
In contrast, when shearing, for example, a rock-salt structure, the nearest-neighbor bond lengths only change in order $\mathcal{O}(\varepsilon_{44}^2)$ so that the shear modulus is the same as that of a rigid-shell potential. 
In other words, $C_{44}>C_{12}$, is also unavoidably obtained for charge-transfer potentials applied to rocksalt crystals, irrespective of whether neutral atoms or ions are taken as reference~\cite{Sukhomlinov2015JCP}. 
Figure~\ref{fig:shell-deformation} pictures the just-made arguments for shell models schematically and explains qualitatively how anisotropic shell models can lead to a Cauchy pressure of either sign.
Closed form expressions for the elastic constants in breathing-shell models are summarized in Ref.~\cite{Basu1974PSS}, which also contains a critical comparison of different breathing-shell potentials.

Ion shrinkage under compressive stress without charge transfer and the unavoidable sublinear scaling of repulsion with the coordination number $Z_0$ in breathing-shell models strikes us as implausible. 
It certainly violates the superlinear scaling of repulsion with $Z_0$ obtained in a first-order DFT-based perturbation theory, in which the kinetic energy density of the electrons increases with $\rho^{5/3}$, where $\rho$ is the electronic density. 
Despite this argument being merely qualitative, it should reflect the proper trend as  the trend is very clear.
Our own quick and dirty analysis of the EOS of MgO in the NaCl and CsCl structures using exponential repulsion plus Coulomb interaction and the respective Madelung constants makes us believe that trends are complicated.
Although the compressive part of the EOS can be described quite well with the most simple Born Mayer potential, the fitted charges turn out close to unity in both cases while Bader analysis finds more intuitive charges close to $1.7$. 
When using the parameters deduced for the NaCl and CsCl-structure for the dimer, differences in the Coulomb energy turn out larger than for the repulsion.
At the same time, the Bader charges barely change with the lattice constants in contradiction to our suspicion that breathing-shell models reflect charge-transfer effects to low order. 
Nonetheless, short-range repulsion certainly induces non-spherical distortions leading to electrostatic multipoles, which then must be included in the overall electrostatic interactions between atoms~\cite{Wilson1996JPC}.

\section{Many-body potentials for open-shell systems}
\label{sec:open-shell}

By definition, open-shell atoms have an incomplete valence shell.
Bonding between them occurs through the formation of either covalent, or, in condensed phases, also through metallic bonds. 
Interactions between open-shell atoms cannot be faithfully described with pair potentials without producing false trends, some of which are discussed in Sect.~\ref{sec:consequences}.
The arguably most important many-body effect in open-shell systems is the weakening of a bond due to the presence of additional atoms.
An extreme example is the onset of repulsion between hydrogen atoms, which attract each other as free radicals but repel each other in the presence of oxygen.
A more subtle effect is the contraction of layers near unpassivated metal surfaces, where missing neighbors of the atoms in the surface layer strengthen their interaction with atoms in the layer underneath. 

\begin{figure}
\begin{center}
\includegraphics[width=\textwidth]{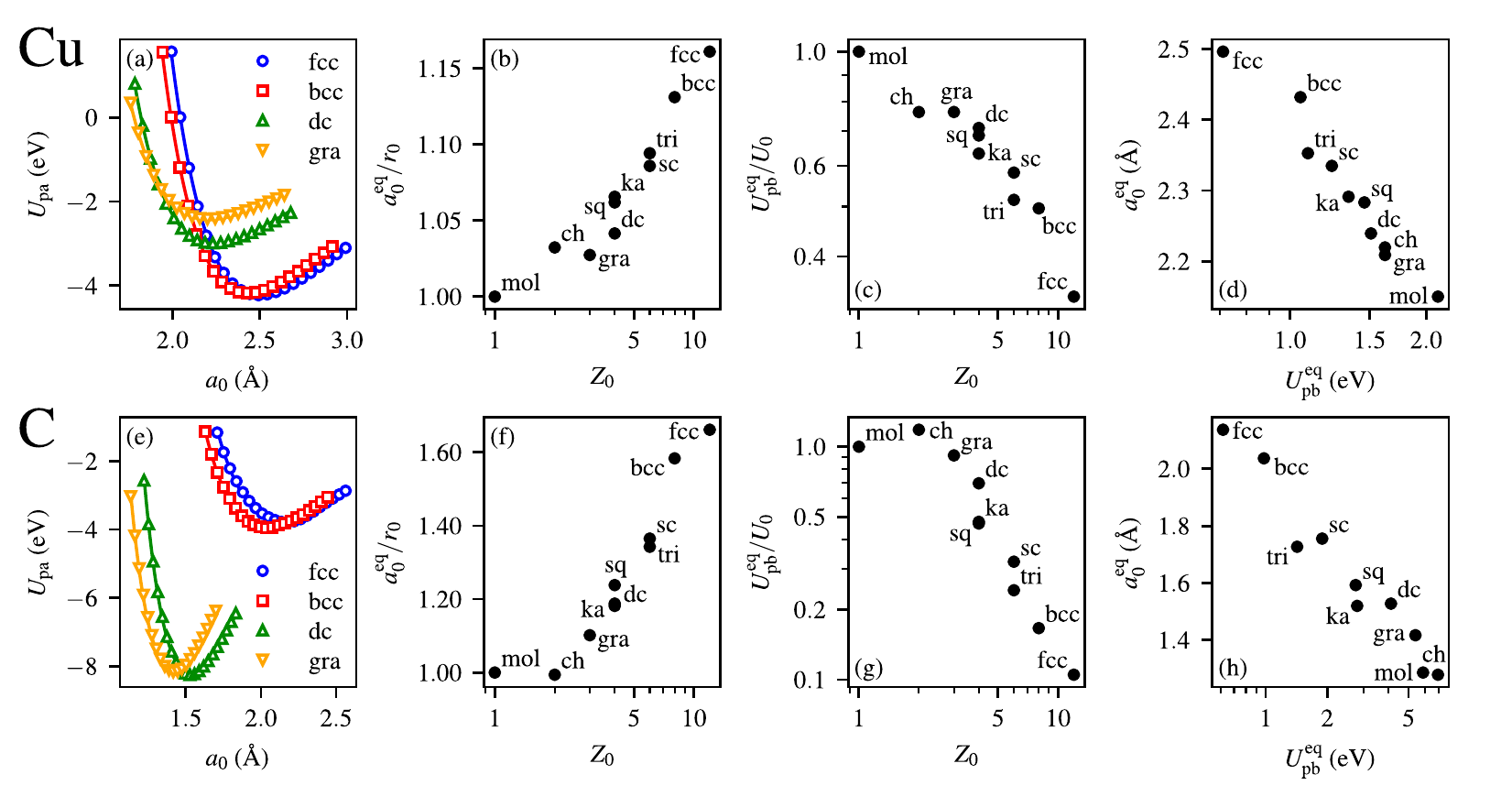}
\end{center}
\caption{\label{fig:structural-trends} (a,e) Energy per atom $U_\textrm{pa}$ in selected crystal structures as a function of the nearest-neighbor spacing $a_0$. Solid lines are fits of the data to a standard Morse potential assuming nearest-neighbor interactions. (b,f) Equilibrium bond length $a_0^\textrm{eq}$ as a function of coordination number $Z_0$.
(c,g) Equilibrium binding energy per bond $U_\text{pb}^\text{eq}$ as a function of $Z_0$.
 (d,h) Equilibrium bond length $a_0^\textrm{eq}$ as a function of equilibrium binding energy per bond $U_\text{pb}^\text{eq}$ (Pauling plot).
Crystal structures range from 1D to 3D include hypothetical ones and are abbreviated as follows: dimer/molecule (mol), chain (ch), graphene (gra), triangular lattice (tri), kagome lattice (ka), square lattice (sq), diamond cubic (dc), simple cubic (sc), body-centered cubic (bcc) and face-centered cubic (fcc). The top row shows results for copper, while the bottom row shows results for carbon. $U_0$ and $r_0$ are the binding energy and bond length of the dimer.
Data, which were produced using consistent DFT-LDA methodology, are merely meant to convey trends.
}
\end{figure}

Figure~\ref{fig:structural-trends} reveals this weakening for copper and carbon in different crystalline structures, including hypothetical structures.
The energy per atom $U_\text{pa}(a_0)$ as a function of nearest-neighbor bond length  (Fig.~\ref{fig:structural-trends}a -- copper, Fig.~\ref{fig:structural-trends}e -- carbon) was obtained from the materials scientist's favorite electronic structure method, DFT~\cite{hohenberg_inhomogeneous_1964,kohn_self-consistent_1965} within the local density approximation~\cite{kohn_self-consistent_1965}
%\cite{Ceperley1980PRL}
using projector-augmented waves~\cite{blochl_projector_1994}.
The minimum of these curves is the cohesive energy of the crystal $U_\text{coh}=U_\text{pa}(a_0^\text{eq})$, where $a_0^\text{eq}$ is the equilibrium bond length shown in Fig.~\ref{fig:structural-trends}b for copper and in Fig.~\ref{fig:structural-trends}f for carbon.
$a_0^\text{eq}$ is approximately a logarithmic function of the coordination number $Z_0$ for both copper and carbon.
The energy per bond in the equilibrium configuration, $U_\text{pb}^\text{eq}(Z_0) = 2 U_\text{coh}(Z_0)/Z_0$, clearly decreases with $Z_0$ and much more so for carbon (once $Z_0 \ge 3$) than for copper. 
For both elements, the decrease of $U_\text{pb}^\text{eq}(Z_0)$ is approximately algebraic in $Z_0$, albeit with a steeper power law for carbon than for copper. 

The results presented in Fig.~\ref{fig:structural-trends} imply that pair potentials cannot describe the energetics of these systems, although each individual $U_\textrm{pa}(a_0)$ at fixed $Z_0$ can be reproduced quite accurately using a regular Morse potential as evidenced by the solid lines in panels (a) and (e) of Fig.~\ref{fig:structural-trends}.
As Abell~\cite{abell1985PRB} emphasized, environment-dependent parameters are needed.
Various families of many-body potentials for open-shell systems have been developed over the years to reflect these trends and, of course, more subtleties.
In this section, we will outline their functional forms and the properties of the solids that they describe. We will also motivate some of them from quantum-mechanical considerations.

A commonality of the most popular, simple open-shell potentials, namely EAM, second-moment tight-binding expansion (TB2M), Stillinger-Weber (SW), and Tersoff, which will all be introduced in this section, is that their total potential energy can be written as a \emph{cluster functional}~\cite{carlsson_beyond_1990}
\begin{equation}
    U(\{\v{r}\}) = \frac{1}{2}\sum_{ij} \left\{ U_2(r_{ij}) + U_\text{m}(r_{ij}, \xi_{ij}) \right\},
    \label{eq:genU}
\end{equation}
where $U_2(r_{ij})$ is \emph{a} pairwise additive interaction---to be distinguished from \emph{the} pair potential---and $U_\text{m}(r_{ij}, \xi_{ij})$ is an effective pair interaction, which, however, depends on a bond-environment variable $\xi_{ij}$, which is not necessarily symmetric in the indices and may contain a non-zero on-site term $\xi_{ii}$.
Its  specific interpretation changes from one potential class to the next but it always depends on a sum over third atoms through
\begin{equation}
    \xi_{ij} = \sum_{k} \Xi(r_{ij}, r_{ik}, r_{jk}).
    \label{eq:genxi}
\end{equation}
In a limiting case, $\Xi(r_{ij}, r_{ik}, r_{jk})$ simply corresponds to the third term,  $U_3(r_{ij}, r_{ik}, r_{jk})$, of the expansion given in Eq.~\eqref{eq:vExpandGalilei}.
However, $U_\text{m}(r_{ij}, \xi_{ij})$ can also implicitly include higher-order terms. 
The way how to express the various potentials in the generic form of Eqs.~\eqref{eq:genU} and \eqref{eq:genxi} are detailed in Sect.~\ref{sec:pastewkaPotential}.
Those summaries may be useful to design prototype templates for functions or to work out elastic properties falling into this general category.

\subsection{Density-dependent potentials}
\label{sec:eddp}

The alkali metals can be seen as a polar opposite to closed-shell systems, because their electrons are delocalized to the extent that their dispersion relation is close to that of free electrons up to the Fermi wave vector $k_F$, albeit with an effective mass.
This is because there is only one electron per atom in the conduction band so that only its minimum is sampled. 
For this reason, early attempts of describing bonding in (alkali) metals assume the Jellium model~\cite{Wigner1934PR,Ceperley1980PRL,Born1954book}, which expresses the energy of an electron gas as a function of its density $\rho = N/V$, where in the case of (neutral) alkali metals the number of atoms $N$ is identical to that of valence electrons. 
The Jellium model is the DFT-based approach to matter, in which the main effect of ions is assumed to provide a charge-neutralizing background to the electrons without explicitly accounting for the relative positions between ions. 
Interestingly, the Jellium model leads to densities and bulk moduli that are relatively close to those of crystals formed by light alkali metals, even when using the true electronic mass~\cite{Wigner1934PR}. 
However, Jellium does not stiffen sufficiently much with increasing density when the Coulomb repulsion between the ionic cores starts to matter.

The potential of the form~\cite{Ascarelli1969PRL,vitek_pair_1996},
\begin{equation}
  U( \{ {\v{r}}_{i} \} )
  =
  N u(\rho) + \frac{1}{2}\sum_{i,j} U_2( r_{ij} ),
  \label{eq:pairplusdens}
\end{equation}
can certainly be motivated from the Jellium model. 
In this potential, which could be coined a global-glue potential, the Jellium model is augmented with a pair-wise additive repulsion, e.g., in the form of damped Coulomb interactions. 
In principle, not only the pair repulsion but also $u(\rho)$ can be made a function of the element, or, elements under consideration. 
While Eq.~\eqref{eq:pairplusdens} can be generalized to non-homogeneous phases and alloys, e.g., via the construction of Wigner-Seitz cells and the request of local charge neutrality, surfaces would cause problems, since surface atoms ``own'' excessive volume, which makes a local density difficult to define. 
Thus, it is clear that potentials footing on Eq.~\eqref{eq:pairplusdens}, or, simple generalizations thereof, are not practicable.
It is yet utmost instructive, to study their properties and further reasons for their failure.

Assuming interactions past the nearest-neighbor shell to be screened, see Sect.~\ref{sec:limiting}, severe restrictions for the elastic constants of simple solids ensue.
The glue effectively acts as a positive, external pressure, $p_\textrm{g}$, squeezing the ions together. 
Combining Eqs.~\eqref{eq:cubicPairSecShell} and \eqref{eq:pressEffectOnCij} then allows the elastic tensor elements of elementary fcc and bcc crystals at zero external stress to be approximated (assuming a smoothly changing $p_\textrm{g}$ with density) with 
\begin{equation}
\label{eq:Cij4glue}
\begin{array}{cccl}
    C_{11}  = 2\sqrt{2} C_0 - p_\textrm{g} \phantom{/} & 
    C_{12}  = \sqrt{2} C_0 + p_\textrm{g} \phantom{3/} & 
    C_{44}  = \sqrt{2} C_0 - p_\textrm{g} \phantom{3/} & \textrm{fcc} \\
    C_{11}  = 2/\sqrt{3} C_0 - p_\textrm{g} & 
    C_{12}  = 2/\sqrt{3} C_0 + p_\textrm{g} & 
    C_{44}  = 2/\sqrt{3} C_0 - p_\textrm{g} & \textrm{bcc},
\end{array}
\end{equation}
where $C_0$ would be an  element-specific constant. 
Equation~\eqref{eq:Cij4glue} predicts $C_{44}<C_{12}$. Check mark for simple metals!
For fcc, the sum rule $c_\textrm{sr} \coloneqq (C_{11}+C_{12})/(C_{12}+C_{44}) = 3/2$ can be read off. 
It is obeyed reasonably well by our favorite metal, $c_\textrm{sr}(\textrm{Cu}) \approx 1.46$ and other fcc metals, $c_\textrm{sr}(\textrm{Pb}) \approx 1.56$, $c_\textrm{sr}(\textrm{Ni}) \approx 1.46$, $c_\textrm{sr}(\textrm{Pd}) \approx 1.66$, $c_\textrm{sr}(\textrm{Ag}) \approx 1.54$, but less so by the favorite metals of our wives $c_\textrm{sr}(\textrm{Au}) \approx 1.73$ and $c_\textrm{sr}(\textrm{Pt}) \approx 1.85$. 

Unfortunately, bcc turns out mechanically unstable, because the pertinent elastic tensor is not positive definite owing to $C_{12}>C_{11}$ for positive glue pressures. 
This inequality can be easily understood.
The global glue attempts to increase density.
This is done most effectively when nuclei keep a maximum mutual distance, which the most closed-packed structures achieve the best. 
But why then do alkali metals condense into bcc? 
Before addressing that question, we note that $C_{11}/C_{12}$ tends to be quite close to unity for simple bcc alkaline metals but not for bcc metals with an approximately half-filled $d$-shell.
Specifically, $C_{11}/C_{12} = 1.2$ within 2\% for the alkali metals from lithium to cesium, while  transition metals assume much larger ratios.
For example, 1.95 (V), 4.37 (Cr), 2.60 (W), and 1.76 (Fe).
Thus, density-dependent potentials fail distinctly more for transition metals than for alkali metals. 
Also, most alkali metals transform into fcc at relative moderate pressures less than 10~GPa, thus, check mark again, before they actually undergo a series of additional phase transformations at even higher pressures~\cite{Degtyareva2010HPR}. 
Thus, despite being highly flawed, Eq.~\eqref{eq:pairplusdens} contains a few elements of truth for alkali metals.
However, the existence of non-closed-packed equilibrium structures at large compression is also a clear indication that the assumption of pairwise repulsion may not be particularly accurate.

In a seminal work, in which the term \emph{electron correlation} appears to have been introduced, Wigner~\cite{Wigner1934PR} demonstrated that free electrons at small density can lower their energy compared to the cheapest, simple constant-density solution of fermions, which is spanned by the product of two Slater determinants, one for spin up and one for spin down, in which all $k$-states are filled up to $k_\textrm{F}$.
At electron densities characteristic for alkali metals and using second-order perturbation theory, he found them to condense into a bcc crystal. 
One possible conclusion is that a reliable potential for alkali metals should implicitly reflect higher-order gradients in the electronic charge density.
Stabilizing the right phase for the wrong reason, e.g., with potentials merely depending on (estimates for) the local electronic charge density and/or by tweaking cut-off distances or functional dependencies, as done, for example by some funky potentials in Sect.~\ref{sec:funky}, cannot lead to accurate, transferable potentials.

\subsection{Embedded-atom method and second-moment tight-binding potentials}
\label{sec:eam}

A simple generalization of Eq.~\eqref{eq:pairplusdens} is
\begin{equation}
  U( \{ \v{r}_{i} \} )
  =
  \frac{1}{2}\sum_{i,j} U_2( r_{ij} ) + \sum_i F(\rho_i),
  \label{eq:EAMgeneric}
\end{equation}
where $\rho_i$ would be an estimate for the electronic charge density near atom $i$.
Equation~\eqref{eq:EAMgeneric} forms the basis of a class of potentials, which has two names, embedded-atom method (EAM) and tight-binding (TB) approximated to second-moment (2M) expansion.
It has been historically pursued by two communities.
The TB2M and EAM camps differ predominantly in how they motivate and later generalize $F(\rho_i)$ for alloys or encode directional bonding, which will be sketched further below.
Yet, neither camp keeps the interpretation of $\rho_i$ as the charge density of \emph{all} valence or conduction-band electrons at or averaged over the vicinity of $\mathbf{r}_i$ and assumes instead
\begin{equation}
    \rho_i = \sum_{j\ne i} f(r_{ij}),
\end{equation}
which excludes the contribution of the valence shell of the atom $i$ itself.
The function $f_j(r_{ij})$ is the square of a bond-integral in the TB interpretation and the charge density from atom $j$ seen by atom $i$ for the EAM camp. 
However, as Finnis and Sinclair~\cite{finnis_simple_1984} rightfully note in their pioneering study: ``\emph{the consequences of the form of the model [...] does not depend on the physical interpretation}.''

Ducastelle~\cite{Ducastelle1970JDP} appears to have been first to suggested Eq.~\eqref{eq:EAMgeneric} building on earlier work by Cyrot-Lackmann~\cite{CyrotLackmann1969SS} and Friedel~\cite{friedel_physics_1976} on TB2M approaches to metals. 
He also proposed what could be seen as the most generic functional form of an EAM or TB2M potential,
simple exponential repulsion and furthermore
\begin{subequations}
\begin{align}
    \label{eq:embedding-tb2m}
    F(\rho_i) & = -A \sqrt{\rho_i} \\
    \rho_i & = \sum_{j\ne i}  e^{-r_{ij}/\sigma_\text{a}}. \label{eq:EAMchargeDensity}
\end{align}
\end{subequations}
For alloys, $A$ and $\sigma_\text{a}$ depend on the atom type. 
Moreover, atom (EAM) or bond (TB2M) specific prefactors must be added to each summand on the r.h.s. of Eq.~\eqref{eq:EAMchargeDensity}.
The central arguments leading to the square-root dependence of $F(\rho)$ are sketched in Sect.~\ref{sec:beyond-second-moments}.
The main reason for the exponential dependence is, as always, the exponential (Slater) type of the atomic orbitals, even though the full TB bond integrals have polynomial prefactors, see, e.g., Ref.~\cite{pettifor_bonding_1995}.
This functional form was optimized for a wide variety of metals and alloys by Cleri and Rosato~\cite{cleri_tight-binding_1993}.

Recast into our generalized Morse form, Eq.~\eqref{eq:genMorse}, the potential energy expression becomes
\begin{equation}
    U( \{ \v{r}_{i} \} )
    =
    \frac{1}{2}\frac{U_0}{m-n}
        \sum_{i} \left\{
            \sum_{j\not=i} m e^{n(1-r_{ij}/r_0)}
            -
            \left[
                \sum_{j\not=i} n^2 e^{2m(1-r_{ij}/r_0)}
            \right]^{1/2}
    \right\}.
    \label{eq:ducastellium}
\end{equation}
We would like to argue that the usefulness of this expression for rationalizing trends in metallic bonding (see Sect.~\ref{sec:properties}) is on par with that of the Lennard-Jones potential for noble gas atoms, which is why we feel that calling systems described by this potential \emph{Ducastellium} is as appropriate as using the well-establsihed term Lennard-Jonesium.

Historically, the EAM/TB2M potential described by Eq.~\eqref{eq:EAMgeneric} was independently discovered several times after Ducastelle's work.
For example, Gupta~\cite{Gupta1981PRB} showed that it reproduces the lattice relaxation of metals near surfaces.
Tom\'anek, Mukherjee, and Bennemann revealed its appropriateness to describe the energetics of small metal clusters~\cite{tomanek_simple_1983} as well as surface and vacancy energies of transition metals~\cite{tomanek_electronic_1985}.
In thermodynamics, a formulation identical to EAM/TB2M has been employed by Pagonabarraga and Frenkel for coarse-grained particle dynamics calculations, where $F(\rho_i)$ becomes a free-energy that is adjusted to the equation of state of a liquid~\cite{Pagonabarraga2001-ey,Warren2003-hj}.

Equation~\eqref{eq:EAMgeneric} can also be motivated from the quasiatom theory of Stott and Zaremba~\cite{stott_quasiatoms_1980} and from the effective-medium approximation theory of Nørskov and Lang~\cite{norskov_effective-medium_1980,norskov_covalent_1982}.
In their approaches, the energy gained when embedding an atom $i$ into a given site $\mathbf{r}_i$, e.g., an interstitial or vacancy site, is argued to be a functional of the electronic density at the embedding site and thereby to depend in leading order on the charge density $\rho_i$ that exists at $\mathbf{r}_i$ \emph{before} the atom is embedded. 
Daw and Baskes~\cite{daw1983PRL,daw1984PRB} built on these ideas and approximated the charge density $\rho_i$ at the embedding site $\mathbf{r}_i$ as a superposition of the charge density from neighboring atoms.
They coined the term \emph{embedded-atom method} and provided  important insight as to what extent many-body terms, as well as hydrogen, affect materials behavior including ductility.
Ercolessi, Parrinello and Tosatti renamed EAM into \textit{glue potential}~\cite{ercolessi1986PRL,ercolessi1986SS,ercolessi1988PMA} while claiming that their specific realization of an EAM potential
``\emph{accounts for all known lattice properties of Au}''~\cite{ercolessi1986PRL}. 

Modern EAM potentials use more complex functional forms than the simple exponentials of Ducastelle.
Those are out of scope for the discussion in this review and the interested reader is referred to the original literature, see for example works by Mishin and coworkers~\cite{mishin_interatomic_1999,Mishin2001PRB,mishin_embedded-atom_2002,williams_embedded-atom_2006,pun_embedded-atom_2012}.
In the spirit of early works by Foiles, Daw, and Baskes~\cite{foiles1986PRB}, parameterizations of such EAM potentials are distributed using tabulated data for $F(\rho)$, $f(r)$ and the pair-potential $U_2(r)$.\footnote{A large set of these tabulated potentials files can be found in the NIST interatomic potentials repository at \url{https://www.ctcms.nist.gov/potentials/}.}
Care has to be taken when interpolating between these data points in the final implementation of a potential, as the choice of spline order is not to be cast aside as a technicality, because it affects the properties of the potential~\cite{wen_interpolation_2015}.

In our own experience on copper~\cite{Jalkanen2015MSMSE}, the simple form first used by Ducastelle~\cite{Ducastelle1970JDP} and Gupta~\cite{Gupta1981PRB} performs best when testing structures with coordination numbers varying systematically from $Z = 2$ to $Z = 12$.
The reason for this trend might be, as so often, that simple, physically well motivated functional forms are less prone to overfitting than elaborate functions having been tweaked to enforce right numbers for selected properties in one or few structures. 
To this we wish to add that we highly doubt the pair-additive repulsion in open-shell systems to be an accurate approximation.
Thus, there is little reason to fiddle around with the 4'th digit of an embedding function, if ``bond-order effects'' on repulsive forces lead to errors in the second digit.  

\subsection{Modified-embedded-atom-method potentials}

A deficiency of EAM potentials is their generic preference for closed-packed structures. 
While funky parameterizations allow bcc to be stabilized, we refer yet again to Wigner's work on alkali metals in Sect.~\ref{sec:eddp} and repeat our claim that stabilizing the right phase for the wrong reason cannot yield a robust potential.
To better encode directional bonding in EAM, 
% which certainly is a consequence of electrons wanting to reduce symmetry, 
Baskes~\cite{baskes_application_1987,baskes1989PRB}, having silicon in mind, suggested to augment the computation of the electronic density with an angular three-body term.
Baskes already hinted~\cite{baskes1989PRB} that his explicit expression can be interpreted as a dependence of the embedding function on the gradient and higher-order spatial derivatives of the density, i.e., to replace 
$F(\rho)$ with $F(\rho,\partial_\alpha\rho,\partial_\alpha\partial_\beta\rho,...)$, where $\partial_\alpha$ is short-hand notation for $\partial/\partial r_\alpha$.
Shortly after, Baskes~\cite{baskes_modified_1992,baskes_modified_1994} suggested to replace (the estimate for) the embedding density according to $\rho_i \to \bar{\rho}_i$ with
\begin{equation}
    {\bar\rho_i}^2 = \sum_l t_i^{(l)} \left( \rho_i^{(l)} \right)^2,  
\end{equation}
the first four $\rho_i^{(l)}$ satisfying
\begin{subequations}
\label{eq:baskesInvariants}
\begin{align}
\rho_i^{(0)}& = \sum_j f_{j}^{(0)}(r_{ij}) \\
\rho_{i\alpha}^{(1)} & = \sum_j n_{ij}^{\alpha}  f_{j}^{(1)}(r_{ij})\\
\label{eq:baskesInvar2}
\left\{\rho_i^{(2)}\right\}^2 & = \left\{ \sum_j n_{ij}^{\alpha}  n_{ij}^{\beta} f_{j}^{(2)}(r_{ij})\right\}^2 - \frac{1}{3} \left\{\sum_{j} f_{j}^{(2)}(r_{ij}) \right\}^2\\
\rho_{i\alpha\beta\gamma}^{(3)} & = \sum_j n_{ij}^{\alpha} n_{ij}^{\beta} n_{ij}^{\gamma} f_{j}^{(3)}(r_{ij})
\label{eq:diamondMEAM}
\end{align}
\end{subequations}
where $t_i^{(l)}$ are weighting coefficients, $\mathbf{n}_{ij}$ is a  unit vector parallel to bond $i$-$j$, and Einstein summation convention is implied on the Cartesian indices for the squared quantities.
Here, the $f_j^{(l)}(r_{ij})$ reflect generalized, partial background electron density from atom $j$ seen by atom $i$.
Through this generalization, semi-explicit angular dependencies result, as for example, through the squaring of the $\rho_i^{(1)}$ term.
It leads to a summand of the form $n_{ij}^\alpha n_{ik}^\alpha f_j^{(1)}(r_{ij}) f_k^{(1)}(r_{ik})$, which is proportional to $\cos\vartheta_{ijk}$. 
With each higher-order term, higher-order sinusoidal dependencies on bonding angles result. 
An appealing aspect of these modifications is that modified EAM (MEAM) potentials pick up information on the local order beyond the coordination number.
For example, the nearest-neighbor shell-tensor that can be associated with Eq.~\eqref{eq:diamondMEAM} disappears for any mirror-inversion crystal but not for the diamond lattice. 
Thus, MEAM-potentials for silicon and germanium contain the square of the left-hand side of Eq.~\eqref{eq:diamondMEAM}.

We now argue how a MEAM potential can stabilize open crystal lattices.
To this end we consider the expression 
\begin{eqnarray} \label{eq:bondAngleEAM1}
    g_i & =  & {\sum_{j,k \ne i,j}}
%   \left\{ \cos\vartheta_0 - n_{ij}^\alpha n_{ik}^\alpha\right\}^2
    \left\{ \cos\vartheta_0 - n_{ij}^\alpha n_{ik}^\alpha \right\}^2
%    \cos\vartheta_{ijk}\right\}^2
    f_j(r_{ij}) f_k(r_{ik}) \nonumber
    \\
    & = & \rho_i^2 \cos^2\vartheta_0 - 2 \rho_{i\alpha} \rho_{i\alpha} \cos\vartheta_0 + \rho_{i\alpha\beta} \rho_{i\alpha\beta} - (1-\cos\vartheta_0)^2 \sum_j  f_j^2(r_{ij}),
\end{eqnarray}
where the upper index on the partial embedding charge densities was omitted.
$g_i$ clearly has a minimum at the preferred bond angle $\vartheta_0$, since $n_{ij}^\alpha n_{ik}^\alpha$ is the cosine of the angle $j$-$i$-$k$.
Now let us proceed with an ideal crystal structure, in which only partial densities from nearest neighbors substantially contribute to the shell.
In the given approximation, we can replace $f(r_{ij})$ with $\rho/Z_0$ so that after executing the restricted double sum correctly
\begin{equation}
    g_i
    \approx \rho^2 \left(
    \cos^2\vartheta_0 - 2 \cos\vartheta_0\nu_0^\alpha\nu_0^\alpha + \nu_0^{\alpha\beta}  \nu_0^{\alpha\beta}\right)  - \frac{\rho^2}{Z_0}(1-\cos\vartheta_0)^2.
    \label{eq:meam-penalty}
\end{equation}
Assigning an energy penalty on $g_i$ allows the degeneracy of different crystalline structures with identical coordination number but different bond angles to be lifted within a nearest-shell approximation.
At the same time, open structures with small $Z_0$ and correct bond angles can be favored over closed-packed structures.
It is clear that Eq.~\eqref{eq:meam-penalty} can be constructed (with the exception of the additive constant) from a linear combination of the MEAM invariants like those given in Eq.~\eqref{eq:baskesInvariants}.

Although MEAM potentials, similar in spirit to the ones just described, are routinely used in molecular simulations, it is easily argued that these MEAMs have issues.
First, there is no physically motivated reason for the existence of the upper indices in the $f_j^{(l)}$ coefficients when justifying the modification from Baskes' argument that the embedding function should also depend on derivatives of the embedding density.
Second, within this picture, the embedding function can generally depend on scalars that can be constructed from derivatives, which, however, must be constructed such that they are invariant w.r.t. a rotation of the coordinate system.
There are many more scalars of a similar order as those contained in Eq.~\eqref{eq:baskesInvariants}~\cite{Jalkanen2015MSMSE}.
For example, if $\rho_{\alpha_1...\alpha_l}(\mathbf{r}) \coloneqq \partial_{\alpha_1} ... \partial_{\alpha_l} \rho(\mathbf{r})$, then in addition to $\rho_{\alpha\beta}^2$ there can be a contribution of $\rho_{\alpha\alpha} \rho_{\beta\beta}$, which, in principle, can be linearly independent of the former, rather than to be constrained to $\rho_{\alpha\beta}^2-\rho_{\alpha\alpha} \rho_{\beta\beta}/3$, which can be loosely associated with the term of Eq.~\eqref{eq:baskesInvar2}.
Yet another invariant, which can be constructed with four Cartesian indices and a square in the densities, would be $\rho_\alpha\rho_{\alpha\beta\beta}$.
Thus, to flesh out our second point of criticism, the $l=2$ and $l=3$ terms in Eq.~\eqref{eq:baskesInvar2} are far from complete and it is not clear if the most relevant ones have been considered.

The ``expansion'' of Eq.~\eqref{eq:baskesInvar2} truncates at the third-order term, which is the lowest-order invariant that can discriminate between nearest-neighbor-shell energies of fcc and hcp.
However, different cubic structures cannot be distinguished with shell tensors of rank three or less. 
Fourth, the coefficients $t_i^{(l)}$ can depend, in principle, on $\rho_i^{(0)}$.
It might be possible to obtain some dependencies through asymptotic, dimensional, or other systematic analyses.
Guessing them correctly appears an almost impossible task.
Thus, we feel that the already impressive performance of MEAMs should be further improvable.
However, one potentially important ingredient missing is the environment dependence of the functions $f_j(r_{ij})$ beyond screening, e.g., an environment-dependence of $\sigma_\textrm{a}$.
Lifting this restriction would allow mediated interaction to take place as they occur, for example, in breathing shell models or in TB potentials going beyond the second-moment approximation, albeit at a potentially much increased computational cost. 

\subsection{Stillinger-Weber potential and extensions}
\label{sec:stillinger-weber}

The binding energy in non-funky pair-, EAM, and TB2M potentials is largest when atoms are most closely packed. 
However, the three lightest group-14 elements (C, Si, and Ge) are (meta-) stable in the diamond lattice, which has a packing fraction less than 50\% of that in fcc or hcp. 
To stabilize the \emph{open} diamond lattice, Stillinger and Weber~\cite{stillinger_computer_1985} proposed to add a rather simple angle-dependent term to a Mie pair potential, which, however, was multiplied with the rather clever cut-off function
\begin{equation}
    f_\textrm{c}(r_{ij},\gamma,r_\textrm{c}) = \exp\left[ \gamma/(r_{ij}-r_\textrm{c})\right]\,\Theta(r_\textrm{c}-r_{ij}).
\end{equation}
It has the nice property that all derivatives continuously approach zero at the cut-off distance $r_\textrm{c}$.
The angular add-on consisted of a penalty quadratic in the deviation from tetrahedral bonding on a given atom $i$, as originally proposed by Keating~\cite{keating_effect_1966}.
The extra-term is cut off when the length of one of the two bonds, $r_{ij}$ and $r_{ik}$, forming a bond angle of $\vartheta_{ijk}$ on atom $i$, approaches $r_\textrm{c}$.
Put together in the notation used throughout this article, the SW potential reads
\begin{equation}
\begin{split}
    U(\{\mathbf{r}\}) = & \sum_{i,j>i} \frac{U_0}{n-m}\left\{ m\left( \frac{r_0}{r_{ij}}\right)^n - n\left( \frac{r_0}{r_{ij}}\right)^m \right\}  f_\textrm{c}(r_{ij},\gamma_1,r_\textrm{c}) \\
    & + \sum_{i,j\ne i,k \ne i,j} U_\textrm{t}
    \left\{ \cos\vartheta_{ijk} - \cos\vartheta_\textrm{0} \right\}^2 \,
    f_\textrm{c}(r_{ij},\gamma_2,r_\textrm{c}) \,
    f_\textrm{c}(r_{ik},\gamma_2,r_\textrm{c})
\end{split}
\end{equation}
with $\cos\vartheta_\textrm{0} = -1/3$ for the tetrahedral angle.
Besides this angle, the SW potential has eight independently adjustable parameters: $U_0$, $r_0$, $m$, $n$, $\gamma_1$, $r_\textrm{c}$ for the pair potential, and in addition $U_\textrm{t}$, $\gamma_2$ for the tetrahedral bonding part. 
Although the SW potential is occasionally used to also model carbon, it should be kept in mind that the original SW potential does not recognize graphite as being energetically favorable at ambient pressure. 

Despite its simplicity and empirical nature, the SW potential can be parametrized for silicon to reproduce (by construction) its elastic properties in the diamond structure, but also fairly well some high-pressure phases, and the melting temperature at ambient pressure, including the anomalous jump from small to large density during melting~\cite{stillinger_computer_1985,Broughton1987PRB,kluge_velocity_1989}.
The latter goes hand in hand with a coordination change from $Z_0=4$ in the diamond structure to $Z\gtrsim 6$ in the liquid~\cite{Waseda1975}.

An appealing aspect of the SW potential is that the competition between tetrahedral and dense packing favored by the angular and the pair part, respectively, can be tuned through the ratio % $\varepsilon_\textrm{t0} \coloneqq 
$U_\textrm{t}/U_0$~\cite{Molinero2006PRL}.
By increasing it from a value characteristic for germanium through that of silicon, it is possible to make the disordered phase adopt a local tetrahedral order at large $U_\textrm{t}/U_0$, which explains why SW is occasionally used for the united-atom modeling of water~\cite{Dhabal2016JCP}.
Using small $U_\textrm{t}/U_0$, SW can describe condensed phases of tin, including that condensing in the $\beta$-tin structure~\cite{hujo_rise_2011}, which can be seen as a compromise between tetrahedral and close packing. 

Of course, as Stillinger and Weber~\cite{stillinger_computer_1985} admitted, their model has quantitative deficiencies.
They were revealed most clearly by Biswas and Hamann~\cite{Biswas1985PRL}, who demonstrated that the equation of state of any crystalline structure other than the diamond structure is highly flawed. 
However, even for the diamond structure, the deficiency of the SW can become qualitative.
For example, SW fails to reproduce that silicon is brittle under tension, although SW can be used to mimic brittle (non-silicon) tetrahedral solids, see Refs.~\cite{holland_ideal_1998,holland_erratum_1998}.
This weakness is shared by other empirical potentials and related to the finite range of the cutoff, which will be discussed in more detail in Sect.~\ref{sec:limiting}.

The SW potential formally looks like a cluster potential.
Given that the three-body term has a zero-energy contribution in the ideal diamond structure, it must be concluded that the pair interaction is an effective $Z_0=4$ pair potential rather than the ``true'' pair potential. 
Improvements of the SW address this issue partially by augmenting SW with an environment dependence in terms of a coordination-dependent equilibrium bond angle $\vartheta_0$.
Such environment-dependent interaction potentials (EDIPs) were first developed for silicon~\cite{bazant_environment-dependent_1997,justo_interatomic_1998} and later for other elements including carbon~\cite{marks_generalizing_2000,marks_modelling_2002}. 
Due to its ability to reflect the quasi-degeneracy of graphite and diamond, an EDIP for carbon was the first empirical interatomic potential that correctly predicted the relation of $3$-fold (graphite-like, sp$^2$-hybridized) and $4$-fold (diamond-like, sp$^3$-hydridized) atoms in amorphous carbon obtained from liquid quenches at varying densities~\cite{marks_modelling_2002}.

Another interesting generalization of the SW potential, which Biswas and Haman~\cite{Biswas1985PRL} proposed to model bond-angle energetics on atom $i$ beyond the  Keating model, reads
\begin{equation}
    U_3(r_{ij}, r_{ik}, r_{jk}) = \sum_l C_l F_l(r_{ij}, r_{ik}) P_l(\cos\vartheta_l),
\end{equation}
where $C_l$ are expansion coefficients, $P_l(\cos\vartheta)$ Legendre polynomials, and the $F_l(r_{ij}, r_{ik})$ functions, which can be defined to reproduce the exact or any general three-body potential.
When assuming them to factorize as $F_l(r_{ij},r_{ik}) = \phi_l(r_{ij}) \phi_l(r_{ik})$, e.g., with simple exponential dependencies, Biswas and Hamann demonstrated the computation of $U_3$ to be reducible to order $Z_\textrm{loc}$, where $Z_\textrm{loc}$ is the number of atoms within the interaction range of atom $i$, while irreducible potentials, such as the ATM potential, require a number of operations proportional to $Z_\textrm{loc}^2$. 

\subsection{Tersoff potentials}

Dissatisfied by the poor transferability of the SW potential~\cite{Biswas1985PRL} and inspired by Abell's analysis of the sensitivity of bond strengths on the local environment~\cite{abell1985PRB}, Tersoff presented a series of papers~\cite{tersoff_new_1986,tersoff_new_1988,tersoff_empirical_carbon_1988,tersoff_empirical_silicon_1988,tersoff_modeling_1989}, in which he introduced the concept of bond order to the world of empirical potentials.
According to Pauling~\cite{pauling_nature_1960}, the bond order is the difference of the number of electrons in bonding and anti-bonding orbitals.
It is a monotonically decreasing function of the coordination number. 
Tersoff's work was an attempt to construct functions that measure (effective) coordination numbers and thereby the bond order.
To do so, he introduced a cut-off function, which is unity up to a distance $r_{\textrm{c}1}$ meant to include nearest neighbors and which then quickly falls off to zero at a distance $r_{\textrm{c}2}$ supposedly less than typical next-nearest neighbor distances:
\begin{equation}
    f_\textrm{c}(r) = \Theta(r_{\textrm{c}1}-r) + \frac{\Theta(r-r_{\textrm{c}1})\Theta(r_{\textrm{c}2}-r)}{2}\left\{ 1 + \cos\left( \pi\frac{r-r_{\textrm{c}1}}{r_{\textrm{c}2}-r_{\textrm{c}1}} \right)\right\}.
\end{equation}
With the help of this cut-off function, the bonds to atom $i$ other than the $ij$ bond can be characterized with
\begin{equation}\label{eq:effZ}
    \xi_{ij} = \sum_{k\ne i,j} g(\cos\vartheta_{ijk}) f_\textrm{c}(r_{ik}),
\end{equation}
where the angular dependence, originally chosen as
\begin{equation}
    g(\cos\vartheta) = 1+\frac{c^2}{d^2} - \frac{c^2}{d^2+(\cos\vartheta_0-\cos\vartheta)^2}
\end{equation}
will later help to make a crystal in the diamond structure resist shear stresses. 
Here $c$ and $d$ are adjustable parameters, while $\vartheta_0$ is, for example, the ideal tetrahedral angle. 
The term $\xi_{ij}$ can be interpreted as an effective coordination number (minus one, since atom $j$ is excluded from the sum) under which atom $i$ ``sees'' atom $j$. 
If all bond angles on atom $i$ are equal to $\vartheta_0$, $\xi_{ij}=Z_i-1$ assuming the cut-off function to assume the values one and zero for nearest and next-nearest neighbors, respectively. 
However, ``unhappy'' angles formed by $ij$ and $ik$ bonds lead to an increase of $\xi_{ij}$.

Tersoff then constructed a prefactor, $b_{ij}$,  to the attractive pair interaction as a function of the (effective) coordination number and called it a bond-order variable.  
It is defined heuristically as 
\begin{equation} 
    b_{ij}(\xi_{ij}) = \left\{ 1 + (\beta_i\xi_{ij})^{\gamma_i} \right\}^{-1/(2 \gamma_i)},
    \label{eq:bond-order-tersoff}
\end{equation}
where $\beta_i$, and $\gamma_i$ are element-specific parameters.
For large $\xi_{ij}$, $b_{ij}$ scales with $1/\sqrt{\xi_{ij}}$, which was shown by Brenner~\cite{brenner_relationship_1989} to be similar to EAM or TB2M potentials.
With these ingredients, a general Tersoff potential reads
\begin{equation}
    U(\{\mathbf{r}\}) = \frac{1}{2}\sum_{i,j\ne i} \left\{ u_\textrm{r}(r_{ij}) - b_{ij} u_\textrm{a}(r_{ij}) \right\} f_\textrm{c}(r_{ij}),
    \label{eq:tersoff-potential}
\end{equation}
where the (pair) contributions $u_\textrm{r}(r_{ij})$ and $u_\textrm{a}(r_{ij})$ are strictly repulsive and strictly attractive, respectively.
Typically, $u_\textrm{a,r}(r_{ij})$ are exponential functions to reflect the universal equation of state of crystals~\cite{rose_universal_1983,abell1985PRB}. 

Using a symmetrized bond-order parameter, $\bar{b}_{ij} = (b_{ij}+b_{ji})/2$, the Tersoff potential can be cast as
\begin{equation}
\label{eq:generalized-tersoff}
    U(\{\mathbf{r}\}) = \sum_{i,j>i} \frac{U_0}{n-m}  \left\{
    m\, e^{n(1-r_{ij}/r_0)} - \bar{b}_{ij}\, n \, e^{m(1-r_{ij}/r_0)} \right\} f_\textrm{c}(r_{ij}).
    % \left(m e^{n(1-r/r_0}} \right)
\end{equation}
Tersoff chose $n=2m$, as to reproduce the original Morse potential~\cite{tersoff_empirical_silicon_1988}.
From the Ducastelle-potential perspective, i.e., when setting $g(\cos\vartheta)$ to unity, this is precisely the choice of exponents, in which the binding energy is insensitive to the coordination number within the nearest-neighbor approximation, see our discussion on this in Sect.~\ref{sec:lattice-props}.
Tweaking the exponents toward larger (smaller) $n/m$ ratios biases structures in favor of larger (smaller) coordination numbers. 

A point that might be particularly important and easy to make in the context of bond-order potentials is that the effect of screening is cast through a cutoff rather than through the analysis of local topology. 
For example, given fixed values for the two cut-off radii $r_{\textrm{c}1,2}$ in the Tersoff potential, it simply seems wrong that the variable $\xi_{ij}$ can change from its generic value of 3 in the diamond structure to 15 upon a relatively minor isotropic compression or to 0 upon a hypothetical homogeneous decompression.
A more elaborate discussion of this issue is given in Sect.~\ref{sec:limiting}.

\subsection{Generic functional form}
\label{sec:pastewkaPotential}

All open-shell potentials discussed in the preceding sections, and the closed-shell ATM potential, can be cast into the universal functional form given by Eqs.~\eqref{eq:genU} and \eqref{eq:genxi}.
The universal form highlights similarities between the construction of these empirical potentials and simplifies the analytic manipulation, such as the computation of properties that depend on derivatives like forces,
\begin{eqnarray}
\v{f}_n = & - & \sum_j \left(\frac{\partial U_2}{\partial r_{nj}} + \frac{\partial U_\text{m}}{\partial r_{nj}}\right) \frac{\v{r}_{ij}}{r_{ij}} \nonumber \\
          & - & \frac{1}{2}\sum_{ij} \frac{\partial U_\text{m}}{\partial\xi_{ij}}\sum_{k\not=i,j}\left\{ \mathbf{\Xi}_{ij}(\delta_{in}-\delta_{jn}) + \mathbf{\Xi}_{ik}(\delta_{in}-\delta_{kn}) + \mathbf{\Xi}_{jk}(\delta_{jn}-\delta_{kn}) \right\}
\end{eqnarray}
with $\mathbf{\Xi}_{ij}\equiv (\partial \Xi/\partial r_{ij})\, \hat{n}_{ij} $, where $r_{ij}, r_{ik}, r_{jk}$ are the first, second, and third argument of $\Xi(...)$. The remaining algebraic operations are partial derivatives in scalar variables that are straightforward to carry out.
Stresses or higher order derivatives like elastic constants are equally straightforward to evaluate.

We summarize the functional forms for the potentials discussed here in Tab.~\ref{tab:genericfunc}.
Both ATM and SW potentials have a many-body contribution of the form
    $U_\text{m}(r_{ij},\xi_{ij})
    =
    \xi_{ij}$,
which reduces the many-body contribution to a true three-body contribution, e.g., $U_3$ in Eq.~\eqref{eq:vExpandGalilei}.
Carlsson~\cite{carlsson_beyond_1990} dubbed these types of expressions \emph{cluster potentials}.
Ducastelle and Tersoff introduce a nonlinear mapping for the functional dependency on $\xi_{ij}$ in $U_\text{m}$, turning the cluster potential into what Carlsson termed a \emph{cluster functional}.
The critical difference is that while cluster potentials go to a finite order in the slowly converging series Eq.~\eqref{eq:vExpandGalilei}, cluster functionals implicitly include higher-order terms.
This can be seen by expressing the many-body contribution as a Taylor series in $\xi_{ij}$.
This Taylor series is truncated for cluster potentials but contains quadratic $\xi_{ij}^2$, cubic $\xi_{ij}^3$ and higher-order terms for cluster functionals.
The quadratic term,
\begin{equation}
    \xi_{ij}^2 = \sum_{k,l} \Xi(r_{ij}, r_{ik}, r_{jk}) \Xi(r_{ij}, r_{il}, r_{jl}),
\end{equation}
clearly contains a four-body contribution to $U_4$ in Eq.~\eqref{eq:vExpandGalilei}.
Similarly, the cubic terms contributes a five-body interaction and so on.
This implicit incorporation of higher-order terms % circumvents issues with 
can alleviate the slow convergence of the formal series expansion given by Eq.~\eqref{eq:vExpandGalilei}, while avoiding computationally expensive, explicit calculations of higher-order terms.

\begin{table}
\begin{align*}
     U(\{\v{r}\}) &= \frac{1}{2}\sum_{ij} \left\{ U_2(r_{ij}) + U_\text{m}(r_{ij}, \xi_{ij}) \right\}
     \quad
     \text{with}
     \quad
     \xi_{ij} = \sum_{k\not=i,j} \Xi(r_{ij}, r_{ik}, r_{jk})
    \\
    U_2(r_{ij}) &\propto \begin{cases}
    \text{$m$-$n$ Lennard-Jones} & \text{ATM, SW} \\
    \exp(-r_{ij}/\sigma_\text{r}) f_\text{c}(r_{ij}) & \text{Ducastelle/Gupta, Tersoff}
    \end{cases}
    \\
    U_\text{m}(r_{ij}, \xi_{ij}) &\propto \begin{cases}
    \xi_{ij}
    &
    \text{ATM, SW}
    \\
    \delta_{ij} \sqrt{\xi_{ij}}
    &
    \text{Ducastelle/Gupta} \\
    ( 1 + \xi_{ij})^{-1/(2 \gamma)} \exp(-r_{ij}/\sigma_\text{a}) f_\textrm{c}(r_{ij})
    &
    \text{Tersoff}
    \end{cases} 
    \\
    \Xi(r_{ij}, r_{ik}, r_{jk}) &\propto \begin{cases}
    (1+3\cos\vartheta_{ijk} \cos\vartheta_{jki} \cos\vartheta_{kij})/(r^3_{ij} r^3_{jk} r^3_{ki})
    &
    \text{ATM}
    \\
    \exp(-r_{ik}/\sigma_\text{a})
    &
    \text{Ducastelle/Gupta}
    \\
    g^\text{SW}(\cos\vartheta_{ijk})
    f^\text{SW}_\textrm{c}(r_{ij})
    f^\text{SW}_\textrm{c}(r_{ik})
    &
    \text{SW}
    \\
    g^\text{T}(\cos\vartheta_{ijk}) f_\textrm{c}(r_{ik})
    &
    \text{Tersoff}
    \end{cases}
\end{align*}
\caption{\label{tab:genericfunc}
Selected many-body potentials as cast into the generic functional form, Eqs.~\eqref{eq:genU} and \eqref{eq:genxi}. We here present simplified expressions for unary systems and have omitted prefactors to highlight functional dependencies, but generalization to alloys and introduction of prefactors is straightforward. $\sigma_\text{r}$ and $\sigma_\text{a}$ are characteristic length-scales of repulsion and attraction, respectively. $f_\text{c}$ are cutoff functions whose expressions differ between Stillinger-Weber and Tersoff, see Eq.~\eqref{eq:cutOffFunctions}. The functions $g^\text{SW}$ and $g^\text{T}$ encode the angular dependence of Stillinger-Weber and Tersoff, respectively. Divergent on-site terms $\xi_{ii}$ are implicitly excluded from summation. Generic forms of the SW and Tersoff potentials differ from their original formulation by a (constant) onsite contribution that is inconsequential for forces.}
\end{table}

\subsection{Brenner potentials}
\label{sect:brenner}

Brenner reparameterized Tersoff's functional form, specifically with $\gamma_i=1$, for modeling hydrocarbon chemistry~\cite{brenner_empirical_1990,brenner_erratum:_1992}. This potential is now known as the \emph{reactive empirical bond-order potential} (REBO). Brenner realized that Tersoff's potential essentially ignores the $\pi$ bond in $sp$-valent system. Thus, in its raw form, it is unable to describe the complex chemistry of hydrocarbons, for example radicals or bond conjugation. To correct for these deficiencies, Brenner introduced lookup tables that correct Tersoff's total energy expression depending on the coordination of the atoms involved in the bond. The many-body term becomes
\begin{equation}
    U_\text{m}(r_{ij}, \xi_{ij}) = -\left[b_{ij}(\xi_{ij}) + \Pi_{ij}(N_i,N_j, N_{ij}^\text{conj})\right]u_\text{a}(r_{ij}) f_\text{c}(r_{ij}),
\end{equation}
where $N_i$ is the coordination of atom $i$ obtained with a smooth cutoff function $f_\text{c}(r)$, $N_i=\sum_j f_\text{c}(r_{ij})$. The function $\Pi_{ij}$ accounts for the missing $\pi$-bond and is a bicubic interpolation between values tabulated for specific coordination of the atoms. $N_{ij}^\text{conj}$ is another integer variable that is $\geq 2$ if the bond is conjugated, which is the case if a neighboring atom (of the bond $i$-$j$) has a coordination less than four. $N_{ij}^\text{conj}$ increases the neighbor shell, which influences the energy of bond $i$-$j$, making the potential longer ranged than Tersoff's.

In addition to the correction for the $\pi$-bond, Brenner also corrected the $\sigma$-bond contribution, using the bond-order expression
\begin{equation}
    b_{ij}(\zeta_{ij}) = \left[ 1+ \xi_{ij} + P_{ij}(N_i^\mathrm{(H)}, N_i^\mathrm{(C)})\right]^{-1/2},
\end{equation}
where $P_{ij}$ is again a lookup table and $N_i^\mathrm{(H)}$ and $N_i^\mathrm{(C)}$ are coordination numbers of atom $i$, but only counting hydrogens or carbons, respectively, such that $N_i=N_i^\mathrm{(H)} + N_i^\mathrm{(C)}$. $P_{ij}$ is only nonzero for C-C bonds if a hydrogen is present as a bonding partner, $N_i^\mathrm{(H)}>0$. The lookup tables were fitted to atomization energies of many different hydrocarbon molecules. More information can be found in Refs.~\cite{brenner_empirical_1990,brenner_erratum:_1992}, including nodal values for the lookup table $\Pi_{ij}$ and $P_{ij}$.

A \emph{second generation reactive empirical bond-order potential} (REBO2) was published twelve years later by Brenner~\etal~\cite{brenner_second-generation_2002}. This potential improves upon the original formulation by adopting new functional forms for $u_\text{r}(r)$ and $u_\text{a}(r)$.
Specifically, the repulsion in REBO2 contains a screened Coulomb (Yukawa) contribution, as discussed in Sect.~\ref{sec:electrostaticScreening}, in addition to the purely exponential repulsion of REBO and Tersoff's formulation. 
REBO2 has two distinct angular functions $g(\vartheta_{ijk})$, depending on the coordination of atom $i$. The potential also added a (four-atom) dihedral potential, allowing proper modeling of rotations around carbon-carbon double bonds. The REBO2 potential was augmented with non-bonded (dispersion) interactions by Stuart, Tutein and Harrison~\cite{Stuart2000-jc} (a development that, despite being published earlier, occurred after the development of REBO2).

We here put forward the bold claim that REBO is the first successful machine-learned potential. Essentially, Brenner did away with the fixed functional form through his lookup tables to correct Tersoff's potential for a specific application. While the sophisticated interpolation techniques that are nowadays used for machine-learned potentials (neural networks, Gaussian processes~\cite{rasmussen_gaussian_2006}) were still under development at the time of Brenner's publications, the spline-interpolated lookup tables essentially fulfill a similar role of extrapolating in a high-dimensional space.

\subsection{Beyond second moments}
\label{sec:beyond-second-moments}

Before talking about potentials \emph{beyond} second moments, we need to take a step back and elucidate the origins of the second-moment expansion.
In particular, we want to emphasize that there is an atom-centered and a bond-centered expansion that lead to different functional forms for binding energies.
We start our discussion with the atom-centered expansion.

The TB2M expressions can be rationalized from orthogonal tight-binding models as follows: The total (band) energy of the electronic system is given by
\begin{equation}
    E_\text{band}
    =
    \tr \hat{\rho} \hat{H}
    =
    \sum_{n} f(\varepsilon_n) \varepsilon_n
    =
    \rho_{i\alpha j\beta} H_{i\alpha j\beta}
    \label{eq:band-energy}
\end{equation}
where the $f(\varepsilon)$ is the Fermi-Dirac distribution, $\varepsilon_n$ are the energy eigenvalues and $\hat{\rho}=f(\hat{H})$ is the density matrix.\footnote{Functions of the Hamitonian are defined through their action on eigenfunctions, $f(\hat{H})|n\rangle=f(\varepsilon_n)|n\rangle$, or through a power series expansion of $f$.}
At low electronic temperature, $f(\varepsilon)$ is simply a step function such that the sum in Eq.~\eqref{eq:band-energy} runs over all states with energy below the Fermi level $\varepsilon_\text{F}$.
The total energy of the system typically includes a pair-wise repulsive contribution in addition to $E_\text{band}$, which absorbs (and approximates) all the things that were implicity or explicitly neglected in the derivation of the orthogonal tight-binding band model, such as three-center terms, overlap repulsion, and others.

A crucial step in turning the tight-binding total energy picture into the atom or bond-centric pictures underlying interatomic potential was the formulation of the local density of states (DOS) based on moments of the Hamiltonian. 
In metals, bonding is determined by the shape of the electronic density of states around the Fermi level. The local density of states is given by~\cite{Finnis2005Book}
\begin{equation}
    n_{i\alpha}(\varepsilon) = \langle i\alpha| \delta(\varepsilon - \hat{H}) |i\alpha\rangle,
    \label{eq:local-dos}
\end{equation}
where $|i\alpha\rangle$ is orbital $\alpha$ on atom $i$.
In terms of this local density of states, the band energy becomes
\begin{equation}
    E_\text{band}=\sum_{i\alpha} \int \dif \varepsilon\, \varepsilon f(\varepsilon) n_{i\alpha}(\varepsilon).
    \label{eq:band-energy-in-iterms-of-local-dos}
\end{equation}
The \emph{moments} of the local density of states are given by~\cite{cyrot-lackmann_sur_1968,ducastelle_moments_1970},
\begin{equation}
    \mu_{i\alpha}^{(p)}
    =
    \int \dif \varepsilon\, \varepsilon^p n_{i\alpha}(\varepsilon)
    =
    \langle i\alpha | \hat{H}^p | i\alpha\rangle
    =
    H_{i\alpha j\beta} H_{j\beta k\gamma} \cdots H_{l\delta i\alpha},
    \label{eq:dos-moment}
\end{equation}
without a sum over $i\alpha$ (although that index is repeated).
The right hand side of Eq.~\eqref{eq:dos-moment} has an intuitive interpretation. For the second moments,
\begin{equation}
    \mu_{i\alpha}^{(2)} = H_{i\alpha j\beta} H_{j\beta i\alpha},
    \qquad\text{(no summation over $i\alpha$!)}
    \label{eq:second-moment}
\end{equation}
we hop from atom $i$ to all neighboring atoms $j$ and back to atom $i$ and the sum of the squared Hamiltonian elements given the moment. For the third moment,
\begin{equation}
    \mu_{i\alpha}^{(3)} = H_{i\alpha j\beta} H_{j\beta k\gamma} H_{k\gamma i\alpha},
\end{equation}
we take all self-returning hopping paths involving three atoms, and so on.
Since the Hamiltonian matrix is short ranged (remember, it decays roughly like an exponential!), the second moment only depends on the close vicinity of atom $i$. Higher moments ``feel'' the atomic structure out to further distances.
The moments are useful because they characterize the shape of the density of states. If we know all moments, we know the density of states; conversely we can approximate the density of states using just the lowest moments.

Systematically approximating the density of states from the lowest moments is highly nontrivial, and a number of approaches have been suggested.
These approaches include the recursion method~\cite{haydock_electronic_1972,haydock_electronic_1975} (that lead to the systematic development of \emph{bond-order} potentials~\cite{pettifor_new_1989,horsfield_bond-order_1996}), the maximum entropy method~\cite{Jaynes1957-sy,Jaynes1957-dv,Mead1984-vm,Brown1985-fx}, or simply approximating the local DOS with a rectangle~\cite{friedel_band_1973,friedel_physics_1976}.
Let us now assume we have an unary (for simplicity $s$-valent) system (e.g. Cu) with bond integral $\beta_{ss\sigma}(r)$ for the ss$\sigma$-bond.
There is no angular dependence and the second moment of the local DOS at atom $i$ -- see Eq.~\eqref{eq:second-moment} -- is given by
\begin{equation}
    \mu^{(2)}_{is}=\sum_j \beta_{ss\sigma}^2(r_{ij}).
    \label{eq:secondmoment}
\end{equation}
Assuming a half-filled rectangular DOS and a single electron per atom in our $s$-valent metal, the contribution of atom $i$ to the band \emph{energy} (the first moment of the local DOS integrated up to the Fermi level) becomes
\begin{equation}
    E_{\text{band},i} = -\sqrt{ 6 \mu_{is}^{(2)}},
    \label{eq:siteenergysvalent}
\end{equation}
which is the origin of the square-root in Eq.~\eqref{eq:embedding-tb2m}.
A maximum entropy estimate would lead to a Gaussian, which changes the numerical prefactor of Eq.~\eqref{eq:siteenergysvalent} but not the functional form~\cite{ackland_validity_1988}.

Semiempirical atom-centered higher-moment methods have also been developed.
Carlsson considered fourth moments (that define something like the kurtosis of the DOS) for models of semiconductors~\cite{carlsson_generalized_1990} and transition metals~\cite{carlsson_angular_1991}.
While higher-moment atom-centered techniques have seen some development since~\cite{hansen_self-diffusion_1991,foiles_interatomic_1993}, they have to the best of our knowledge not been widely employed in molecular calculations.
The breakthrough of higher-moment methods came with the development of systematic bond-centered techniques, briefly described in the following.

The arguments present above imply an atom-centered view, as the local density of states is a per-atom quantity.
For a slight shift in perspective, we abandon this band-centric view and define
$E_\text{band} = E_\text{bond} + E_\text{prom}$
where the \emph{bond energy}
\begin{equation}
    E_\text{bond} = \frac{1}{2} \sum_{i\not=j} 2 \rho_{i\alpha j\beta} H_{i\alpha j\beta}
    \label{eq:bondenergy}
\end{equation}
and the \emph{promotion energy} $E_\text{prom}$ contains the diagonal terms, i.e. the on-site energies of the tight-binding model.
This construction is the basis of the tight-binding \emph{bond} model~\cite{sutton_tight-binding_1988}.
Equation~\eqref{eq:bondenergy} for the bond energy is appealing, because it introduces the bond $i$-$j$ as the central quantity: It has the appearance of a pair interaction with pair energy $E_{ij}=2 \rho_{i\alpha j\beta} H_{i\alpha j\beta}$ -- compare it for example to the attractive part of the Tersoff potential given by Eq.~\eqref{eq:tersoff-potential}.
The term $2 \rho_{i\alpha j\beta}$ is called the \emph{bond order}.
As we know from Tersoff's potential, this bond order -- or rather the bond order variable $b_{ij}$, Eq.~\eqref{eq:bond-order-tersoff} -- depends on the environment of the specific bond and modulates its strength.
The promotion energy is constant if local charge neutrality is imposed and can be ignored for the computation of forces and stresses~\cite{horsfield_bond-order_1996}.

We can write the density matrix as
\begin{equation}
    \rho_{i\alpha j\beta} = \int \dif \varepsilon \, f(\varepsilon) n_{i\alpha j\beta}(\varepsilon)
    \quad\text{with}\quad
    n_{i\alpha j\beta}(\varepsilon) = \langle i\alpha | \delta(\varepsilon - \hat{H}) | j\beta \rangle,
\end{equation}
which is easily seen by inserting this expression into Eq.~\eqref{eq:band-energy}.
Note that written this way, the density matrix has a form similar to Eqs.~\eqref{eq:local-dos} and \eqref{eq:band-energy-in-iterms-of-local-dos}.
The onsite terms $\rho_{i\alpha i\alpha}$ correspond to the integral over the local density of states $n_{i\alpha}(\varepsilon)\equiv n_{i\alpha i\alpha}(\varepsilon)$ up to the Fermi level, which are the number of electrons $N_{i\alpha}$ in that orbital.
We now consider the linear combination of orbitals $|\pm\rangle = (|i\alpha\rangle \pm |j\beta\rangle)/\sqrt{2}$ that correspond to bonding ($+$) and anti-bonding ($-$) states~\cite{pettifor_bonding_1995}.
Then $n_\pm = \frac{1}{2}\left\{ n_{i\alpha} + n_{j\beta} \right\} \pm n_{i\alpha j\beta}$
and hence $n_{i\alpha j\beta}=(n_+-n_-)/2$.
The bond-order can therefore be written as
\begin{equation}
    2\rho_{i\alpha j\beta} = N_+ - N_-,
\end{equation}
with $N_\pm=\int\dif \varepsilon f(\varepsilon) n_\pm(\varepsilon)$.
This is the difference of electrons in anti-bonding ($N_-$) and bonding ($N_+$) orbitals.
We conclude that Pauling's interpretation~\cite{pauling_nature_1960} of the bond-order was indeed correct.

Exploiting the symmetry of the $i$-$j$ bond, the off-diagonal Hamiltonian block $H_{ij}$ can be diagonalized (by rotating into the frame of the local bond $i$-$j$), allowing the pair energy to be written as~\cite{pettifor_exact_1990},
\begin{equation}
    E_{ij} = 2 H_{\sigma}(r_{ij}) \rho_{\sigma,ij} + 2 H_{\pi}(r_{ij}) \rho_{\pi,ij} + 2 H_{\delta}(r_{ij}) \rho_{\delta,ij},
    \label{eq:bondenergy-decomposition}
\end{equation}
the sum of $\sigma$, $\pi$ and $\delta$ contributions to the bond energy.
Pettifor, Aoki, Horsfield and coworkers~\cite{pettifor_new_1989,Pettifor1991-or,Nishitani1994-jt,horsfield_bond-order_1996} developed a systematic expansion of the bond-order in the moments computed on atom $i$ and $j$.
To second order, they find $\rho_{\sigma,ij}\propto (\mu_{i\sigma}^{(2)} + \mu_{j\sigma}^{(2)})^{-1/2}$, leading to an expression similar to Tersoff's total energy (if $\pi$-bonds are neglected).
Since this pioneering work, higher-order bond-order potentials have been developed for a number of elements and elemental combinations, such as hydrocarbons~\cite{oleinik_analytic_1999,mrovec_atomistic_2007,zhou_analytical_2015}, molybdenum~\cite{mrovec_bond-order_2004}, gallium-arsenide~\cite{murdick_analytic_2006}, silicon~\cite{gillespie_bond-order_2007}, tungsten~\cite{mrovec_bond-order_2007}, iron~\cite{mrovec_magnetic_2011}, cadmiun-telleride~\cite{ward_analytical_2012}, and aluminum-copper-hydrogen~\cite{zhou_analytical_2015_2,zhou_analytical_2016,zhou_bond-order_2018}.

\section{Properties of pair and many-body potentials}
\label{sec:properties}

\subsection{Binding energy, lattice constants, and equation of state}
\label{sec:lattice-props}

The properties of crystals with a single atom in the primitive cell that interact through a smooth, short-range two-body potential $U_2(r)$ [with a single (steep) minimum in $U_2(r)$ and a single extremum in relevant higher-order derivatives] are severely restricted. 
We start by discussing the cohesive energy, arguably the most important property of any structure.
For a perfect mono-atomic and thus non-ionic crystal, in which each atom is equivalent, the energy per atom in the (effective) pair-potential approximation can be given by 
\begin{equation}
\label{eq:energyCrystPa}
    U_\textrm{pa}(a_0) = \frac{1}{2} \sum_{s=0} Z_s U_2(a_s).
\end{equation}
Values for $Z_0$, $Z_1$, $a_1/a_0$ are stated in Table~\ref{tab:ntensor} for mono-atomic cubic crystals along with other characteristic, dimensionless numbers, which will be introduced further below. 
Given that $Z_0=12$ is the maximal coordination number, face-centered cubic (fcc) and hexagonal closed packed (hcp), which has the same $Z_0$, $Z_1$, and $a_1/a_0$ as fcc, are the generally preferred structures of mono-atomic systems with short-range interactions. 
What phase wins depends on the precise functional form of the potential but also where and how the potentials are cut off, as discussed in more detail in Sect.~\ref{sec:cutting}.

To lowest order, i.e., when only including interactions within the first shell, the nearest-neighbor spacing satisfies $a_0 = r_0$, where $r_0$ is the location of the energy minimum of the dimer. 
Like nearest neighbors, more distant neighbor shells are located at a negative potential energy.
As a result, the total binding energy per atom in the solid will exceed $Z_0$ times the energy per atom in the dimer, which is $U_0/2$.
Due to the further shells sitting in the attractive part of the potential, the lattice will contract. 
As a consequence of this contraction, nearest neighbors are pushed more deeply into their repulsive interaction, which then leads to an increase of the bulk modulus $B$, despite counteracting effects of other shells.
The inequalities
\begin{subequations}
\begin{align}
\label{eq:inequaltiesPP} 
    a_0^\textrm{eq}/r_0 & \lesssim 1 \\
    U_\textrm{pa}^\textrm{eq}(Z_0)/U_0   & \gtrsim
    Z_0/2  \\
    B^\textrm{eq} & \gtrsim {Z_0} \, \frac{k_\textrm{d}r_0^2}{18\,v_\textrm{pa}^\textrm{eq}}
    % Z_0 / \left( 18 v_\textrm{pa}^\textrm{eq} \right)
\end{align}
\end{subequations}
encapsulate this discussion, $U_\textrm{pa}^\textrm{eq}$ being the binding energy per atom (in the ground state with nearest-neighbor distance $a_0^\textrm{eq}$) and $v_\textrm{pa}^\textrm{eq}$ is the volume per atom. %, both in the respective solid. 
Values of $v_\textrm{pa}$ for different structures are tabulated in Tab.~\ref{tab:ntensor}. These equations are expected to hold for non-ionic solids interacting through pair potentials.
$k_\text{d}$ is the curvature of the pair potential in the minimum, which could be called the dimer spring constant. 
It takes the value
\begin{equation}
    k_\textrm{d} = m n U_0/r_0^2
    \label{eq:pair-potential-stiffness}
\end{equation}
for the pair potentials considered here, at least in their most simple form without amendments. 
The presented treatment is easily generalized to simple binaries as long as Coulomb interactions are negligible.

Ionic crystals with long-range Coulomb interaction must be treated separately, because $Z_s$ can be of order $Z_0 a_s/a_0$ so that the sum over shells is not convergent. 
Summing up Coulomb interactions (in periodically repeated systems) is an entire industry by itself~\cite{Fennell2006JCP,Cisneros2013CR}. 
However, as long as crystallographic positions are fixed relative to the unit cell, it is clear that the sum over $1/r^m$ terms is a geometric factor depending on the lattice times $1/a_0^m$.
To keep the treatment analytically amenable, we only consider nearest-neighbor repulsion plus Coulomb interaction and approximate the energy per atom as
\begin{equation}
\label{eq:energyIonicCryst}
    U_\textrm{pa}(a_0) \approx \frac{Z_0}{2}\,\frac{U_0}{n-1}\,e^{n(1-a_0/r_0)} - \frac{\alpha_\textrm{M}}{8\pi\varepsilon_0}\,\frac{Q^2}{a_0},
\end{equation}
where $\alpha_\textrm{M}$ is the geometric factor for $m=1$, 
the Madelung constant. 
Since $\alpha_\textrm{M}<Z_0$ for all (simple) crystal structures, it is clear that the absolute ratio of Coulomb and repulsive potential is reduced compared to that of the dimer.
This leads to a reduction of the energy per atom in the crystal and consequently to an inversion of all three inequalities listed in Eq.~\eqref{eq:inequaltiesPP}.
Eq.~\eqref{eq:energyIonicCryst} is not necessarily sufficiently accurate to determine what phase an ionic crystal assumes as a function of the parameter $n$~\cite{Jansen1966AQC,Sangster1976AP}.
To this end, it is necessary to also include dispersive interaction and repulsion between anions plus potentially many-body terms~\cite{Jansen1966AQC}, which  are frequently cast through so-called breathing and deformable shell models as described in Sect.~\ref{sec:shellModels}.

Using Mie instead of exponential repulsion and restricting repulsion to nearest neighbors allows simple closed-form expressions to be derived~\cite{AshcroftMermin76},
\begin{subequations}
\begin{align}
a_0^\textrm{eq}/r_0 = & \sqrt[n-1]{Z_0/\alpha_\textrm{M}}  \\
U_\textrm{pa}^\textrm{eq}/U_0 = & \sqrt[n-1]{\alpha_\textrm{M}/Z_0} \, \alpha_\textrm{M} \\
%B^\textrm{eq} = & \frac{n^2}{n-1}\frac{r_0}{a_0^\textrm{eq}} \alpha_\textrm{M} \frac{U_0}{18v_\textrm{pa}^\textrm{eq}}
B^\textrm{eq} = &  \sqrt[n-1]{\alpha_\textrm{M}/Z_0} \, \alpha_\textrm{M} \,\frac{n}{n-1}\, \frac{k_\textrm{d}r_0^2}{18\,v_\textrm{pa}},
% k_\textrm{d}r_0^2  \frac{\alpha_\textrm{M}}{18v_\textrm{pa}^\textrm{eq}}\frac{n}{n-1}\frac{r_0}{a_0^\textrm{eq}}
\end{align}
\end{subequations}
where $k_\textrm{d}$ follows from Eq.~\eqref{eq:pair-potential-stiffness}.
As expected, the gain in energy and stiffness obtained through the condensation from molecules into crystals is much reduced compared to that of short-range potentials.

\begin{figure}[hbtp]
    \centering
    \includegraphics[width=0.75\textwidth]{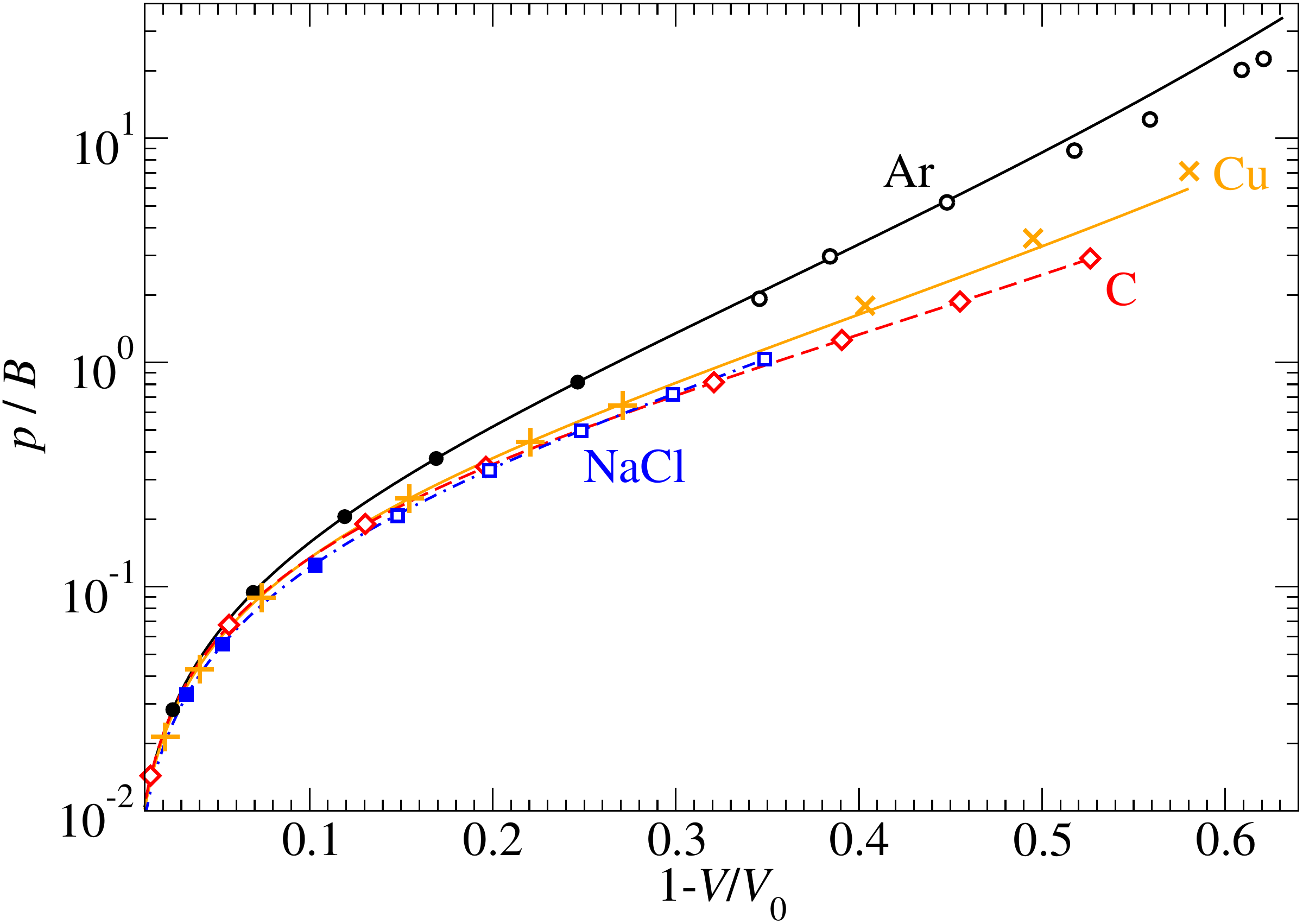}
    \caption{Equation of state, $p(V)$, for C, Cu, Ar, and NaCl as determined experimentally and by pertinent fits assuming pair-additivity of potentials of the same functional form as in Fig.~\ref{fig:allPot} and with parameters listed in Table~\ref{tab:compMolSolParam}. The pressure $p$ and the volume $V$ are normalized to the zero-pressure bulk modulus $B$ and volume $V_0$, respectively, which the fits were constrained to reproduce. Black full circes~\cite{Anderson1975JPCS} and open circles~\cite{Ross1986JCP} show reference data for fcc argon, orange plus signs~\cite{Dewaele2004PRB} and crosses~\cite{Wang2000PRL} for fcc copper, red diamonds~\cite{Occelli2003NM} for carbon in the cubic-diamond structure as well as blue, full~\cite{Boehler1980JPCS} and open~\cite{Dorogokupets2007HPR} squares for NaCl in the rock salt structure. 
    % Except for NaCl, the computed EOSs were determined with cutoffs exceeding $n$ bond lenths. (\textcolor{blue}{(MM2SS: Remind me that I ask you to work out further neighbor shells.) Only nearest neighbor repulsion but all Coulomb interactions were included for NaCl.})
    }
    \label{fig:EOSvarious}
\end{figure}

Given $U_\textrm{pa}(a_0)$, the equation of state (EOS) at small temperature follows from
$p = -\partial U_\textrm{pa} / \partial v_\textrm{pa}$, which can be evaluated quite easily through
\begin{equation}
    \label{eq:paireos}
    p(v_\textrm{pa}) = -\frac{\partial U_\textrm{pa}}{\partial a_0}\,
    \frac{\partial a_0}{\partial v_\textrm{pa}}
\end{equation}
when using the volumes per atom $v_\textrm{pa}(a_0)$ listed in Tab.~\ref{tab:ntensor}.
The lean pair potentials introduced here allow the EOS of our four reference compounds to be reproduced quite accurately in their thermodynamically stable phase, as is revealed in Fig.~\ref{fig:EOSvarious}. 
However, for copper and carbon, the obtained parameters differ quite substantially from those describing the dimers, while those for NaCl and Ar differ only by roughly 10--20\%.
Pertinent numbers are listed in Tab.~\ref{tab:compMolSolParam}.

The apparent dimer binding energies $U_0$ deduced from the fits are decreased by a factor of approximately 2.2 for carbon ($Z_0 = 4$) and by a factor of 4.6 for copper ($Z_0 = 12$).
Both numbers can be crudely rationalized using the EAM/TB2M potential proposed by Ducastelle (see Sect.~\ref{sec:eam}), which predicts the binding energy per atom to increase a little less slowly than with $\sqrt{Z_0}$.

Frequently, a measured or computed EOS is also fitted to the so-called ``universal equation of state''~\cite{rose_universal_1983,ferrante_diatomic_1983,Vinet1989JPCM} given by 
\begin{equation}
\label{eq:EOSuniversal}
    p(V) = 3\,B_0\,\frac{1-\eta}{\eta^2}  e^{3\,(B_0'-1)\,(1-\eta)/2}.
\end{equation}
Here, $B_0$ is the bulk modulus at $p = 0$ and $\eta = \sqrt[3]{V/V(p=0)}$,
while $B'_0$ is the change of the bulk modulus $\partial B/\partial p$ at $p=0$ and constant temperature, which is ideally zero temperature when gauging interaction parameters from lattice sums. 
In fact, the reference data for carbon~\cite{Occelli2003NM} in Fig.~\ref{fig:EOSvarious} was produced using Eq.~(\ref{eq:EOSuniversal}) whose adjustable coefficients had been gauged on experimental data. 

As can be seen, also an ionic solid satisfies the universal EOS under compressive stress reasonably well, although its two-body potential is markedly distinct from that of carbon or copper.
Thus, inverting an EOS, in particular when obtained solely under compression, allows potentials to be tested but not to be parametrized. 
Using first-principle methods, the EOS can be extended to tensile stresses, thereby providing useful information also on the attractive part of the potential.
Careful parameterizations of potentials make use of such data~\cite{albe_modeling_2002,albe_modeling_2002-1,nord_modelling_2003,erhart_analytical_2005}.

Even though pair potentials can be used to fit the EOS of individual crystalline structures, they are not transferable between them.
As described in detail in Sect.~\ref{sec:open-shell}, many-body terms are required for transferability.
We will now analyze the Ducastelle potential as one of the simplest many-body formulations for metals.
In the nearest-neighbor approximation, the per-bond energy of crystalline Ducastellium, Eq.~\eqref{eq:ducastellium}, can be approximated as
\begin{equation} 
U_\textrm{pb}(a_0, Z_0) = \frac{U_0}{m-n}\left\{m e^{n(1-a_0/r_0)} - \frac{1}{\sqrt{Z_0}} n e^{m(1-a_0/r_0)} \right\}.
\label{eq:ducastelle-per-bond}
\end{equation}
It follows for 
the equilibrium bond length that
\begin{equation} 
    \frac{a_0^\textrm{eq}(Z_0)}{r_0} = 1 + \frac{1}{2}\frac{\ln Z_0}{n-m}
    \label{eq:ducastellium-bond-length}
\end{equation}
and for binding energy
\begin{equation} \label{eq:bindingEnergInEAM}
    U_\textrm{pa}^\text{eq}(Z_0) = U_0 Z_0^\mu/2
    \quad\text{or}\quad
    U_\textrm{pb}^\text{eq}(Z_0) = U_0 Z_0^{\mu-1}
\end{equation}
with $\mu = (n/2-m)/(n-m)<1/2$.
Moreover, the generalized stiffness of the potential turns out to be
\begin{equation}
    k^\text{eq}(Z_0) = Z_0^{\mu-1} k^\text{eq}(1),
    \label{eq:ducastellium-stiffness}
\end{equation}
where $k^\text{eq}(1)$ is the stiffness of the bond in the dimer.
Corrections due to next-nearest and more distant shells alter these results.
However, all trends are robust and reflect the results for copper shown in Fig.~\ref{fig:structural-trends}---as well as for a few other metals condensing in fcc. 
These trends are a logarithmic increase of $a_0^\text{eq}$ with $Z_0$ and an algebraic increase of $U_\text{pa}^\text{eq}$ with a power less than $1/2$.
It is important to note that for $n=2m$ we obtain $\mu=0$, and Eq.~\eqref{eq:bindingEnergInEAM} shows that the cohesive energy is then independent of crystal structure.

Note that the same trends are also observed by carbon (also shown in Fig.~\ref{fig:structural-trends}), despite the fact that it cannot be described by Ducastelle's potential, as carbon's ground-state structures are open crystal lattices.
To understand covalent bonding, it is instructive to consider the energy per bond
\begin{equation}
    U_\text{pb}(a_0) = \frac{U_0}{n-m}  \left\{
    m\, e^{n(1-a_0/r_0)} - \bar{b}\, n \, e^{m(1-a_0/r_0)} \right\} f_\textrm{c}(r_{ij})
\end{equation}
of nearest-neighbor Tersoffium, Eq.~\eqref{eq:generalized-tersoff}.
We directly see that we recover Eq.~\eqref{eq:ducastelle-per-bond} of Ducastellium for $\bar{b}=1/\sqrt{Z_0}$, illustrating the relationship between the bond-order variable and the coordination number, see also Ref.~\cite{brenner_relationship_1989}.
The bond-order variable $\bar{b}$  typically does not depend on bond-length, but only on crystal structure.
We can therefore read-off the equilibrium bond length as,
\begin{equation}
    \frac{a_0^\text{eq}}{r_0} = 1 - \frac{\ln\bar{b}}{n - m}.
    \label{eq:tersoffium-bond-length}
\end{equation}
Similarly, the energy per bond scales as
\begin{equation}
    U_\text{pb}^\text{eq}(\bar{b}) = U_0 \bar{b}^{2-2\mu},
    \label{eq:tersoffium-bond-energy}
\end{equation}
explaining the rough structural trends in Fig.~\ref{fig:structural-trends} for carbon.
We can combine Eqs.~\eqref{eq:tersoffium-bond-length} and \eqref{eq:tersoffium-bond-energy} to eliminate $\bar{b}$, yielding
\begin{equation}
    \frac{a_0^\text{eq}}{r_0} = 1 - \frac{1}{n} \ln\frac{U_\text{pb}^\text{eq}}{U_0},
    \label{eq:pauling-relation}
\end{equation}
which is displayed in Fig.~\ref{fig:structural-trends}h.

These analytical equations are useful in parameterization of these potentials.
Albe~\cite{albe_modeling_2002,albe_modeling_2002-1,erhart_analytical_2005} developed a parameterization procedure, based on data of ground-state crystalline structures such as the one shown in Fig.~\ref{fig:structural-trends}.
The first step is fitting the parameters $n$ to the universal energy-bond relation given by Eq.~\eqref{eq:pauling-relation}.
Albe calls the corresponding plot, such as Fig.~\ref{fig:structural-trends}d for copper and \ref{fig:structural-trends}h for carbon, a Pauling plot.
The remaining (pair) parameters $U_0$, $m$ and $r_0$ are then adjusted to the binding energy, bond length and vibrational frequency of the dimer.
In a final step, the parameters that enter the specific calculation of $\bar{b}$ are fitted to reproduce detailed structural information in the database.

\subsection{Defects, melting, and boiling}

Reproducing defect energies correctly constitutes an important benchmark for potentials, as their proper description is required for the atomistic modeling of thermodynamic and non-elastic mechanical properties.
Defects are broadly characterized into point, line, and planar defects.
Reproducing their energies accurately is more difficult than reproducing an EOS, because atoms near defects lack the symmetry of ideal crystallographic positions.
Parameters having no effect on the EOS for reasons of symmetry and a small effect on lattice vibrations can become important for the proper description of defects, mostly because defects imply a coordination change and might be accompanied by the generation of radicals or ions in covalently bonded systems.

Vacancies are the most important point defects in simple crystals. 
Two values for their energy are associated with it, i.e., the vertical vacancy energy, $U_\textrm{vac}^{(0)}$, for which the lattice in the vicinity of the removed atom is kept undistorted, and the true defect energy, $U_\textrm{vac}$, which is obtained after lattice relaxation.
% so that $\Delta U_\textrm{vac} \lesssim \Delta U_\textrm{vac}^0$.
%
The $\mathcal{O}(10\%)$ reduction of the defect energy due to relaxation can be deemed irrelevant when studying trends, in contrast to the 10 to 10$^5$ fold increase in computing and human time to estimate that correction to within 10\%. 
Different system sizes would have to be selected, each structure be relaxed to the energy minimum, and results be extrapolated to the thermodynamic limit, while assessing a vertical defect energy requires merely one calculation in addition to the ones finding the crystal's ground state. 

\begin{figure}[htb]
\begin{center}
\includegraphics[width=0.75\textwidth]{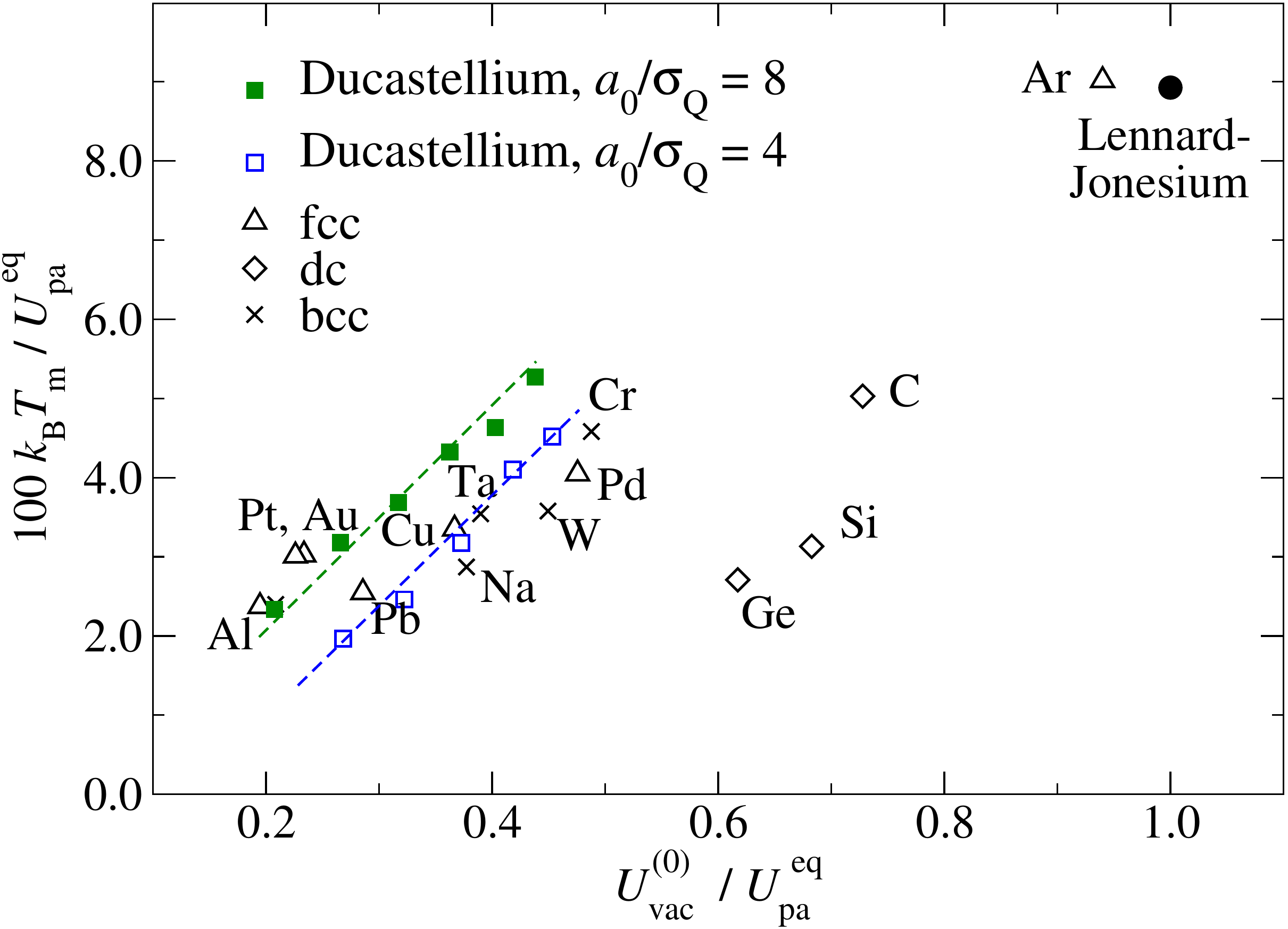}
\end{center}
\caption{\label{fig:melt_vacancy}
Melting temperature in units of cohesive energy as a function of $U_{\rm{vac}}^{(0)}/U_\text{pa}^\text{eq}$ for a variety of elemental systems condensing in bcc (crosses), fcc (triangles), and diamond-cubic structure (dc, diamonds).
Melting temperatures and cohesive energies were taken from Kittel's book~\cite{KittelBook}.
Vertical defect energies are provided for Al~\cite{Schaefer1987PSSA}, Au~\cite{Schaefer1987PSSA}, Pt~\cite{Schaefer1987PSSA}, Cu~\cite{Schaefer1987PSSA}, Pb~\cite{Schaefer1987PSSA}, Pd~\cite{Schaefer1987PSSA}, Na~\cite{March1973JPFMP}, Ta~\cite{Schaefer1987PSSA}, W~\cite{Schaefer1987PSSA}, Cr~\cite{Schaefer1987PSSA}.
For C, Si and Ge, $U_\text{pa}^\text{eq}$ values were calculated using DFT-LDA.
In addition, the vertical defect energy reduction of Ar was deduced from the ATM parameters from Ref.~\cite{Tang2012JCP}.
}
\end{figure}

For a short-range potential without explicit angular dependence, the energy per atom can be estimated with $U^\text{eq}_\textrm{pa} = U_0 Z_0^\mu/2$ -- see also Eq.~\eqref{eq:bindingEnergInEAM} --, where, however, $\mu = 1$ for pair potentials and $\mu \lesssim 1/2$ for Ducastellium.  
Removing an atom from an ideal lattice site and placing it hypothetically into another ideal lattice site
makes its $Z_0$ ditched neighbors lose one bond each, leads to an estimate of, $U_\textrm{vac} = Z_0\{ U_\text{pa}^\textrm{eq}(Z_0)-U_\text{pa}^\textrm{eq}(Z_0-1)\}$, which becomes
\begin{equation}
    U_\textrm{vac}^{(0)} = \frac{U_0}{2}   Z_0 \left\{ Z_0^\mu-(Z_0-1)^\mu\right\}
    \approx \mu U_\textrm{pa}^\textrm{eq}
\end{equation}
where the latter approximation  applies if $(1-\mu)/Z_0\ll 1$. 
Thus, $U_\textrm{vac}^{(0)} = U^\text{eq}_\textrm{pa}$ for a pair potential ($\mu =1$), while for $\mu < 1$, our back-of-the envelope calculation provides a lower, yet good estimate for $U_\textrm{vac}^{(0)}$. 
Taking into account that the bond lengths of the ditched neighbors is not at the perfect bond length for their new $Z_0-1$ coordination would lead to a relative increase of $U_\textrm{vac}^{(0)}$ of typically less than 5\% in a Ducastelle potential. 
Fig.~\ref{fig:melt_vacancy} confirms the predicted trend that $\mu'=U_\textrm{vac}^{(0)}/U_\textrm{pa}^\textrm{eq}$ is distinctly less than unity for simple metals, in fact their $\mu'$ falls slightly below 0.5, 
which is the upper bound for $\mu$ approached in the limit $n\gg m$, mentioned in the text below Eq.~\eqref{eq:bindingEnergInEAM}.
Group~14 elements have substantially larger $\mu'$ than metals, however, $\mu'$ and $\mu$ cannot be expected to be similar, whenever bond-angle corrections matter.
Only argon out of the elements considered in Fig.~\ref{fig:melt_vacancy} has a vertical defect energy reasonably consistent with the pair-potential assumption.

One reason why defect energies in metals are important is that they are believed to correlate with the melting temperature $T_\textrm{m}$:
G{\'{o}}recki found the vacancy concentration just below $T_\textrm{m}$ in many different metals to be $n_\textrm{vac}^* \approx 0.0037$~\cite{Gorecki1977SM} .
Estimating the relative number of vacancies with $n_\textrm{vac}\approx \exp(-\beta U_\textrm{vac})$ would yield 
$k_\textrm{B} T_\textrm{m}/U_\textrm{vac} = -1/\ln n_\textrm{vac}^* \approx  2/11$,
which roughly overestimates the melting temperatures reported in Fig.~\ref{fig:melt_vacancy} by a factor of two.
Thus, the simple picture laid out here is only half the story, or, depending on viewpoint, it is already half the story.
Given the utmost simplicity of the arguments, we would consider the melting-temperature glass to be half full. 

To extend the discussion to the boiling (or sublimation) temperature, $T_\textrm{b}$, the temperature has to be identified at which the Gibb's free energy $G$ of the condensed phase equals that of the gas. 
At low temperature, $G$  can be crudely estimated with the cohesive energy of the crystal and at high temperature with $-T_\textrm{b}$ times the entropy of an ideal gas.
This gives a quasi-linear relation between $U_\textrm{pa}^\text{eq}$ and $T_\textrm{m}$ with corrections logarithmic in $T_\textrm{m}$ and $p$. 
When checking data for metals, we found them to obey a $k_\textrm{B}T_\textrm{b} \approx 0.06~U_\textrm{pa}^\text{eq}$, surprisingly well. 
Unfortunately, we did not find a name for this relation, although a similar rule certainly exists in the literature.
Even water does not stray too far from this finding with $k_\textrm{B}T_\textrm{b}/ U_\textrm{pa}^\textrm{eq} \approx 0.054$, 
when expressing its cohesive energy of $U_\textrm{pm} \approx \SI{0.6}{\eV}$~\cite{Thierfelder2006PRB} in ice Ih per molecule (pm) rather than per atom. 

Given the discussions on melting and boiling temperatures, their ratio $r_\textrm{mb} \coloneqq T_\textrm{m}/T_\textrm{b}$ must be expected to correlate with the exponent $\mu$.
For our representative pair-potential element argon, melting and boiling temperature differ only by 4\% at atmospheric pressure.
%
% For water, one would obtain $r_\textrm{mb} \approx 273.15/373.15 \approx 0.73$.
%
In contrast, $r_\textrm{mb}(\textrm{alkali}) \approx 0.3$,  $r_\textrm{mb}(\textrm{earth alkali}) \approx 0.6$, and most transition metals being somewhere in between. 

Of course, much more detailed analyses are required to estimate reliably melting temperatures and boiling points from defect and cohesive energies than the crude arguments presented above.
The trouble is, as with potentials, that improving estimates even marginally requires a substantially increased effort.
Thus, for the sake of brevity, we cannot go into the required level of detail and instead invite the reader to conduct the fun/instructive exercise of melting and vaporizing a two-dimensional crystal in a simulation cell having many times times the crystal's volume: one time with a simple pair potential and one time with a generic Ducastelle or related potential.
If done correctly, interesting insights can be gained, for example, on long-range positional and orientational order in two spatial dimensions, or the exceedingly small densities required to make the metal gas consist of monomers rather than of dimers, revealing one more reason why the above presented reason for the correlation between $T_\textrm{b}$ and $U_\textrm{c}$ should be bad.  
For reasons of brevity, we do not only abstain from a more detailed analysis of the effects of the $U_\textrm{pa}^\text{eq}\propto Z^\mu$ scaling, but also from reporting trends for surface or grain boundary energies, although both are central to the properties of materials.

\subsection{Elastic properties}
\label{sec:elasticGeneralForm}

Deforming a solid from its equilibrium geometry requires the application of external stress.
For a macroscopically affine deformation it can be deduced from elastic tensors.
A brief introduction to them including their relation to pair potentials is given in Sect.~\ref{sec:Cauchy}.
%
% There are three independent elastic tensor elements for cubic materials, on which we focus in this section.
%
% They are $C_{11}$, $C_{12}$, and $C_{44}$ in Voigt notation.
%
While we have already elaborated on the Cauchy relations, it seems much less appreciated that the pair-potential approximation also places rather tight bounds on $C_{11}/B$.

The bulk modulus $B$ is easy to evaluate given the equation of state, such as Eq.~\eqref{eq:paireos},
\begin{equation}
    B
    =
    -v_\text{pa} \frac{\partial p}{\partial v_\text{pa}}.
\end{equation}
The first constraint to be discussed is the ratio $C_{44}/B$ for fcc crystals interacting with short-range potentials. 
Using the results compiled in Eq.~\eqref{eq:cubicPairSecShell}, a first estimate of $C_{44}/B = 3/4$ is obtained using the Cauchy relation $C_{44}=C_{12}$ and $B = (C_{11}+2C_{12})/3$ valid for cubic systems interacting through pair potentials. 
This is already close for to the value of $C_{44}/B \approx 0.8$ for fcc Lennard-Jonesium and, by fortuitous error cancellation, even closer to the experimental value for argon.
%
% Numbers are listed in , which is discussed in reference to many-body interactions in Sect.~\ref{sec:transferability}.
%

There are two leading-order corrections to the nearest-shell approximation, which both decrease $C_{11}$ relative to $C_{12}$ and thus $C_{44}$ so that the exact value for $C_{44}/B$ is increased compared to the nearest-shell approximation. 
First, the effective spring constant $k_1$ associated with the next-nearest neighbor shell is slightly negative, because next-nearest neighbors are located well past the inflection point of the LJ potential at zero external pressure and even further beyond the point at which $k(r)$ becomes negative.
Numerically, we obtain $k_0 \approx 129\,U_0/r_0^2$ and $k_1 \approx -4.8\,U_0/r_0^2$ when using $a_0/r_0 = 0.971$ for fcc Lennard-Jonesium.
This accounts for more than 50\% of the difference between the nearest-shell approximation and the exact value for fcc Lennard-Jonesium. 
Second, the dominant effect of more distant neighbors is similar to an isotropic compressive stress $p_\textrm{c}$, pushing the nearest neighbors more deeply into the repulsion and leading to a reduction of $C_{11}$ by $p_\textrm{c}$ and an increase of $C_{12}$ and thus $C_{44}$ by the same amount, see also Eq.~\eqref{eq:pressEffectOnCij}.

The second constraint to be discussed for a short-range potential is the stability of the bcc phase.
In the nearest-neighbor approximation, $C_{11}=C_{12}$, which means that the tensor of elastic constants is not positive definite thereby violating the Born mechanical stability criteria~\cite{Born1940MPCPS}.
Including the second shell helps stabilizing bcc, but only if the interaction is sufficiently long ranged so that $k_1$ is positive. 
For a Morse potential, we find $C_{12}$ to be less than $C_{11}$, once $n \lesssim 9$ and the binding energy per atom $E_0(\textrm{bcc})>E_0(\textrm{fcc})$ for $n \lesssim 6.1$, at which point not even 20\% of the interaction energy is associated with the nearest-neighbor shell.
Thus, stabilizing bcc over fcc using a two-body potential straying not too far away from a physically meaningful, \emph{atomistic} potential does not appear to be possible.
This can differ for positively charged colloids repelling each other through Yukawa potentials~\cite{Robbins1988JCP}.
% As the exponent is increased, the ratio $C_{12}/C_{11}$ grows and reaches unity at $n\approx 9$, and bcc phase becomes unstable according to Born mechanical stability criteria~\cite{Born1940MPCPS}.
%
% When the exponent $n\lessapprox 6.1$, $E_0(\rm{bcc}) < E_0(\rm{fcc})$ and $C_{12}/C_{11} \approx 0.80$. This means that bcc phase is elastically stable and is energetically favorable over fcc.

Some solid-state physics textbooks argue that a positive next-nearest neighbor coupling $k_1$ stabilizes the simple cubic structure against shear, polonium being the single element in the periodic table adopting it as ground state.
However, for $k_1$ to be positive at a next-nearest-neighbor distance of $r_1 = \sqrt{2}\,a_0$, the Morse exponent has to be as small as $n \approx 2.34$, which is much smaller than any reported value for a Morse type potential that we have come across.

\begin{figure}[htb]
\begin{center}
\includegraphics[width=0.75\textwidth]{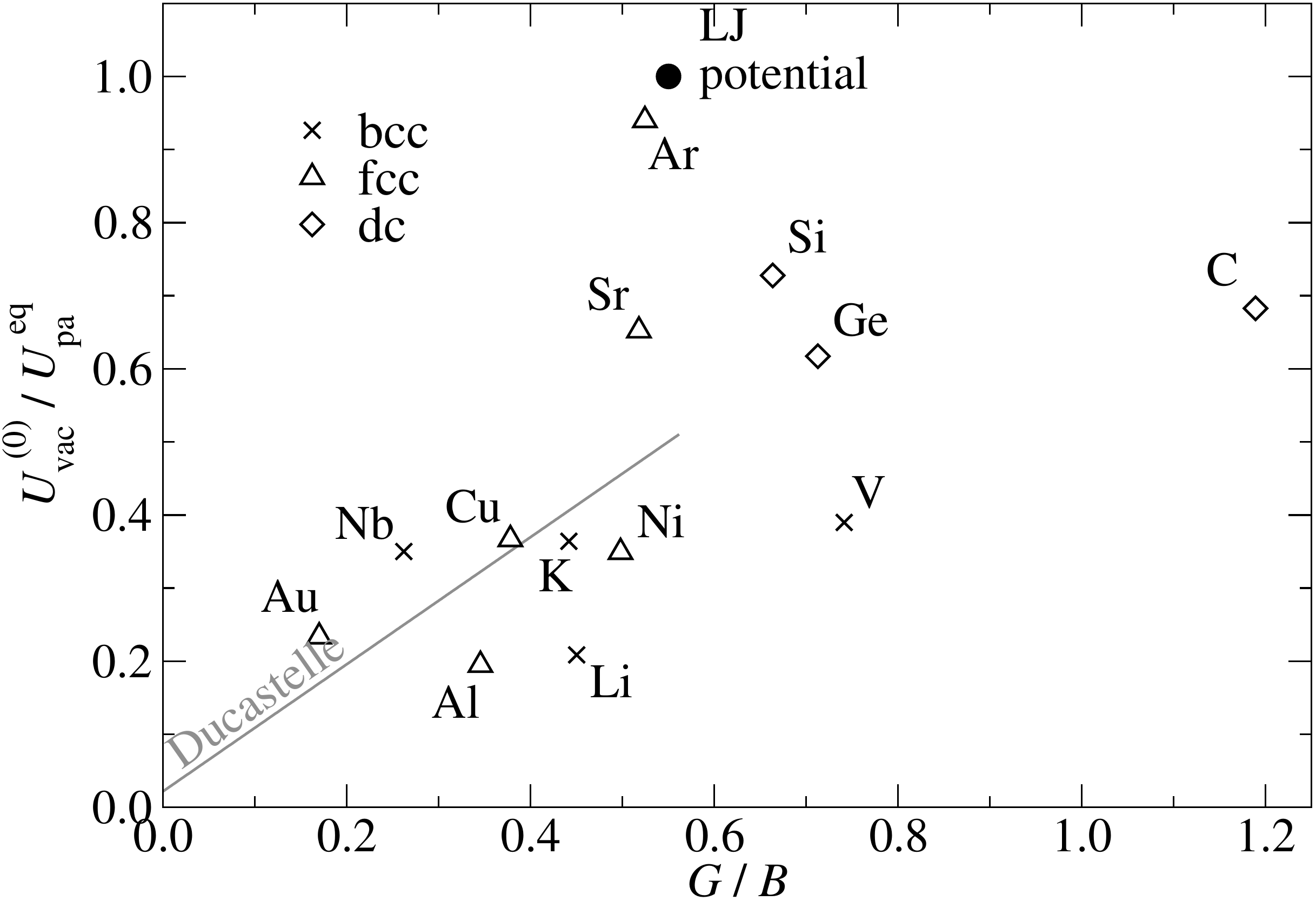}
\end{center}
\caption{\label{fig:Vacancy_Cauchy}
Vertical defect energy in units of cohesive energy as a function of the symmetrized, dimensionless shear modulus $G/B$ for a variety of elemental systems condensing in bcc (crosses), fcc (triangles), and diamond-cubic structure (dc, diamonds). 
Elastic constants data for all chosen elements but Ar were taken from Ref.~\cite{Simmons1971}.
Data for $C_{44}/C_{12}$ for Ar from Ref.~\cite{Fujii1974PRB}.
Vertical defect energies are provided for Al~\cite{Schaefer1987PSSA}, Au~\cite{Schaefer1987PSSA}, Ni~\cite{Schaefer1987PSSA}, Cu~\cite{Schaefer1987PSSA}, Li~\cite{March1973JPFMP}, K~\cite{Schaefer1987PSSA}, V~\cite{Schaefer1987PSSA}, Nb~\cite{Schaefer1987PSSA}.
The vertical defect energy reduction of Ar was deduced from the ATM parameters from Ref.~\cite{Tang2012JCP}.
For C, Si, Ge and Sr, vertical defect energies were produced using DFT-LDA, with $4\times4\times4$ super-cell to minimize the interaction between a vacancy site and its image.
The full and dashed gray lines show the prediction of the Ducastelle potential for fcc and bcc, respectively.
}
\end{figure}

Given the general form of an EAM or TB2M potential of Eq.~\eqref{eq:EAMgeneric}, elastic constants of simple crystals can be found quite easily with the formalism laid out in Sects.~\ref{sec:Cauchy},
\begin{eqnarray}
    \label{eq:elasticGenericEAM}
    v_\text{pa} C^{U\eta}_{\alpha\beta\gamma\delta}
    & = &
    \sum_s \left\{
        2
        \tilde{U}_2''(a_s^2)
        +
        F'(\rho_0) \tilde{f}''(a_s^2)
    \right\}
    a_s^4
    \nu_s^{\alpha\beta\gamma\delta}\nonumber\\
    & & +
    F''(\rho_0)
    \sum_{s,t}
    \tilde{f}'(a_s^2)
    \tilde{f}'(a_t^2)
    a_s^2 a_t^2
    \nu_s^{\alpha\beta}
    \nu_t^{\gamma\delta},
\end{eqnarray}
where $F'(\rho_0)$ and $F''(\rho_0)$ are the first and second derivative of $F(\rho)$ evaluated at $\rho_0$, respectively, $\tilde{U}_2(S)=U_2(r)$, $\tilde{f}(S)=f(r)$, and $S = r^2$.
In the nearest-neighbor approximation, this expression simplifies to
\begin{equation}
\label{eq:elasticGeneric}
    v_\text{pa} C^{U\eta}_{\alpha\beta\gamma\delta}
    =
    \left\{
        2
        \tilde{U}_2''(a_0^2)
        +
        F'(\rho_0) \tilde{f}''(a_0^2)
    \right\}
    a_0^4
    \nu_0^{\alpha\beta\gamma\delta}
    +
    F''(\rho_0)
    \left(
        \tilde{f}'(a_0^2)
        a_0^2
    \right)^2
    \nu_0^{\alpha\beta}
    \nu_0^{\gamma\delta}
\end{equation}
Similar results for elastic tensor elements have been identified numerous times~\cite{Ducastelle1970JDP,daw1984PRB} including a derivation using the shell-tensor concept~\cite{Sukhomlinov2015JCP}, however, mostly for stress-free references. 
The violation of the Cauchy relations results exclusively from the last summand on the rhs of Eq.~\eqref{eq:elasticGeneric}, e.g., for systems with inversion symmetry,
\begin{equation}
    C^{U\eta}_{12}-C^{U\eta}_{66}
    \equiv
    C^{U\eta}_{1122}-C^{U\eta}_{1212}
    =
    \frac{F''(\rho_0)}{v_\text{pa}}  \left\{\tilde{f}'(a_0^2) a_0^2\nu_0^{11}  \right\}^2
\end{equation}
in the nearest-neighbor approximation.
Thus, if the embedding function $F(\rho)$ has a positive curvature at $\rho_0$, as is the case for a Ducastelle potential with $F(\rho)\propto -\sqrt{\rho}$, then $C_{12}>C_{44}$ follows for cubic systems. 
The symmetrized shear modulus, $G \coloneqq (C_{44}+C_\textrm{s})/2$, where the so-called tetragonal shear modulus is defined as $C_\textrm{s}\coloneqq (C_{11}-C_{12})/2$, can be shown to obey $G = \alpha \mu B$ in a nearest-shell Ducastelle potential, where $\alpha=9/8$ for fcc and $\alpha = 1$ for bcc~\cite{Sukhomlinov2016JPCM}.
Thus, the most generic EAM or TB2M potential predicts $G/B$ to scale linearly with $\mu$, just like the vacancy energy. 
In fact, simple metals obey this trend, as demonstrated in Fig.~\ref{fig:Vacancy_Cauchy}.
Nonetheless, this correlation must be taken with a grain of salt, since bcc is not elastically stable in the nearest-neighbor approximation. 

In contrast to the embedding term, angular terms as those used by Keating, Stillinger-Weber, or Tersoff, counteract shear, so that they lead to a Cauchy violation having the opposite sign as that in Ducastellium and related potentials. 

\subsection{Plasticity}

Plasticity and ductility hinge on the ability of solids to introduce low-dimensional structural defects, in particular dislocations, whose presence is needed to prevent metals from brittle fracture under loading. 
The more easily (non-planar) defects are inserted into a crystal, the more ductile it will be. 
In a very microscopic picture, metals can be argued to be  ductile, because  \textit{``delocalized electrons allow metal atoms to slide past one another without being subjected to strong repulsive forces that would cause other materials to shatter''}~\cite{wikiDuctility}.
% https://en.wikipedia.org/wiki/Ductility
%
However, going directly from the delocalized nature of the electrons in metallic bonds is somewhat of a big step and does not provide any guidelines allowing the degree of ductility to be assessed. 

In this section and the literature reviewed therein, vacancy energies in units of the cohesive energy and shear elastic constants in units of the bulk modulus are found to be relatively small if interactions are isotropic and bonds weaken with increasing coordination number.
Both ratios turn out to be of similar order, and, generally less than 1/2 for most metals.  
Given that the Poisson's ratio of an isotropic material satisfies $\nu = (3/2-G/B)/(3+G/B)$, it can be seen that the low limit of $G/B \to 0$ bring the Poisson's ratio close to that of a liquid, i.e., to $\nu = 1/2$, while the lower limit for $\nu$ is just a little above the value valid for pair potentials $\nu = 1/4$.
Thus, the smaller $G/B$, the more ``liquid-like'' the elastic tensor, which, of course, also implies that the energy to introduce a dislocation, which is linear in $G$, is relatively small.
In fact, the elemental solids with a small $\mu':=G/B$ ratio, but also the ones with a small $U_\textrm{vac}/U_\textrm{c}$, are by and large more ductile than the ones with a larger $\mu'$. 
A nice example is niobium, which happens to be quite ductile despite condensing in bcc, which, unlike fcc, lacks close-packed glide planes. 
Of course, assessing the effect of defects on the plasticity of a material requires the knowledge of many more properties than the dimensionless ratios $G/B$, $k_\textrm{B}T/U_\textrm{vac}$, and so on~\cite{hirth_theory_1982, Cai2016book},
however, the trends reflected in our examples, which were not cherry picked, can certainly be understood from these numbers.

\begin{figure}
    \centering
    \includegraphics[width=0.45\textwidth]{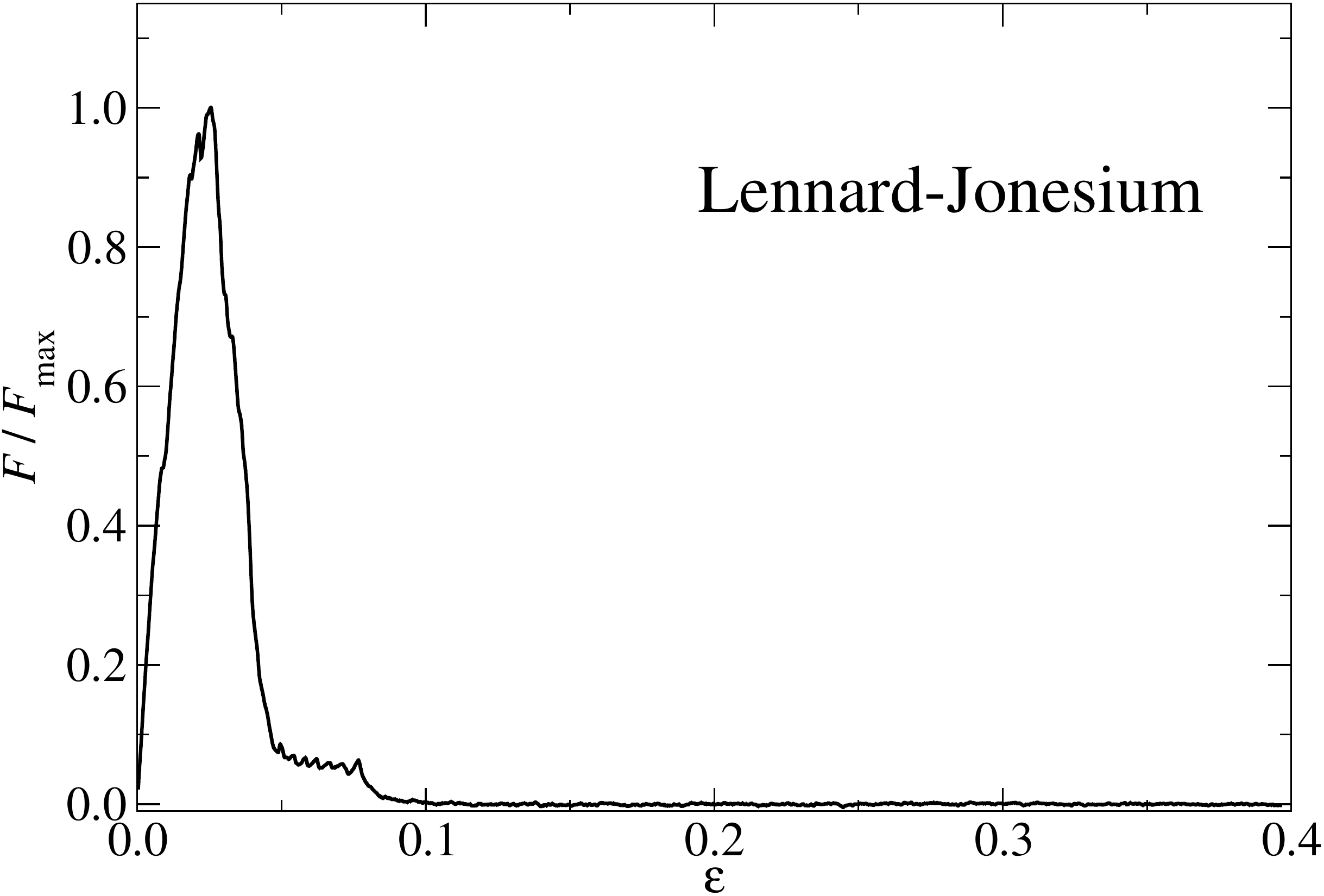}
    \includegraphics[width=0.45\textwidth]{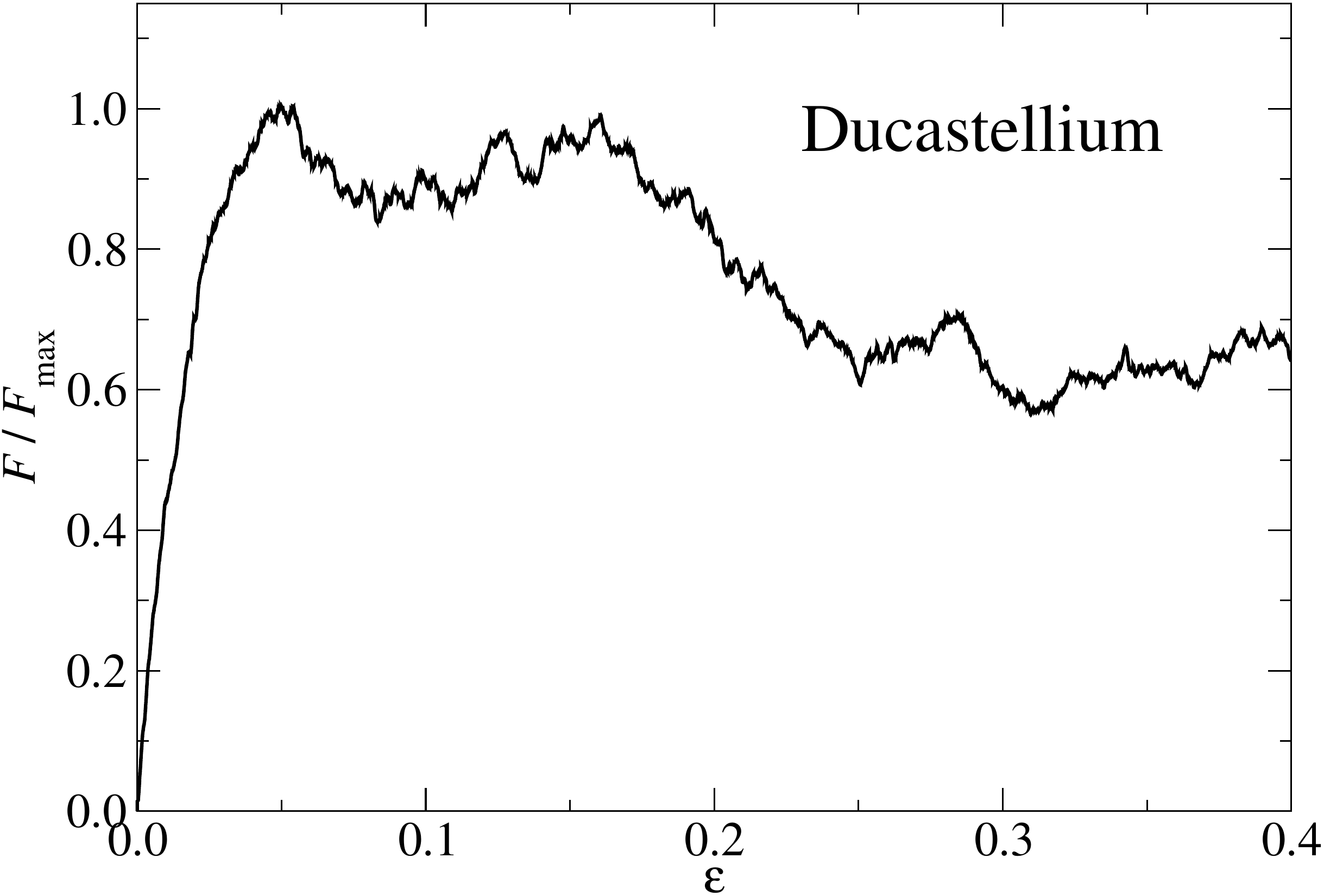}
    \includegraphics[width=1.0\textwidth]{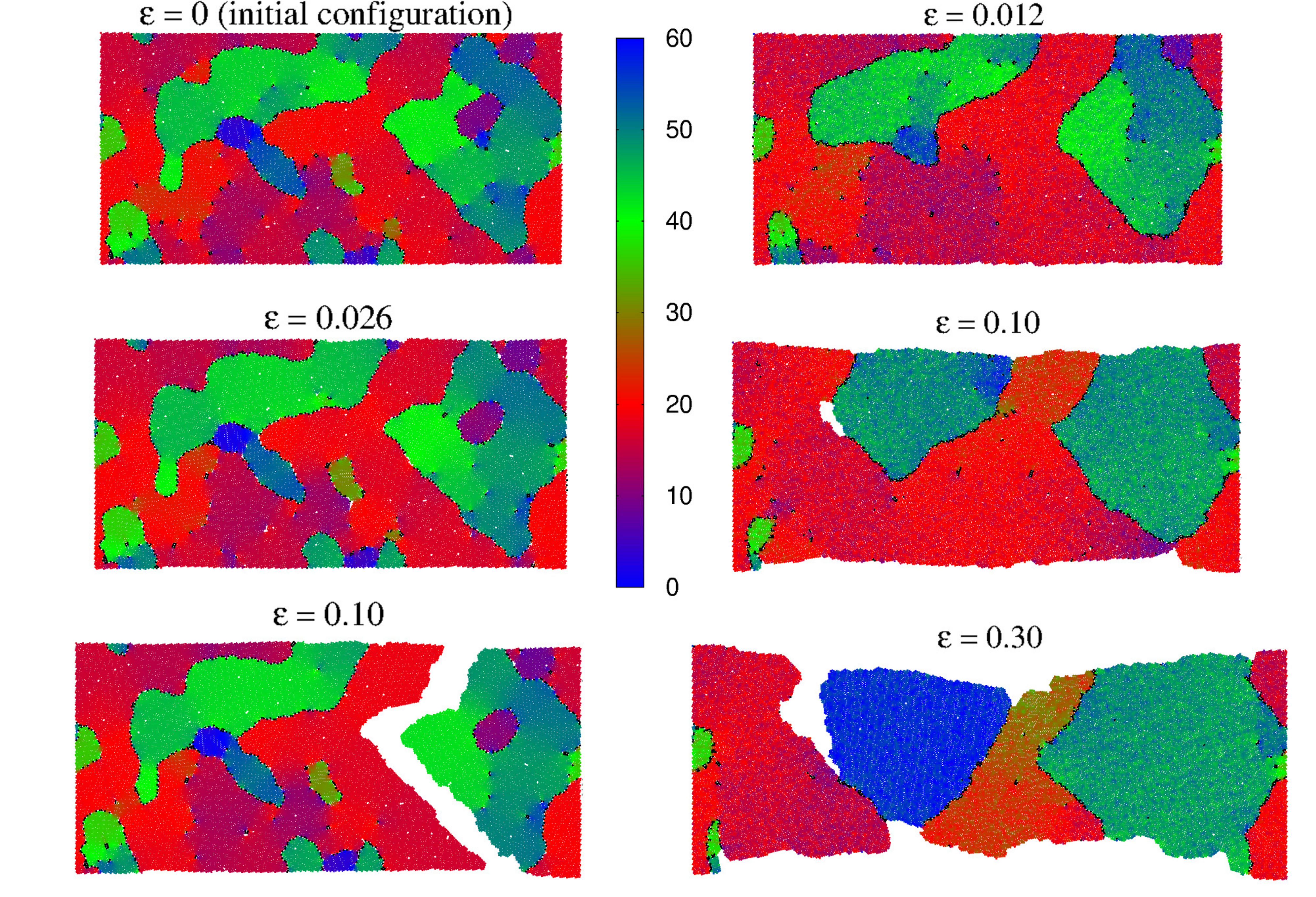}
    \caption{Comparison of two \textit{in-silico} tensile tests. Both simulations started from identical microstructures. The left row assumes Lennard-Jones interactions, the right - a generic model originally proposed by Ducastelle, which later formed the basis for embedded-atom and Finnis-Sinclair potentials.
    Parameters of Ducastelle potential were taken from Ref.~\cite{Jalkanen2015MSMSE}.
    Simulations were run using LAMMPS~\cite{plimpton_fast_1995,Thompson2022-qg}. 
    Colors indicate the orientation of grains, i.e., the angle $\phi_i$ obtained when averaging $\Phi_i \exp(\mathrm{i}6\phi_i) := \langle \exp(\mathrm{i}6\varphi_{ij}) \rangle$ over all atoms within a cutoff, where $\varphi_{ij}$ denotes the angle of a bond w.r.t. $x$-axis. 
    Atoms, for which, $\Phi_i<0.75$, are depicted in black. 
    }
    \label{fig:tensileTest}
\end{figure}

The objective of much atomistic simulation work on metals is to reproduce  thermal and mechanical properties of specific metals and their alloys~\cite{tadmor_modeling_2011,Cai2016book}.
The attempt to rationalize trends has become a secondary objective, which, however, is central to our overview article.
Since we are not aware of a work asking the question how the choice of the potential affects the outcome of a typical tensile-load experiment given a certain micro structure, we took the liberty and produced such data.
The purpose is to demonstrate that the trends, which are discussed in regard to ductility, are indeed borne out.
To this end, we melted two-dimensional Ducastellium ($\mu=0.36$, typical for copper, with $N \approx 33,000$ atoms) and quenched it to small temperature, whereby crystalline grains of different orientation were produced.
The such produced samples, were then subjected to a tensile loading, where the velocity of a few hundred atoms in the outermost layer were constrained to assume a constant value in the direction of loading, and a uniform deformation was simultaneously applied to the rest of the sample.
The exercise was repeated for Lennard-Jonesium, in which case the configurations were rescaled and allowed to relax before loading.
The orientation of the grains remained unaltered during the relaxation.
Results are shown in Fig.~\ref{fig:tensileTest}.

Fig.~\ref{fig:tensileTest} reveals very clearly that the pair-additive Lennard-Jonesium shows the characteristics of brittle fracture, while the isotropic, non-pair-additive Ducastellium produces a force strain relation akin of a ductile metal.
In addition, the pair-additive Lennard-Jonesium breaks along existing grain boundaries without a change of internal grain boundary structure, while Ducastellium undergoes massive reconstruction, the precise nature of which cannot be predicted, as it changes with the random seed for the thermostat.
Of course, these two-dimensional simulations are merely a cartoon for real solids, among other reasons because stress-induced recrystallization in two dimensions occurs much more easily than in three, due to a small (hyper-) surface to volume ratio. 

\section{Charge-transfer potentials}
\label{sec:ctp}

It is sometimes insufficient to treat atoms as if they had a constant (integer) charge. 
This calls for methods allowing charge assignments to be made on the fly.
Adjustable coefficients entering the charge-transfer part of a potential must be gauged on given reference data as any other force-field parameter. 
One possibility is to identify them such that inter-molecular forces or interaction energies from a training set are matched as closely as possible, which, however, is rarely advisable since errors in certain parts of the regular part of the potential may be compensated by false charge assignments, which then jeopardizes the transferability.
This is why reasonable, directly determined reference charges are needed in the training of charge-transfer potentials.

One reason why knowing partial charges accurately -- whatever this means -- is important, is that ionic interactions, so they are present in a system, are on par with covalent interactions.
This is easily seen by comparing their magnitude.
The covalent bond in H$_2$ has a binding energy of 4.74~eV.
The Coulomb interaction in a NaCl molecule with a bond length of $a_0 = 2.36${~\AA} is about 6.1~eV, after subtracting sodium's ionization energy, $I_\textrm{Na} = 5.14$~eV, and adding chlorine's electron affinity, $A_\textrm{Cl} = 3.61$~eV, an energy gain of 4.57~eV remains. 
This needs to be contrasted to typical LJ interaction parameters of order 0.01~eV. 
Thus, even minor changes in partial charges easily alter energies more than relatively large changes of LJ prefactors.

\subsection{Determination of reference charges}

In the absence of periodic boundary conditions, a charge distribution can be rigorously characterized in terms of its net charge, dipole moment, quadrupole tensor, and so on. 
% higher-order electrostatic multipoles.
%
Nonetheless, there is no unique way to assign \emph{atomic} charges, \emph{atomic} dipoles, etc., in a many-body system. 
Yet, the concept of formal and partial charge is central in chemical physics and physical chemistry. 
%, because it often is a useful simplification despite being a sever reduction of reality. 
%
As a consequence, many different charge-assignment schemes have arisen to achieve the ill-defined. 
In fact, any pragmatist simply \emph{must} attempt to assign atomic charges allowing interatomic forces to be computed accurately while providing values that satisfy chemical intuition.
For example, when trying to match the vibrational frequency, the bond length, and the binding energy of a NaCl molecule with a three-parameter potential,
be it Mie or simple Buckingham with two parameters for repulsion plus one independent partial charge, the latter turn out in the vicinity of unity so that chemical intuition is satisfied.
Even more, the such obtained parameters allow quite reasonable predictions to be made, for example, for the lattice constant, bulk modulus and in fact, all three independent elastic tensor elements of NaCl in the rocksalt structure, even if truly accurate potentials require (many-body) corrections. 
We note in passing that producing numbers supporting this claim is a great exercise for students and instructors alike.
At the same time, deducing charges from molecular dipoles, which yields $Q_\textrm{Na} =  0.75$~e when ignoring the induced dipole on the Cl$^-$ ion and making students deduce that a bond in NaCl to be 25\% covalent is a task, which we deem halfway between useless and insane, even if countless chemistry lectures and quite a few text books provide discussions along such lines. 
The above back-of-the-envelope calculation simply neglects the high polarizability of anions~\cite{Rittner1951JCP,Madden1996CSR}.

Complications in assigning partial charges arise as soon as partial charges are not close to an integer value, which happens when atoms with electronegativity differences of, say, $\Delta I \gtrsim 0.5$~eV, are present, simple alkali halides being the major exception to this rule. 
Unfortunately, very few partial charges are assigned as easily as those for sodium or chlorine atoms in the NaCl molecule. 
As argued at the beginning of Sect.~\ref{sec:ctp}, it is usually beneficial to separate charge assignment from the construction of the potential.
% and to identify reference charges in a learning set.
%
Existing methods to compute reference charges can be crudely categorized into those in which atomic charges are deduced from an analysis of 
(i) a representation of the wave function (Mulliken~\cite{Mulliken1955JCP} and Löwdin~\cite{Lowdin1950JCP}),
(ii) the electronic density (Bader~\cite{Bader1991CA} and Hischfeld~\cite{Hirshfeld1977TCA}),
(iii) the electrostatic potential (ESP), 
(iv) displacement-induced changes in polarization (Born~\cite{Born1954book}),
and frequently forgotten
(v) experiment~\cite{Meister1994JPC}.
These methods will be sketched next.
For more detailed overviews, see Refs.~\cite{Meister1994JPC,Mao2014JPDR,Choudhuri2020JCTC}. 

The Mulliken~\cite{Mulliken1955JCP} definition of the partial charge starts from the representation of the total wave function as a linear combination of atomic orbitals.
The charge associated with each orbital is assigned to the atom, however, results are {far} from unique, in particular when using large basis sets.
Löwdin~\cite{Lowdin1950JCP} removed a certain degree of ambiguity from Mulliken's definition that originated from the non-orthogonality of the functions spanning the basis set.
However, uncertainties remain.
Bader~\cite{Bader1991CA} partitions the space into regions such that each atom is assigned the charge density in its vicinity up to the points at which the charge density passes through a local minimum.
However, this procedure ignores the possibility of atomic charge distributions to interpenetrate.
It yields finite charges for the promolecules, whose charge density is defined to be the superposition of the atomic charge densities~\cite{Maslen1985AJP}. 
Hirshfeld~\cite{Hirshfeld1977TCA} originally assumed each atom to own space with a weight proportional to the electronic density of the neutral atom, which, however is clearly inappropriate for ions.
For example, in an alkali hydride, H$^-$ would be assigned a smaller radius than the metal cation. 
Thus, in more advanced methods, the atomic charge densities are replaced with radially symmetric weighting function that are determined self-consistently~\cite{Hirshfeld1977TCA, Lillestolen2008CC}.
In such ``iterative stockholder approaches'', it remains challenging to identify good trade-offs between the ESP accuracy and the transferability of the charges~\cite{Verstraelen2013JCTC}.
In conventinal ESP schemes, atomic charges are adjusted to best match the ESP a save distance away from the nuclei, e.g., outside the atomic van-der-Waals radii.
This becomes problematic for condensed phases, in which such safe distances are sparse or even non-existent, but also for large molecules with hidden atoms.
Even worse, the ESP appears to be ill-defined up to a constant in periodic systems so that partial changes can only be gauged on ESP differences~\cite{Campana2009JCTC}.
Born effective charges reflect the change of polarization occurring in response to atomic displacements, which, for periodic structures would be analyzed in terms of the Fourier transform of the dipole moment at small wave numbers.
However, since both dipoles and displacements are vectors, Born charges are not scalars but tensors of rank two, unless the system is so highly symmetric that all eigenvalues of the tensor are identical.  
In addition, Born charges can cast changing atomic dipoles as effective charge, which can lead to charges exceeding the formal oxidation number. 
Central ways to deduce partial charges from experiment~\cite{Meister1994JPC} include (i) the matching of electrostatic multipoles, which ultimately is a limiting case of an ESP method, (ii) Bader-type analysis of the charge density deduced from x-ray diffraction patterns, but also (iii) a combined analysis of phonons and dielectric response functions.

Quite a few authors claim to have identified the most meaningful way to assign partial charges.
This can scarcely be true, since partial charges are ill-defined so that the optimal charge-assignment scheme depends on the relative weight of target properties. 
Moreover, the optimum method can also depend on the symmetry of the problem.
Problems arising due to ``falsely'' assigning an induced dipole on an atom as a transferred charge vaporize for highly symmetric structure, in which the lowest-order allowed multipole can be a hexadecupole.
Unfortunately, separating atomic dipoles from charge transfer is even difficult for diatomic molecules, which could otherwise serve as a well-defined reference system on which charge-assignment schemes can be gauged. 

An optimum choice for a charge to be used in a force-field based simulation may also depend on whether polarizable dipoles are included or neglected so that the optimum reference-charge-determination method may depend on the precise nature of the force field wanting to be parametrized.
As a consequence, we are neither willing nor able to express an opinion if any of the sketched methods is all in all the optimum choice and believe that it has to be made case by case. 
%
% However, we believe that the analysis of symmetric structures, if any possible, may sometimes help to remove 
% removes quite a few of the disadvantages particularly ESP-based but also Bader-based charge assignment schemes, since dipoles or even higher-order multipoles would be symmetry forbidden.

\subsection{Regular charge-equilibration approaches}

In the ground-state of a full quantum-mechanical system, the functional derivative of the energy $E$ w.r.t. the electron density $\rho(\mathbf{r})$ disappears, because $\rho(\mathbf{r})$ minimizes $E$. 
In a coarse-grained description of the electron density in terms of a set of atomic charges, $\{Q\}$, this minimization principle translates into what Mortier and varying co-authors~\cite{Mortier1985JACS,Mortier1986JACS} called the electronegativity equalization principle.
Assuming that $E$ can be expanded into powers of atomic charges, the ground state energy can be written as
\begin{subequations}
\label{eq:Mortiers}
\begin{eqnarray}
    U & = & U(\mathbf{Q}=0) + \sum_i \frac{\partial U}{\partial Q_i}\,Q_i + \frac{1}{2}\sum_{i,j} \frac{\partial^2 U}{\partial Q_i \partial Q_j} \, Q_i\,Q_j + ... \\
    & = & U(\mathbf{Q}=0) + \sum_i \chi_i \,Q_i + \frac{\kappa_i}{2}\,Q_i^2 +  \sum_{i,j\ne i} \frac{J_{ij}}{2} \,Q_i \, Q_j + ...,
\end{eqnarray}
\end{subequations}
while nuclear coordinates remain fixed.
Charges adopt themselves such that they minimize $U$ usually with the constraint of a fixed total charge. 
In Eq.~\eqref{eq:Mortiers}, the electron density still minimizes the total energy for any $\{Q_i\}$, albeit with the constraint imposed through the assigned partial charges. 
Thus, the parameters $\kappa_i$, $\chi_i$, and $J_{ij}$ are implicitly defined from a mathematical point of view in Eq.~\eqref{eq:Mortiers}, even if their numerical values depend on the method used to relate charge density and partial charges. 

The chemical meaning of the parameters  $\kappa_i$, $\chi_i$, and $J_{ij}$ is best ascertained in limiting cases.
Consider atom $i$ to be neutral and isolated. 
If one electron is added to it so that $Q_i = -e$, the energy decreases by its electron affinity $A_i$, while it increases by its ionization energy $I_i$ when one electron is taken away from it. 
Thus, 
\begin{subequations}
\label{eq:QEterms}
\begin{eqnarray}
    I_i & = & \frac{\kappa_i}{2}\,e^2 + \chi_i\,e \\
    A_i & = & \frac{\kappa_i}{2}\,e^2 - \chi_i\,e 
\end{eqnarray}
\end{subequations}
so that the so-called chemical hardness~\cite{Parr1983JACS} % {Ghosh2009IJQC} 
turns out to be $\kappa_i = (I_i+A_i)/e^2$ and the electronegativity~\cite{Mulliken1934JCP,Parr1978JCP} $\chi_i = (I_i-A_i)/e$.
Meanwhile, $J_{ij}$ can be associated with the Coulomb potential given the charges $Q_i$ and $Q_j$ of two distant atoms $i$ and $j$:
\begin{equation}
\label{eq:Coulomb}
    J_{ij} = \frac{1}{4\pi\varepsilon_0\,r_{ij}} \textrm{ for } r_{ij} \to\infty.
\end{equation}
The effect of an external electrostatic potential would have to be added to  $\chi_i$.
The bold hope now is that partial atomic charges can be predicted using Eq.~\eqref{eq:Mortiers} and parameterizations, which do not stray too far away from Eqs.~\eqref{eq:QEterms} and \eqref{eq:Coulomb}.

Rapp\'e and Goddard~\cite{Rappe1991JPC} pioneered the use of the electronegativity equalization principle for the on-the-fly determination of atomic charges in molecular simulation and renamed the approach to charge-equilibration (QEq) method.
To keep charges in (partially) ionic systems from blowing up, they suggested to shield the Coulomb interactions at short distances through two-electron Coulomb integrals of the form
\begin{equation}
    J_{ij} = \frac{1}{4\pi\varepsilon_0}\, \int\, \mathrm{d}^3r_i\, \mathrm{d}^3r_j \frac{\rho(\mathbf{r}_i)\,\rho(\mathbf{r}_j)}{r_{ij}},
\end{equation}
where the charge densities $\rho(\mathbf{r})$ are those associated with $s$-type Slater orbitals. 
This way, the Hessian, i.e., the matrix formed by $\partial^2U/\partial Q_i \partial Q_j$ remains positive definite even when atoms approach each other closely. 
Nonetheless, their claim that QEq ``\textit{leads to charges in excellent agreement with experimental dipole moments and with the atomic charges obtained from the electrostatic potentials of accurate ab initio calculations}'' can be seen skeptically:
Can Coulomb shielding be responsible for charges in the alkali metal halides to be so darn close to unity? 
And how can QEq reproduce the experimental observation that partial charges of atoms and ions in the gas phase adopt integer multiples of the elementary charge? 
In fact, QEq obviously predicts  the ions of a neutral, dissociated NaCl molecule to each carry a charge of magnitude $(\chi_\textrm{Cl}-\chi_\textrm{Na})/(\kappa_\textrm{Cl}+\kappa_\textrm{Na}) \approx 0.4$~elementary charges when truncating the expansion of Eq.~\eqref{eq:Mortiers} after the quadratic term~\cite{Chen2007CPL}. 

QEq has multiple practical shortcomings, which all originate from the ease with which \emph{partial} charge can be transferred from one atom to another one even over large distances.
To name a few, it leads to the wrong dissociation limit of molecules~\cite{Chen2007CPL}, DFT having related issues, QEq makes the polarizability of polymers grow superlinearly rather than linearly in the chain length~\cite{Chelli1999JCP,LeeWarren2008JCP}, and, similarly, solids adopt the dielectric response function of a metal~\cite{Nistor2009PRB}, i.e., the dielectric constant of a QEq solid is infinitely large. 
Finally, QEq makes the dipole of alcohols be linear in the length of the hydrocarbon chain~\cite{Mikulski2009JCP}.

One fundamental problem of Eq.~\eqref{eq:Mortiers} is that the curve of lowest average energy of an \emph{isolated} system versus the number of electrons turns out to be a series of straight lines when generalizing the Hohenberg-Kohn theorem to fractional particle numbers~\cite{Perdew1982PRL}. 
This property should be reflected in any coarse-grained approximation to DFT, which is built on the Hohenberg-Kohn theorem.
However, Eq.~\eqref{eq:Mortiers} fails to do so.

Despite the fundamental issues of QEq, metals can be discussed, since partial charges can be readily transferred over large distances.
In a parallel-plate capacitance geometry, the length over which an external-field induced charge density decays into the solids scales approximately with $\sqrt{\kappa}$~\cite{Nistor2009PRB}.
In other words, QEq implicitly contains corrections to the behavior of ideal metals.
Thus, mirror charges, or, rather the charge distributions producing the same electrostatic field outside of the metal as mirror charges do, are more smeared out in a QEq solid than in an ideal metal.
To reflect this, the term $\kappa_i$ to be used in a QEq simulation of metals may  best be parameterized through an appropriate choice of the Thomas-Fermi screening length, which can benefit, e.g., the description of the interfacial interactions between a metal and an electrolyte~\cite{Scalfi2020JCP}.
For nanostructured electrodes, the finite density of states at the Fermi level leads to an additional contribution to the capacity of the device~\cite{Pomorski2004-rj}, which can be modeled by replacing the quadratic term in Eq.~\eqref{eq:Mortiers} with an appropriate nonlinear function~\cite{Pastewka2011-qi}.

\subsection{Charge-equilibration methods for non-metallic systems}

Various ideas have been pursued to suppress the non-local charge transfer in QEq for dielectric systems.
One of the first and most influential approaches, coined the fluctuating-charge (fluc-Q) model~\cite{Rick1994JCP}, imposes a charge neutrality constraint on individual molecules. 
Unfortunately, this bug fix does not remedy the superlinear scaling of the polarizability with the chain length of polymers.
Moreover, the bonding topology of molecules needs to be defined making the simulation of bond breaking and formation difficult to describe.

To suppress long-range charge transfer, Chelli~\etal~\cite{Chelli1999JCP} introduced atom-atom charge transfer (AACT) variables $q_{ij} = -q_{ji}$ so that the charge of an atom is given by
\begin{equation}
\label{eq:sumSQs}
    Q_i = \sum_j q_{ji},
\end{equation}
where $q_{ji}$ is the charge transferred from atom $j$ to atom $i$.
In the AACT method, the bond hardness, $\kappa_{ij}$, replace the atomic hardness used in regular QEq so that a charge transfer costs a potential energy of $\kappa_{ij}\,q_{ij}^2/2$.
The AACT model remedies all shortcomings in QEq approaches originating from non-local charge transfer, since the AACT method allows the latter to be (completely) suppressed with (infinitely) large bond hardnesses.
However, for the Hessian of the AACT model to be positive definite, the bond hardness terms have to be made so large that the dielectric constants of condensed phases, $\varepsilon_\textrm{r}$, turn out to be barely exceed unity~\cite{Nistor2006JCP}.
Moreover, AACT suppresses the observed chain-length dependent polarizability of short oligomers~\cite{LeeWarren2008JCP,Nistor2009PRB} and, similarly, produces a zero screening length in dielectrics~\cite{Nistor2009PRB}.
Last but not least, AACT cannot be used to model the polarization of metals that occurs in response to an electric field. 

\begin{figure}[hbtp]
    \centering
    \includegraphics[width=0.95\textwidth]{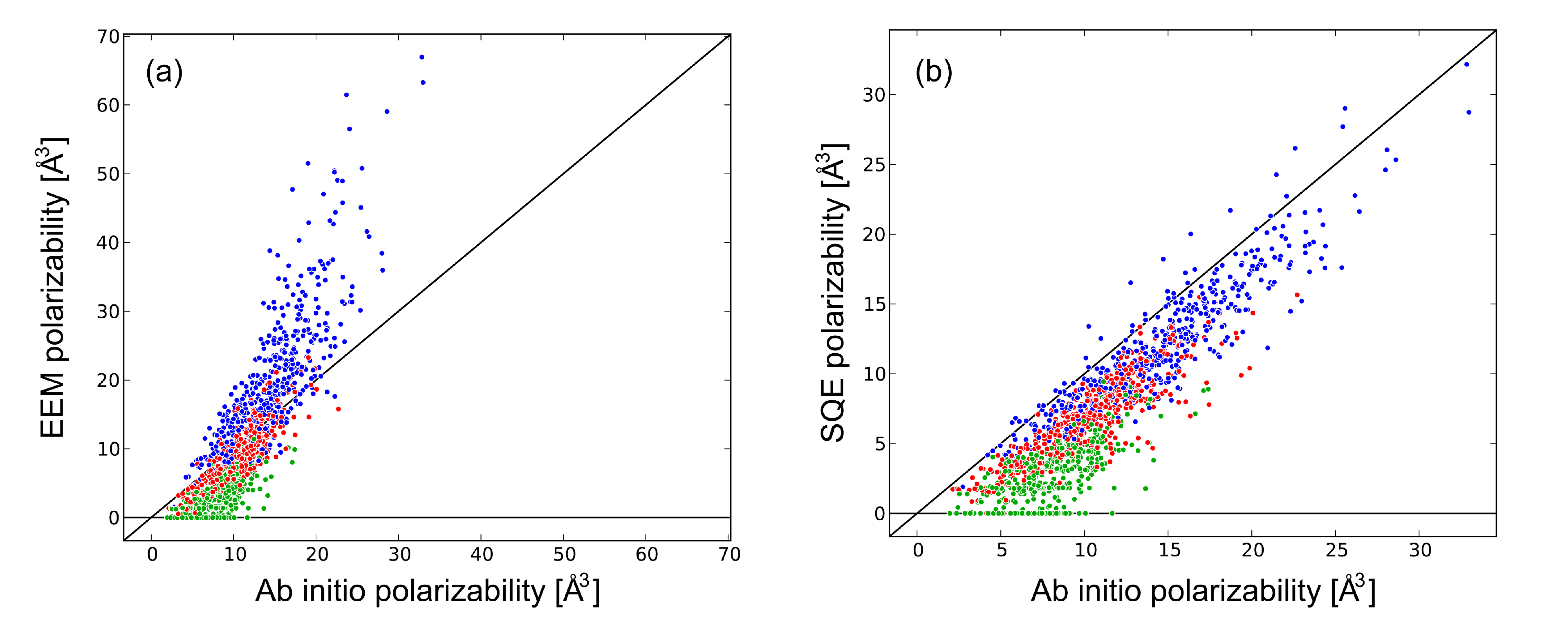} 
    \caption{
    Polarizability of (a) electronegativity equilization model (EEM) (equivalent to the QEq model)  and (b) SQE as a function of the polarizability deduced from quantum mechanical calculations on a variety of molecules.
    Green, red, and blue data points reflect the smallest, the medium, and the largest eigenvalue of the polarizability tensor for different molecules (including varying conformations).
    Reprinted from Verstraelen, T., Van Speybroeck, V., and Waroquier, M. \textit{J. Chem. Phys.} \textbf{131}, 044127, (2009), with the permission of AIP Publishing.
    %From Ref.~\cite{Verstraelen2009JCP}.
    }
    \label{fig:toon}
\end{figure}

Reintroducing the atomic hardness term to the AACT model leads to a new model, which Nistor~\etal~\cite{Nistor2006JCP} first meant to call fluc-Q~2.
Ultimately, they found the term split-charge equilibration (SQE) to be more appropriate, since the model is a split between two models and a ``split charge'', $q_{ij}$, formerly known as AACT variable, is split between two atoms. 
Thus, their model reads
\begin{equation}
    U = U(\textrm{QEq}) + \sum_{i,j>i} \frac{\kappa_{ij}}{2}\,q_{ij}^2,
\end{equation}
where the $Q_i$ entering $U(\textrm{QEq})$, i.e., Eq.~\eqref{eq:Mortiers}, are 
 defined as in Eq.~\eqref{eq:sumSQs}.
Including finite bond hardness suppresses the superlinear scaling of molecules with linear size~\cite{Verstraelen2009JCP}, as is shown in Fig.~\ref{fig:toon}.
The observable underestimation of the polarizability is supposedly due to the neglect of atomic polarizability. 

Other charge-transfer models exist ensuring neutral separation limits.
In one line of approach, chemical potential differences between distant atoms are screened with functions motivated  from overlap integrals between orbitals on different atoms~\cite{Chen2007CPL}. 
However, this procedure implicitly claims that the altitude difference of two objects becomes less when moving them apart laterally, which is not meaningful.
In contrast, a very appealing approach is a model in which electrons are treated as shells, which are not tight to one individual atom~\cite{Leven2019JPCL}, whereby, in principle, history dependence can be mimicked along the lines of the NaCl dissociation described in the introduction.
%MM2SS: Please check if I did not misunderstand. 
%
There are also approaches using non-local softness matrices, the arguably most advanced one being the atom-condensed Kohn-Sham approximated to second order (ACKS2)~\cite{Verstraelen2013JCP}.
Its purpose is twofold:
First, defining the expressions for the softness matrix, which is the inverse of the hardness matrix, from expressions accessible from DFT calculations, and second to employ the result in simulations as a mean to deduce partial charges. 
Ultimately, ACKS2 contains SQE model as a limiting case. 
The downside of ACKS2 is that a $2N\times 2N$ matrix must be defined, where $N$ is the number of charges, while the number of split charges is only of order $Z N$.
A frequently raised argument against SQE is that the large bond hardnesses occurring at large separation cause numerical difficulties.
However, they can be addressed using appropriate pre-conditioners in square-gradient minimization or large split-charge inertia in extended Lagrangian approaches.

One of the main limitations of the original SQE model is that it cannot produce isolated ions as the $\kappa_{ij}$ diverges at large $r_{ij}$.
This limitation can be overcome by introducing oxidation numbers $n_i$
so that an atomic charge is given by
\begin{equation}
    Q_i = n_i\, e + \sum_j q_{ji}
\end{equation}
with $n_i \in \mathbb{Z}$ in a so-called redox-SQE formalism~\cite{Dapp2013EPJB,Dapp2013JCP}.
The remaining part of the SQE formalism is kept, except that force-field parameters, including chemical hardness and electronegativity, must now be assigned individually for each oxidation number. 
The redox-SQE approach allows a system to effectively move on different potential energy surfaces (if you're a chemist) or different Landau-Zener levels (if you're a physicist).
Jumping between them requires the oxidation number of one atom to be increased by an integer, typically by one, while that of another, nearby atom must be decreased by the same number.
The corresponding discrete dynamics can be implemented in practice, for example, in terms of Monte Carlo dynamics. 
%

% ACKS2:
\subsection{Properties resulting from charge-transfer potentials}
\label{sec:chargeTransferPots}

A central aspect of polarizable and charge-transfer potentials is to reproduce the dielectric response of (condensed) media correctly, the most important quantity from a continuum perspective being the dielectric constant $\varepsilon_\textrm{r}$.
The relation between point-dipole polarizability and $\varepsilon_\textrm{r}$ is text-book material and reflected in the Clausius-Mossotti (CM) relation, Eq.~\eqref{eq:ClausiusMossotti}, while that of a bond-polarizable model without point dipole polarizability was worked out by Nistor and Müser~\cite{Nistor2009PRB}.
Combining the treatment of point and charge-transfer dipoles does not appear to have been presented in the literature hitherto.
However, it can be easily achieved when following Hannay's  derivation of the CM relation~\cite{Hannay1983EJP}.
It  starts from the insight that the full electrostatic field of a dipole contains a contribution $\mathbf{E}_\textrm{int} = - \mathbf{p}\,\delta(\mathbf{r})/(3\varepsilon_0)$ acting inside of it in addition to its usually stated far field. 

Since each charge leads to a mean electrostatic field of zero, the mean electrostatic field of a charge distribution is also zero, however, outside the dipoles, it is reduced by $\mathbf{E}_\textrm{int}$ averaged over the volume that each point dipole occupies.
Thus, the average field that a test charge sees outside of dipoles is the external electric field $\mathbf{E}_\textrm{ext}$ (which is assumed to be constant) minus $\langle \mathbf{E}_\textrm{int}\rangle = - \rho\,\mathbf{p}/(3\varepsilon_0)$, where $\rho$ is the density of point dipoles. 
For a rocksalt lattice, or, when  discretizing a homoegeneous material into cubes of linear size $a_0$ and arranging the local charge transfer variables, or, split charges in a vector $\mathbf{q}$, the dipoles obey associated with the charge-transfer, $a_0 \mathbf{q}$ and the point dipoles obey
\begin{subequations}
\begin{eqnarray}
\frac{a_0\mathbf{q}}{\kappa_\textrm{s}}  & = & \mathbf{E}_\textrm{ext} + \frac{\rho}{3\varepsilon_0}\,\mathbf{p} \\
    \alpha\, \mathbf{p} & = & \mathbf{E}_\textrm{ext} + \frac{\rho}{3\varepsilon_0}\,\mathbf{p},
\end{eqnarray}
\end{subequations}
respectively. 
The $x$-component of $\mathbf{q}$ is the split charge donated from an atom or grid point to its right neighbor, and so on, while $\kappa_\textrm{s}$ is the corresponding bond hardness. 
%
% where $\textbf{E}_\textrm{ext}$ is a constant, external electric field.
% while $\rho = 1/a_0^3$.
%
The total polarizability is nothing but the total dipole density $\mathbf{P} = \rho (\mathbf{p}+a_0\,\mathbf{q})$, while $\varepsilon_\textrm{r}$ is defined through $\mathbf{P} = (\varepsilon_\textrm{r}-1) \varepsilon_0 \mathbf{E}_\textrm{loc}$ within linear response.
After some minor algebra
\begin{equation}
    \varepsilon_\textrm{r} - 1 = 
    \frac{\rho/(a_0\,\kappa_\textrm{s})+\rho\alpha}{\varepsilon_0- \rho\alpha/3}
\end{equation}
can be obtained, which contains the CM relation in the limit of $\kappa_\textrm{s}\to \infty$ as well as $\varepsilon_\textrm{r} - 1 = \rho/(a\kappa_\textrm{s})$ valid for bond-polarizable models ignoring point dipoles~\cite{Nistor2009PRB}.
Once bond polarizability is ignored in a charge-transfer model, $\varepsilon_\textrm{r}$ diverges, which means that the system effectively turns metallic. 

Continuum corrections to the above treatment would require the wave-number dependence of the charge-dipole and dipole-dipole coupling but also the chemical hardness coupling to be taken into account. 
The latter would augment the bond hardness in the presented treatment with $\kappa (qa_0)^2$.
A wave-number dependent dielectric constant and thus a dielectric smearing or decay length $\delta$ ensues.
For a pure bond-polarizable model and the geometry considered above, $\delta = \sqrt{\varepsilon_0 a_0 \kappa/(1+\varepsilon_0 a_0 \kappa_\textrm{s})}$~\cite{Nistor2006JCP}. 
One of the consequences of this result is that ideal mirror charges do not accurately represent 
the dielectric response of real metals~\cite{Scalfi2020JCP}.

\begin{figure}[hbtp]
    \centering
    \includegraphics[width=0.465\textwidth]{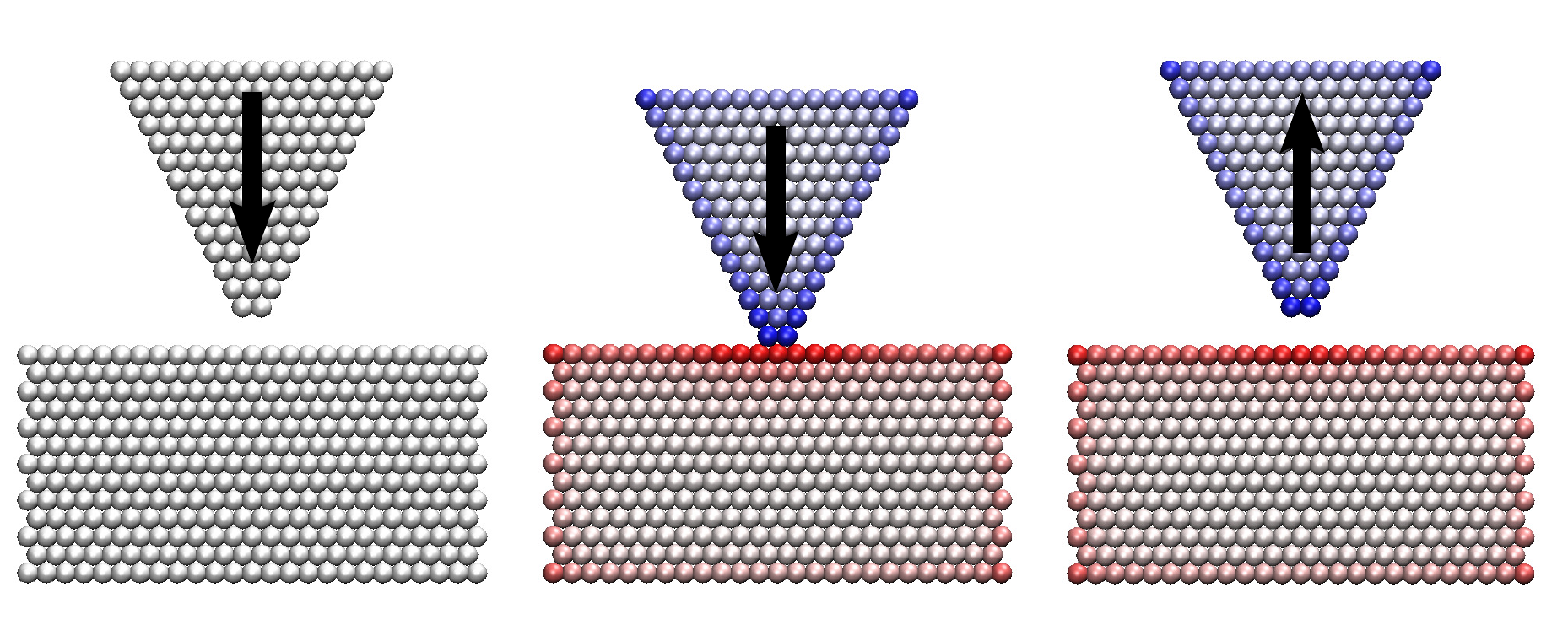} \hspace*{4mm}
    \includegraphics[width=0.465\textwidth]{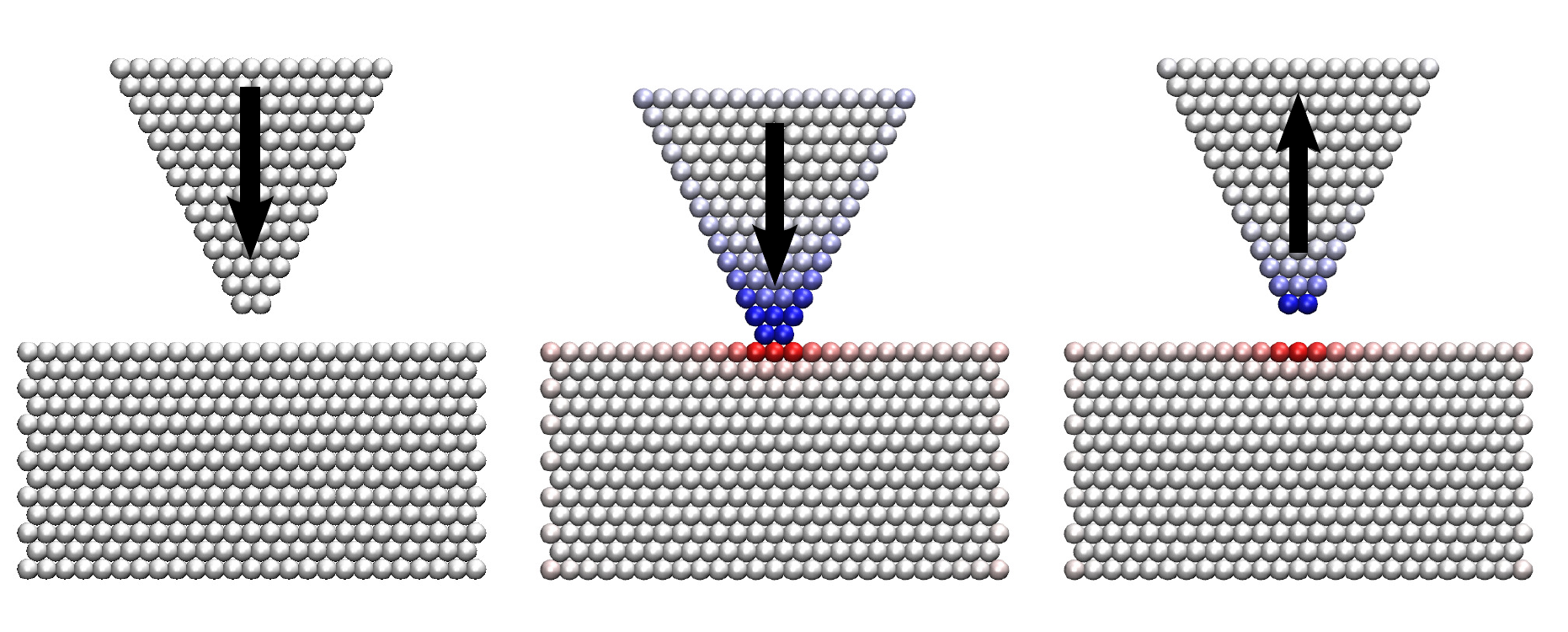}
    \caption{Contact electrification with history-dependent potentials between two metals (\textbf{left}) and two dielectrics (\textbf{right}).
    Reprinted by permission from Springer: \textit{Eur. Phys. J.} \textbf{86}, 337, Towards time-dependent, non-equilibrium charge-transfer force fields, Dapp, W. B. and  M\"{u}ser M. H., [COPYRIGHT] (2013)
    %From Dapp and Müser, Eur. J. Phys. B (2013) ~\cite{Dapp2013EPJB}
    }
    \label{fig:contactElectrification}
\end{figure}

The potential importance of modeling dielectric response correctly is also demonstrated in Fig.~\ref{fig:contactElectrification}.
It shows two pairs of two dislike solids, or, clusters, which can exchange charge, specifically, oxidation numbers, when they are sufficiently close to each other. 
In reality, this distance would be a weak function of the approach and retraction velocity.
In one pair, both materials are metals, in the other, they are insulators. 
A lot of ``electrons'' transfer in the metallic couple and swapped charge delocalizes over the entire samples, mostly on their surfaces.  
In the insulating couple, only one charge exchanges, which remains localized near the former contact points after retraction. 

Note that Fig.~\ref{fig:contactElectrification} is a proof-of-principle demonstrator for the redox-SQE method~\cite{Dapp2013EPJB}.
It does not allow one to conclude that contact electrification between dielectrics primarily occurs through electron transfer.
There are many good reasons to believe that ion exchange matters most~\cite{Lacks2019NRC}.

\section{Machine-learned potentials}

In the past two decades, machine learning has conquered the world of interatomic
potentials, a development that is ongoing and accelerating.
The main idea is to equip a machine learning (ML) technique with a large database of structures for which energies, forces, stresses, etc. are 
computed using well-converged first-principle methods\footnote{See \url{https://libatoms.github.io/GAP/data.html} for links to high-quality \emph{ab-initio} databases for C, Si, and other materials.}.
Energies (or sometimes forces) are then extrapolated (or ``predicted'') based on (local) structural similarity.
Structural similarity is either encoded by directly comparing to configurations stored in this database, or it can be implicit in the weights of a neural network or the points of a sparse Gaussian process.
Recent methods supplement such computational approaches with experimental data~\cite{sivaraman_experimentally_2021}.
Additionally, predictions are not limited to energies and forces. Examples of other properties obtained from ML regression are core energies and X-ray photoelectron spectra~\cite{Aarva2019-ba,Aarva2019-xw,Aarva2021-zl,Golze2021-ri}.

Machine-learned potentials (MLPs) are sometimes characterized as being parameter-free, however,
one may also claim that the number of parameters is gigantic.
Occam's razor suggests that models should have few parameters.
Yet, overfitting of high-dimensional models can be avoided by using parameterization procedures that are based on sound statistical principles, such as Bayesian inference.
Additionally,
% -- as we have shown throughout this review --  (das haben wir ja zum Großteil rausgeworfen, Born-Oppenheimer wir nie erwähnt) 
all physically-justified potential forms (with few parameters) are based on a chain of approximations, e.g., going from the Born-Oppenheimer approximation, to DFT, neglecting three-body integrals in the tight-binding approximation, neglecting nonorthogonality and so on in the bottom-up construction of bond-order potentials for semiconductors.
% ... to DFT ... or to tight-binding
%
It is often unclear to what extend the final form is still a good approximation for interatomic interactions.
To say it with the
words of Bazant, Kaxiras and Justo~\cite{bazant_environment-dependent_1997}: ``\emph{In the case of Si, the abundance of potentials in the literature illustrates the difficulty of the problem and lack of specific theoretical guidance. In spite of the
wide range of functional forms and fitting strategies, all proposed models possess comparable and insufficient overall
accuracy. It has proved almost impossible to attribute the successes or failures of a potential to specific features of its
functional form.}''

MLPs target reproducing the high-fidelity calculations used to construct the database. In that sense, they are higher accuracy (with respect to the reference DFT database) than more empirical or semiempirical models, but this may mean a lack of transferability to structures not explicitly considered in the database.
This pitfall is of course not limited to MLPs: As an example, the first incarnation of the Tersoff potential for silicon~\cite{tersoff_new_1986} turned out to have bcc and not diamond cubic as the ground state~\cite{dodson_development_1987}.
Yet, potentials constructed on carefully curated databases give excellent transferability even to obscure crystal phases, see e.g. Ref.~\cite{Bartok2018-ie} for a current example, again for elemental silicon.
This highlights the specific importance of database quality for MLPs.

Most MLPs consist of three basic components:
First, a database of structures, already outlined above.
Second, a ``descriptor'' that encodes the chemical environment of an atom into a unique fingerprint.
In the absence of external fields, this fingerprint must be objective, i.e., it must be invariant with respect to translation and rotation.
It must also be invariant with respect to the exchange of atoms of the same species.
This allows the data set to be much reduced compared to the situation, where the ML technique has to learn objectivity by itself.
An additional complication is that the database itself may not be perfectly objective, due to numerical errors in the underlying electronic structure calculations.
Since we want forces, fingerprints also must be continuously differentiable at least once, and
ideally, they also have continuous second derivatives.
The third component is a regression framework, the ``learning'' aspect of the potential.
Two broad directions of development have occurred here:
Using neural networks, which can be regarded as a model having many parameters, or, Gaussian processes, which can be regarded as a scheme directly extrapolating between points in the database.
The regression framework takes the fingerprints as input and predicts per-atom energies as output.

Before continuing with more details, we would like to point out that many reviews by researchers more qualified than us to discuss MLPs have appeared recently, see Refs.~\cite{bartok_gaussian_2015,behler_perspective_2016,deringer_machine_2019,deringer_gaussian_2021,Musil2021-lm}.
This review focuses on simple functional forms that allow analytical calculations to be made.
Yet, not discussing at least the basic ideas behind MLPs could be a severe omission in present times.
More importantly, some of the discussion highlighting similarities between classical and MLPs might help the development of either one. 

\subsection{Chemical fingerprints}

Chemical fingerprints come in many forms, and we here discuss just a select few. The interested reader is referred to other literature, e.g. Refs.~\cite{himanen_dscribe_2020,Musil2021-lm}, for more comprehensive descriptions of the wide variety of fingerprints.

%\subsubsection{Coordination numbers}

The
arguably first and one of the simplest 
example of a chemical fingerprint 
consists of 
the coordination number and the bond-conjugation variable employed by Brenner in his  potential for hydrocarbons~\cite{brenner_empirical_1990,brenner_erratum:_1992,brenner_second-generation_2002}.
Given a bond $i$-$j$, Brenner used the coordination numbers $N_i$, $N_j$ and a bond conjugation parameter $N_{ij}^\text{conj}$ as a fingerprint for the state of the bond (see Sect.~\ref{sect:brenner} for more details).
These variables are made continuous by computing the coordination numbers with a smooth cutoff, $N_i=\sum_j f_\text{c}(r_{ij})$ where $f_\text{c}(r)$ is one of the cutoff functions discussed in Sect.~\ref{sec:cutting}.
This fingerprint is then used to correct the bond-energy of a Tersoff potential through a lookup table.
The lookup procedure continuously interpolates between integer values of the triplet $(N_i,N_j,N_{ij}^\text{conj})$ using splines.
The table entries were fitted to a small database of experimental properties of hydrocarbon molecules.

%\subsubsection{Two- and three-body terms}

Brenner's tables are used to correct a baseline potential, a procedure that is sometimes called a $\Delta$-potential, e.g. in Ref.~\cite{deringer_gaussian_2021}. Other descriptors are needed for full-fledged machine-learned energies. As a simple instructive example, we consider two and three-body terms in the sense of Eq.~\eqref{eq:vExpandGalilei}. An objective descriptor for two-body interaction is simply the bond distance $r_{ij}$. For a three-body term, we can use the triplet of distances $(r_{ij}, r_{ik}, r_{jk})$ as a descriptor. While this descriptor satisfies objectivity, it still has permutational symmetries with respect to the exchange of the positions of like atoms $k$ and $j$. Bart\'ok and Cs\'anyi~\cite{bartok_gaussian_2015} suggested the use of the descriptor
\begin{equation}
    \v{q}_{ijk} = (r_{ik} + r_{ij}, (r_{ik}-r_{ij})^2, r_{jk})
    \label{eq:permuttriplet}
\end{equation}
which is equivalent to $(r_{ij}, r_{ik}, r_{jk})$ but symmetric with respect to the permutation $j\leftrightarrow k$.

We note that common potentials, like Axilrod-Teller-Muto (Sect.~\ref{sec:atm}) and Stillinger-Weber (Sect.~\ref{sec:stillinger-weber}), which in their original formulation depend on the $(r_{ij}, r_{ik}, r_{jk})$ triplet, can also be written in terms of the descriptor $\mathbf{q}_{ijk}$. These analytical potentials obey the above permutational symmetries by construction. However, in a machine-learning context, where an energy is essentially obtained by extrapolating from a given set of database entries, there is no guarantee that such permutational symmetry will be fulfilled by the extrapolated quantities. This requires reformulation in a form such as Eq.~\eqref{eq:permuttriplet}.

The specific descriptor of Eq.~\eqref{eq:permuttriplet} was used in the Gaussian approximation potential (GAP)~\cite{bartok_gaussian_2010} for amorphous carbon~\cite{deringer_machine_2017}, later extended for improved accuracy of crystalline phases~\cite{Rowe2020-qn}. This potential followed a specific three-step construction: In the first step, a (nonparametric, i.e. machine learned) pair-potential was fitted to a database using the scalar $r_{ij}$ as a descriptor.
The second step involved correcting this potential using a three-body term employing the descriptor given by Eq.~\eqref{eq:permuttriplet}.
Finally, this potential was again corrected using the many-body SOAP descriptor described further below. Incorporating simple descriptors (including a pair descriptor) into the construction of the potential gave improved transferability.
This basic potential without the SOAP descriptor can be regarded as a machine-learned variant of the cluster expansion, see Eq.~\eqref{eq:vExpandGalilei}, in an embedding medium.
% Gabor is probably going to kill me for that last sentence
% Yes, but why SW? I'd say, it's a machine-learned variant of a cluster expansion in an embedding medium rather than in vacuum. 
% Changed it

%\subsubsection{Atom-centered symmetry functions}

A successful strategy for obtaining more generic descriptors is the use of atom-centered symmetry functions introduced by Behler and Parrinello~\cite{behler_generalized_2007,behler_atom-centered_2011}. The atom-centered fingerprints consist of radial symmetry functions
\begin{equation}
\label{eq:atomsymG1}
    q_i^{(1,\alpha)} = \sum_{j\not=i} e^{-\eta^{(\alpha)}(r_{ij}-R_s^{(\alpha)})^2} f_\text{c}^{(\alpha)}(r_{ij}).
\end{equation}
These functions are Gaussians centered at a pair distance $r_s^{(\alpha)}$, which are forced smoothly to zero at a cutoff distance via $f^{(\alpha)}_\text{c}(r)$.
They are similar to coordination numbers, but put specific weight on atoms at the distance $R_s^{(\alpha)}$. In addition to these radial functions, there are angular functions
\begin{eqnarray}
   q_i^{(2,\beta)} & = & 2^{1-\zeta^{(\beta)}} \sum_{j,k\not=i} (1+\lambda^{(\beta)} \cos \vartheta_{ijk})^{\zeta^{(\beta)}} \nonumber \\
   & & e^{-\eta^{(\beta)}(r_{ij}^2+r_{ik}^2+r_{jk}^2)}
    f_\text{c}^{(\beta)}(r_{ij})
    f_\text{c}^{(\beta)}(r_{ik})
    f_\text{c}^{(\beta)}(r_{jk}).
\end{eqnarray}
We have added the superscripts $(\alpha)$ and $(\beta)$ to emphasize, that those functions are not evaluated for a single parameter vector $(\eta,R_s,\lambda,\zeta)$ and cutoff function but for a set of these parameters.
The fingerprint vector $\v{q}_i$ is then constructed by combinations of $q_i^{(1,\alpha)}$ and $q_i^{(2,\beta)}$ over these parameters sets.
This allows one to construct fingerprint vectors of arbitrary dimension but those fingerprints may contain redundant information.
In the following, we will refer to the components of the fingerprint vector as $q_i^{(\mu)}$ and drop the double notation $(1,\alpha)$, $(2,\beta)$, implying that the index $\mu$ also refers to the type of symmetry function and not just its parameters.

To emphasize connection to classical potentials, the $q_i^{(1)}$ look like the onsite $\xi_{ii}$ terms occurring in our general functional form, Eq.~\eqref{eq:genxi}, while the $q_i^{(2)}$ look like offsite $\xi_{ij}$ terms.
Indeed, if we replace the Gaussian in Eq.~\eqref{eq:atomsymG1} with an exponential, the expression becomes identical to the local density of Ducastellium, see Eq.~\eqref{eq:EAMchargeDensity}.
Similarly, $q_i^{(2)}$ are constructed like the many-body contributions in the SW and MEAM potentials.

%\subsubsection{Smooth overlap of atomic positions}
%\label{sec:soap}

Another important strategy for obtaining chemical fingerprint is the smooth overlap of atomic positions (SOAP) descriptor, introduced by Bart\'ok, Kondor and Cs\'anyi~\cite{bartok_representing_2013}. It can be regarded as a generalization of the bond-orientational order parameter of Steinhardt, Nelson and Ronchetti~\cite{steinhardt_bond-orientational_1983}. The idea is to map the atomic configurations onto an atomic density field $\rho(\v{r})$, which is given by atom-centered functions,
\begin{equation}
    \rho(\v{r}) = \sum_i \delta(\v{r}-\v{r}_i),
    \label{eq:delta-density}
\end{equation}
where $\delta$ is broadened to a Gaussian with a characteristic length $\ell_\text{at}$.
Note that $\rho(\v{r})$ is invariant to permutations of atoms by construction and resembles the density constructed for EAM potentials.
To obtain angular information, this density field is then expanded into a set of spherical harmonics $Y_{lm}(\theta, \phi)$ for each atom $i$,
\begin{equation}
    \rho_i(\v{r}) = \sum_{nlm} C_{i,nlm} g_n(r) Y_{lm}(\theta_i,\phi_i)
\end{equation}
with local angles $\theta_i$ and $\phi_i$ of the vector $\v{r}-\v{r}_i$.
The functions $g_n(r)$ are orthogonal radial basis functions.
The power-spectrum of the angular coefficients $C_{i,nlm}$,
\begin{equation}
    p_{i,nn'l} \propto \sum_m C_{i,nlm}^* C_{i,n'lm}
\end{equation}
is objective and hence fulfills symmetry considerations.
Selected elements of this power spectrum, typically chosen up to an upper value of $n$, form the SOAP descriptor.

SOAP taken to higher body order -- and with $\ell_\text{at}\to 0$ as in Eq.~\eqref{eq:delta-density} -- forms the basis of the atomic cluster expansion (ACE)~\cite{drautz_atomic_2019,Dusson2022-tr}.
We recommend study of Ref.~\cite{drautz_atomic_2019} also because it outlines the relationship between current machine-learning approaches and more traditional EAM and bond-order potentials.
First parameterizations of (performant) ACE potentials are starting to appear in the literature~\cite{Lysogorskiy2021-ed}.

\subsection{Regression}

Once we have a fingerprint vector, we need to interpolate (or since we are in a high-dimensional space, rather extrapolate) atomic energies. This is the task of the regression framework.
Regression frameworks take as input a fingerprint vector and predict \emph{per-atom} energies $U_i$, such that the total energy is simply $U=\sum_i U_i$.
This decomposition corresponds to the one used in all other potentials discussed in this review.

Regression can be as simple as a linear (ridge) regression.
This is for example employed in spectral neighbor analysis potential~\cite{Thompson2015-oi}, which uses the fingerprints introduced in the original GAP~\cite{bartok_gaussian_2010}.
A more flexible regression strategy is the use of Gaussian process regression~\cite{rasmussen_gaussian_2006}, which lends its name to the GAP potential~\cite{bartok_gaussian_2010}.
An advantage of Gaussian processes is that there is a clear statistical interpretation of Gaussian process regression, which allows one to predict confidence intervals on energies in addition to expectation value or maximum probability estimates.
In an active learning framework, this can help to identify regions in a database for which additional first-principles calculations (or experiments) are necessary~\cite{Vandermause2020-yo}.
The final expression for the per atom energies takes the simple form,
\begin{equation}
    U_i = \sum_t \alpha_t K(\v{q}_i, \v{q}_t),
    \label{eq:gpr}
\end{equation}
where the sum over $t$ runs over all elements of the training database.
The fingerprints $\v{q}_i$ of the current environment of atom $i$ are compared to all fingerprints $\v{q}_t$ in the training database.
The function $K(\v{q}_i, \v{q}_t)$ is the kernel that encodes a measure of distance in fingerprint space (and implicitly includes a mapping into a high-dimensional representation, typically called ``feature space'' in the ML literature~\cite{rasmussen_gaussian_2006}).
From Eq.~\eqref{eq:gpr} it becomes immediately clear that the numerical effort of a GAP is linear in the number of points in the reference database.
It is therefore crucial to \emph{sparsify} the database, i.e., to remove entries that are close together in fingerprint space and do not improve the predictive power of the potential.
Sparsification is highly nontrivial and we refer to the original literature on details~\cite{deringer_gaussian_2021}.

A key feature of Gaussian process regression is that a notation of smoothness of the potential energy landscape is build-in.
A common choice for the kernel function $K$ is a squared exponential (Gaussian)~\cite{deringer_machine_2017,rasmussen_gaussian_2006}, which containsa set of distance scales $\ell_\alpha$ as a hyper parameter ,
\begin{equation}
    K(\v{q}_i, \v{q}_t) \propto \exp\left\{\frac{(q_{i,\alpha} - q_{t,\alpha})^2}{2 \ell_\alpha^2}\right\}.
\end{equation}
The final potential energy is smoothed over
$\ell_\alpha$, which avoids parasitic local energy minima or artifacted phonon spectra, which are sometimes found when regressing with neural networks~\cite{Wen2020-bx}, or even when simply interpolating with splines~\cite{wen_interpolation_2015}.
GAPs additionally make use of a dot-product kernel for the many-body (SOAP) contribution to the energy~\cite{deringer_machine_2017}.
The dot-product kernel has no intrinsic length-scale, but the SOAP descriptors contain intrinsic smoothing through the width $\ell_\text{at}$ of the underlying atom-centered Gaussians.
The expansion coefficients $\alpha_t$ are obtained using a Bayesian regression scheme that is described in detail in the pertinent literature~\cite{rasmussen_gaussian_2006}.

An interesting use of Gaussian process regression is the on-the-fly parameterization~\cite{Csanyi2004-vc} of a force field based on electronic structure calculations carried out on snapshots throughout a molecular dynamics trajectory.
Li, Kermode and de Vita~\cite{Li2015-fi} used a small database of reference configurations, generated from past simulation steps of a molecular dynamics trajectory, to extrapolate forces to the future.
Their scheme differs qualitatively from the ones discussed above by predicting forces rather than energies.
The reason for this is that unlike the per-atom energy, a per-atom force is an observable that can be computed straightforwardly within first-principles techniques for a subset of atoms in a system.
This leads to a nonconservative force field, yet use of this method to accelerate a quantum-mechanical region in concurrent multi-scale modeling~\cite{Csanyi2004-vc}, e.g., of fracture~\cite{Kermode2008-xe,Kermode2013-mx}, introduces no additional complication as the coupling scheme itself is typically non-conservative.

We note that GAPs are linear models (in feature space), as the final energies are linear combinations of Gaussians.
ACE potentials~\cite{drautz_atomic_2019,lysogorskiy_performant_2021} introduce an additional nonlinear mapping.
Using a square-root (as in EAM or TB2M potentials) reduced the number of basis functions required for high-accuracy regression for copper~\cite{lysogorskiy_performant_2021}.
Fully nonlinear regression can be achieved with neural networks.
Behler and Parrinello~\cite{behler_generalized_2007} employed the specific form 
\begin{equation}
    U_i = f_\text{a}^{(2)}\left\{w_{01}^{(2)} + \sum_{\nu} w_{\nu 1}^{(2)} f_\text{a}^{(1)}\left(w_{0\nu}^{(1)} + \sum_\mu w_{\mu \nu}^{(1)} q_i^{(\mu)}\right)\right\},
    \label{eq:neural}
\end{equation}
for regression in their neural network potential.
Here, $f_\text{a}^{(1)}(x)$ and $f_\text{a}^{(2)}(x)$ are sigmoidal ``activation functions'' and $w_{\nu 1}^{(1)}$ and $w_{\nu 1}^{(2)}$ are weights. The weights are numbers that are adjusted in the process of fitting the neural network. The sum over $\mu$ runs over all entries of the feature vector $\v{q}_i$ of atom $i$. This network has a single ``hidden layer'' with activation function $f_\text{a}^{(1)}(x)$ and the index $\nu$ runs over the outputs of this layer. The original fit for silicon by Behler and Parrinello employed three nodes in the hidden layer and $48$ entries in the fingerprint vector, amounting to a total of $196$ weights. Equation~\eqref{eq:neural} resembles the many-body contribution $U_\text{m}$ of classical potentials for open-shell systems, see Eq.~\eqref{eq:genU}.

\section{Limiting the interaction range}
\label{sec:limiting}
\hyphenation{sys-te-ma-tic}

Chemistry is local for the most part~\cite{Anderson1984PR}. 
As a consequence, forces between distant atoms are minor, unless they are charged or carry dipoles.
By neglecting interactions with distant atoms, much computing time can be saved at the expense of minor systematic errors, which can often be made marginal with mean-field or other coarse-graining corrections~\cite{allen_computer_1989,intVeld2007JCP,IseleHolder2013JCTC}.
Unfortunately, the locality of chemistry cannot necessarily be specified in terms of cut-off radii, because to what extent two atoms ``see'' each other depends on various factors.
Screening and cutting are somewhat related but the bonding topology and density must be known to express screening through cutting.

A metal atom will scarcely be exposed to the charge density of another atom if a third atom is sandwiched in between them.
Baskes incorporated such screening conceptually into MEAM potentials~\cite{baskes_modified_1992,baskes_atomistic_1994} motivated by the observation that ``\emph{the second-nearest neighbors do not play a part even in the bcc structure},'' despite the ratio of nearest to next nearest-neighbor shell being as small as $1.15$. He continued that ``\emph{this surprising result gives some support to the procedure we have used here in ignoring all but first-neighbor interactions}''~\cite{baskes_modified_1992} (in the description of crystals).
Defects and disorder, both central to materials properties, place large demands on the assessment of bonding topology and make it necessary to define gray shades between bonded and non-bonded. 
In realistic descriptions, those cannot be judged using only the distance between two atoms.

Dispersive interactions between noble-gas atoms are not screened so that limiting their interaction range (without accounting for long-range corrections) is merely an exercise in finding a good compromise between accuracy and computational efficiency.
Therefore, the task of cutting potentials might appear technical or even trivial at first sight.
Doubling the cutoff radius $r_\textrm{c}$ beyond which interactions are ignored requires roughly an eightfold computational effort in three spatial dimensions to construct neighbor lists and to evaluate pair potentials, which can be unjustifiable if the total dispersive interaction is already accurate. 
However, undesired \textit{qualitative} artifacts can arise when cutting improperly.
Brittle materials can turn ductile~\cite{mattoni_atomistic_2007,pastewka_describing_2008} and thereby make it on the cover of contemporary text books~\cite{griebel_numerical_2007}, or, 
nano-tube-based water pumps work endlessly without external driving due to improper cut-offs~\cite{Wongekkabut2010NN}, the report of which~\cite{Gong2007NN}  having made it through the exquisite reviewing process of Nature journals. 
Nonetheless, cutting potentials is a prerequisite for the linear scaling of numerical effort per time step with the number of atoms $N$ and for the efficient parallelization on parallel, high-performance computing systems using domain-decomposition~\cite{bruge_concurrent_1990,liem_molecular_1991,chynoweth_simulation_1991,pinches_large_1991,brown_domain_1993,plimpton_fast_1995,Thompson2022-qg}.
Even long-range Coulomb potentials can be computed with a numerical effort scaling linearly or quasi-linearly with $N$~\cite{rokhlin_rapid_1985,greengard_fast_1987,greengard_fast_1994}, which requires atom-atom distance evaluations to be cut off at a finite distance. 

Unfortunately, choosing optimum cut-off radii and procedures depends not only on the potential but also on the property of interest.
Someone studying phase equilibria between different thermodynamic phases wants the potential depths to be reproduced as accurately as possible, which can be achieved with a brutal truncation of the potential at a given $r_\textrm{c}$ without shifting. 
Implementing such a potential is readily done for Monte Carlo, however, for a molecular dynamics simulation, one would have  to account for the $\delta$-function singularity in the interaction force at $r_\textrm{c}$, which no sympletic integrator in the world would be pleased about. % ~\cite{TuckermanBook}. 
On the other hand, someone interested in mechanical properties can care less about the depth of energy minima but should instead be careful about potential curvature and the smoothness of the force at the cutoff. 
A bond whose tensile force disappears only linearly at $r_\textrm{c}$ cannot be broken adiabatically, which risks to exaggerate loss moduli or viscosities: 
a spring attached to a bond would snatch into or out of bonding whenever the bond length $r$ passes through $r_\textrm{c}$, whereby energy is dissipated~\cite{Prandtl1928-cu}. 

Three more general points are worth discussing before cutting to the chase:
First, when two phases like fcc and hcp have a similar thermodynamic potential, the choice of cutting can affect what phase is preferred~\cite{Smit1992JCP}.
However, at that point it might always be wise to consider other effects like many-body interactions and thermal or quantum fluctuations of the nuclei. 
Second, changing the cut-off procedure or cut-off radius may require a new parameterization of the potential.
Third, and perhaps most importantly, not only potentials can be cut or smoothed but all operations discussed here below can also be applied to forces.
However, cutting forces generally leads to non-conservative forces unless they originate from central, two-body potentials.

\subsection{Cutting}
\label{sec:cutting}

\subsubsection{Pair functions}

Not only pair potentials must be cut off but also expressions that many-body potentials depend on, such as the charge density in EAM potentials, or the interaction between point multipoles. 
Cutting procedures for pair functions can be broadly categorized into cut-and-shift and smoothing.

Cut-and-shift procedures~\cite{Toxvaerd2011JCP} redefine a pair potential $U_2(r)$ such that all its derivatives up to $l$'th order approach zero continuously at $r_\textrm{c}$ and remain zero for $r>r_\textrm{c}$:
\begin{equation}
    U_{2,l}^\textrm{shift}(r) = \left\{ U_2(r) - \mathcal{T}[U_2(r),r_\textrm{c},l] \right\}\Theta\left(r_\textrm{c}-r\right).
\end{equation}
Here, $\mathcal{T}[U_2(r),r_\textrm{c},l]$ is the $l$'th order Taylor expansion of $U_2(r)$  about $r_\textrm{c}$. 

\begin{figure}[hbtp]
    \centering
    \includegraphics[width=0.465\textwidth]{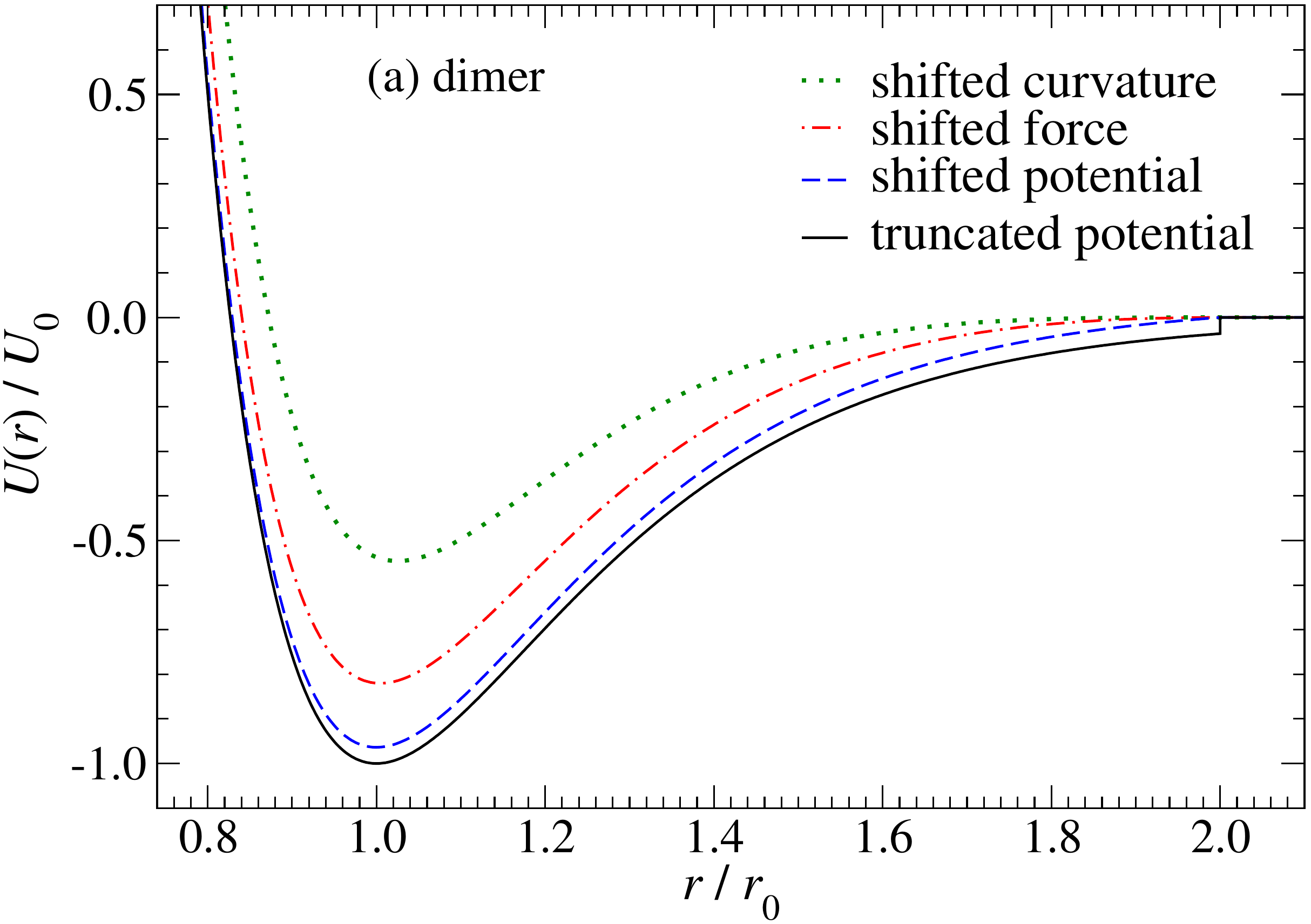}
    \includegraphics[width=0.465\textwidth]{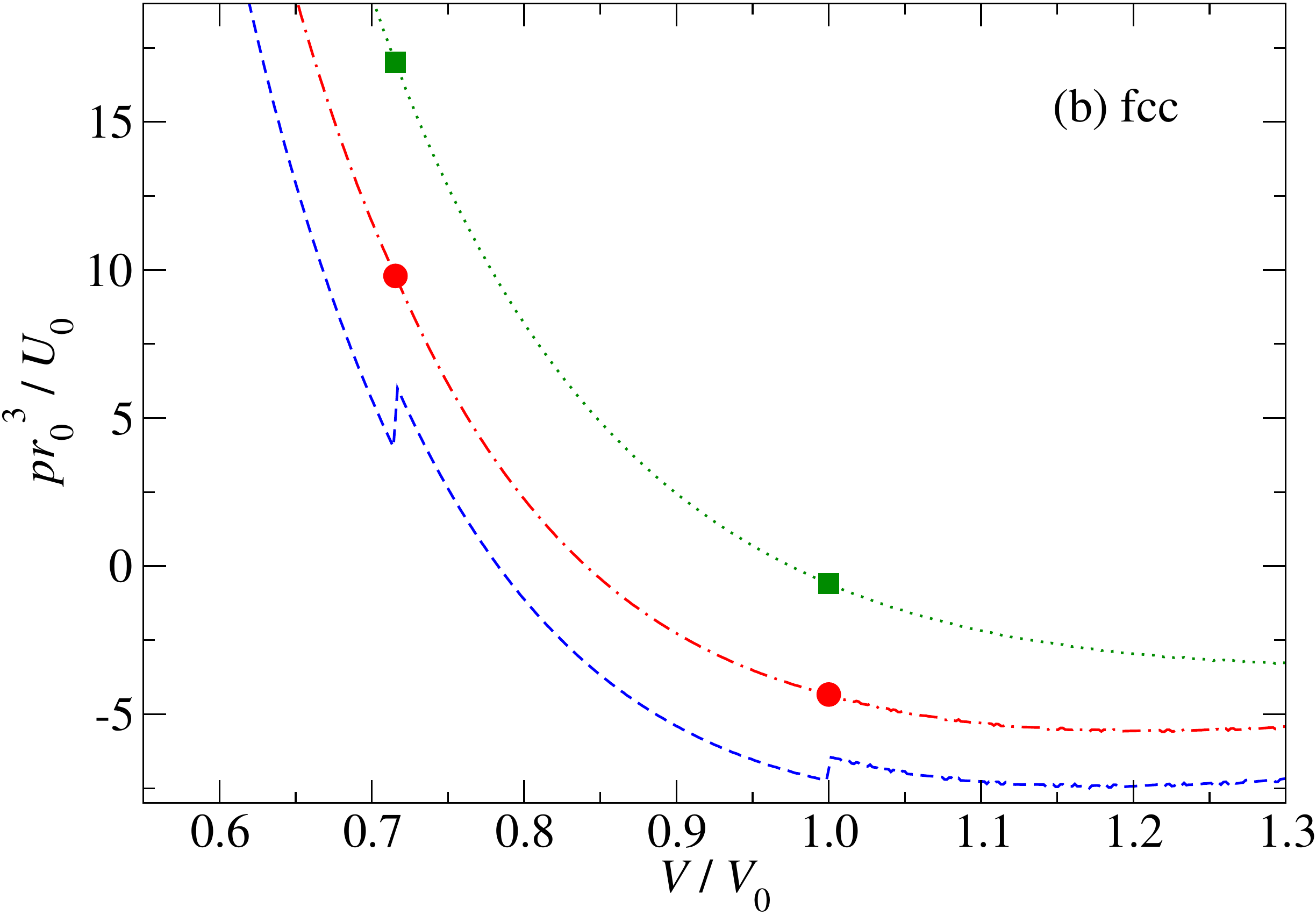}
    \caption{Left panel: Truncated (black), cut-and-shift (blue), shifted-force (red), and shifted-curvature (green) Morse potential with exponent $n = 8$ and a cut-off distance of $r_\textrm{c} = 2.0\,r_0$.
    The cut-and-shift procedures include $l=0$ (blue), $l=1$ (red), and $l=2$ (green).
    Right panel: energy per atom of an ideal fcc crystal as a function of the nearest-neighbor distance $a_0/r_0$ using the shifted potential (blue), shifted force (red), and shifted curvature (green) for the Morse potential with $n=8$ and a cutoff radius $r_\textrm{c} = 2.0\,r_0$.
    Dots and squares denote reduced volumes at which a neighbor-shell radius is equal to $r_\textrm{c}$. 
    }
    \label{fig:ShiftedPotential}
\end{figure}

The effect of cut-and-shift procedures on $U_2(r)$ and on the equation of state is shown exemplarily for a Morse potential ($n=6$, $r_\textrm{c}/r_0 = 2$, $l=0,...,2$) in Fig.~\ref{fig:ShiftedPotential}a.
It becomes evident that ``large'' $l$ lead to a poor representation of the binding energy unless $r_\textrm{c}$ is large, which, however costs computing time. 
However, serious artifacts can occur when potentials are truncated but also when they are merely shifted with $l =0$.
The EOS of an athermal crystal is discontinuous even for $l=0$ as revealed in Fig.~\ref{fig:ShiftedPotential}b. 
Thermal fluctuations smear out and potentially eliminate this discontinuity.
Nonetheless, caution remains advisable. 
In principle, a $l$'th-order shifted potential can have a discontinuity in $\partial^l p/\partial V^l$ so that a force-shifted potential risks to cause a discontinuity in the compressibility or any elastic tensor element including its imaginary, i.e., viscous part, as just discussed.

An alternative to cut-and-shift potentials is the use of cut-off or smoothing functions $f_\text{c}(r)$, which are multiplied with the true pair potential to yield
\begin{equation}
    U_{2}^\textrm{cut}(r) = U_2(r) f_\text{c}(r).
\end{equation}
The general property of smoothing functions is that they are unity below an inner radius, $r_\textrm{i}$, which may be zero or even negative.
They decrease past $r_\textrm{i}$, approach zero at $r_\textrm{c}$, and remain zero ever after. 
Formally they can be expressed as
\begin{equation}
    f_\textrm{c}(r) = 
    \Theta(r_\textrm{i}-r) + \varphi_\textrm{c}(x) \Theta(r-r_\textrm{i})  \Theta(r_\textrm{c}-r)
%    \begin{cases}
%    1 & r < r_\textrm{i} \\
%    0 & r > r_\textrm{c} \\
%    \varphi_\textrm{c}(x) & \textrm{else with } x = (r-r_\textrm{i})/(r-r_\textrm{c}). 
%     \end{cases}
\end{equation}
with $ x = (r-r_\textrm{i})/(r_\textrm{c}-r_\textrm{i})$.
Many of the lessons learned for cut-and-shift applies to smoothing, in particular w.r.t. what artifacts are produced depending on the order $l$ with which derivatives of the potential disappears at $r_\textrm{c}$.
However, an additional risk of smoothing functions is that using ``large'' $l$ does not guarantee artifacts to be suppressed in particular when $r_\textrm{i}$ is close to $r_\textrm{c}$. 
Prominent smoothing functions $\varphi_\textrm{c}(x)$ are:
\begin{subequations}
\label{eq:cutOffFunctions}
\begin{eqnarray}
\left\{ 1 + \cos(\pi x) \right\}/2 & l_\textrm{i} = 1;\; l_\textrm{c} = 1 & \textrm{Tersoff~\cite{tersoff_new_1986}} \\
(1-x)^3(1+3x+6x^2) & l_\textrm{i} = 2;\; l_\textrm{c} = 2 & \textrm{MEAM~\cite{baskes1989PRB}} \label{eq:meam_cut}\\
\left\{8 + 9\cos(\pi x) - \cos(3\pi x) \right\}/16 &  l_\textrm{i} = 2;\; l_\textrm{c} = 2 & \textrm{Murty \& Atwater~\cite{ramana_murty_empirical_1995}} \\
\exp\left\{ - (2x)^m\right\} & l_\textrm{i} = m-1 & \textrm{(unknown)} \\
\exp(\lambda+\lambda/(x-1)) & l_\textrm{i} = 0;\; l_\textrm{c} = \infty &
\textrm{Stillinger \& Weber~\cite{stillinger_computer_1985}}\;\;\;\;\;\;\;\;\\
2(1-x)^m/(1+(1-x)^n) & l_\textrm{i} = 0;\; l_\textrm{c} = m-1 &
\textrm{Mishin~\cite{mishin_interatomic_1999,mishin_embedded-atom_2002},}
\end{eqnarray}
\end{subequations}
where integer Mishin exponents are commonly selected as $m = n = 4$. %
In Eq.~\eqref{eq:cutOffFunctions},
$l_\textrm{i}$ and $l_\textrm{c}$ specify which highest $l$'th-order derivative of the smoothed function approaches zero at $r_\textrm{i}$ and $r_\textrm{c}$, respectively.
Moreover,  $l_\textrm{i}=0$ means that $\varphi_\textrm{c}(r_\textrm{i}) = 1$ while $\varphi'_\textrm{c}(0) \ne 0$, while omitting $l_\textrm{c}$ means that $\varphi_\textrm{c}(x=1^-)\ne 0$.
Recently, it was found that cutting with polynomials constructed such that $l_\textrm{i} = l_\textrm{c} + 1$ can simultaneously improve binding energies while reducing cut-off artifacts~\cite{Muser2022MS}.
This is because cut-off induced discontinuities of (the derivatives of) short-range potentials are distinctly larger at $r_\textrm{i}$ than at $r_\textrm{c}$.

\subsubsection{Cutting Coulomb potentials}

Wolf \textit{et al.}~\cite{Wolf1999JCP} suggested that the Coulomb interaction in an overall neutral system can be cut off without producing systematic errors.
The idea is to represent the Coulomb interaction through the Ewald summation~\cite{Ewald1921-fa}, keep the $q=0$ contribution (which could be seen as a kind of mean-field correction), and to ignore any contribution at non-zero wave vector $\mathbf{q}$.
Some authors find support for the correctness of Wolf's optimistic assessment~\cite{Fennell2006JCP,Rahbari2018MS}, while others don't~\cite{Patra2003BJ,Vlugt2008JCTC}.
In fact, small cutoff radii can be used for homogeneous systems when using appropriate cut-off functions, e.g., clearly less than $10$~\AA~ in silica melts~\cite{Muser2022MS}. 
However, problems arise when structural heterogeneities extend on  length scales $\lambda>r_\textrm{c}$~\cite{Muser2022MS}.
They induce undulations in chemical potentials with wave vectors of order $q = 2\pi/\lambda$, which favor a charge separation. 
Coulomb interactions suppress that separation with terms scaling as $1/q^2$.
The pertinent restoring forces are neglected for the most part in the Wolf summation.

Another argument for why the Wolf summation cannot be cutoff in heterogeneous systems can be made when considering a  gold nugget separated by a large distance from a chunk of lithium.
If electrons can equilibrate, gold will acquire a negative charge due to its larger electronegativity. 
While the charge density on each lump will decrease with their size, the absolute force attracting them will increase.
This mechanism also matters for discharge simulations of batteries or related devices~\cite{Dapp2013JCP}, despite the presence of a Helmholtz double layer near the electrodes. 
If a Maxwellian demon were to keep all electrolyte atoms in place and closed an electrical switch allowing some charge to flow between anode and cathode, a current would flow, which would extend much beyond an elementary charge.
Thus, local charge neutrality cannot be obeyed during the entire time of the discharge process.
Once the demon releases the electrolyte particles, discharge can continue after the electrolyte rejuvenated the double layers.
Modeling the described process with a Wolf summation properly would require one to use a cut-off exceeding the anode-cathode separation.
Thus, the Wolf summations, or better generalizations thereof avoiding a discontinuous force at $r_\textrm{c}$~\cite{Fennell2006JCP,Muser2022MS}, can be made for homogeneous systems, but steer clear of it otherwise, in particular in any biophysical context~\cite{Cisneros2013CR}.

\subsubsection{Many-body potentials}

\begin{figure}
    \centering
    \includegraphics[width=0.45\textwidth]{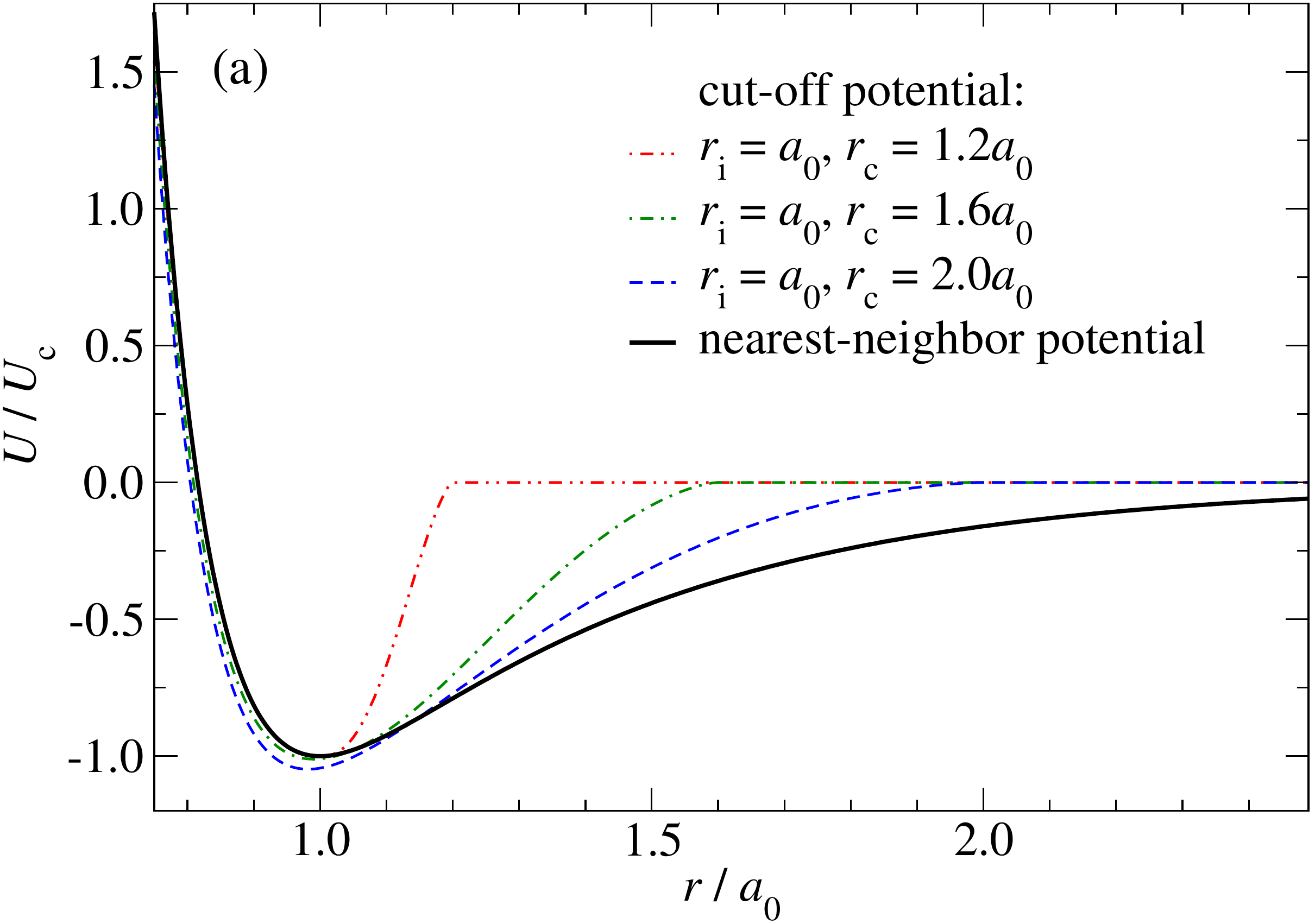}
    \includegraphics[width=0.45\textwidth]{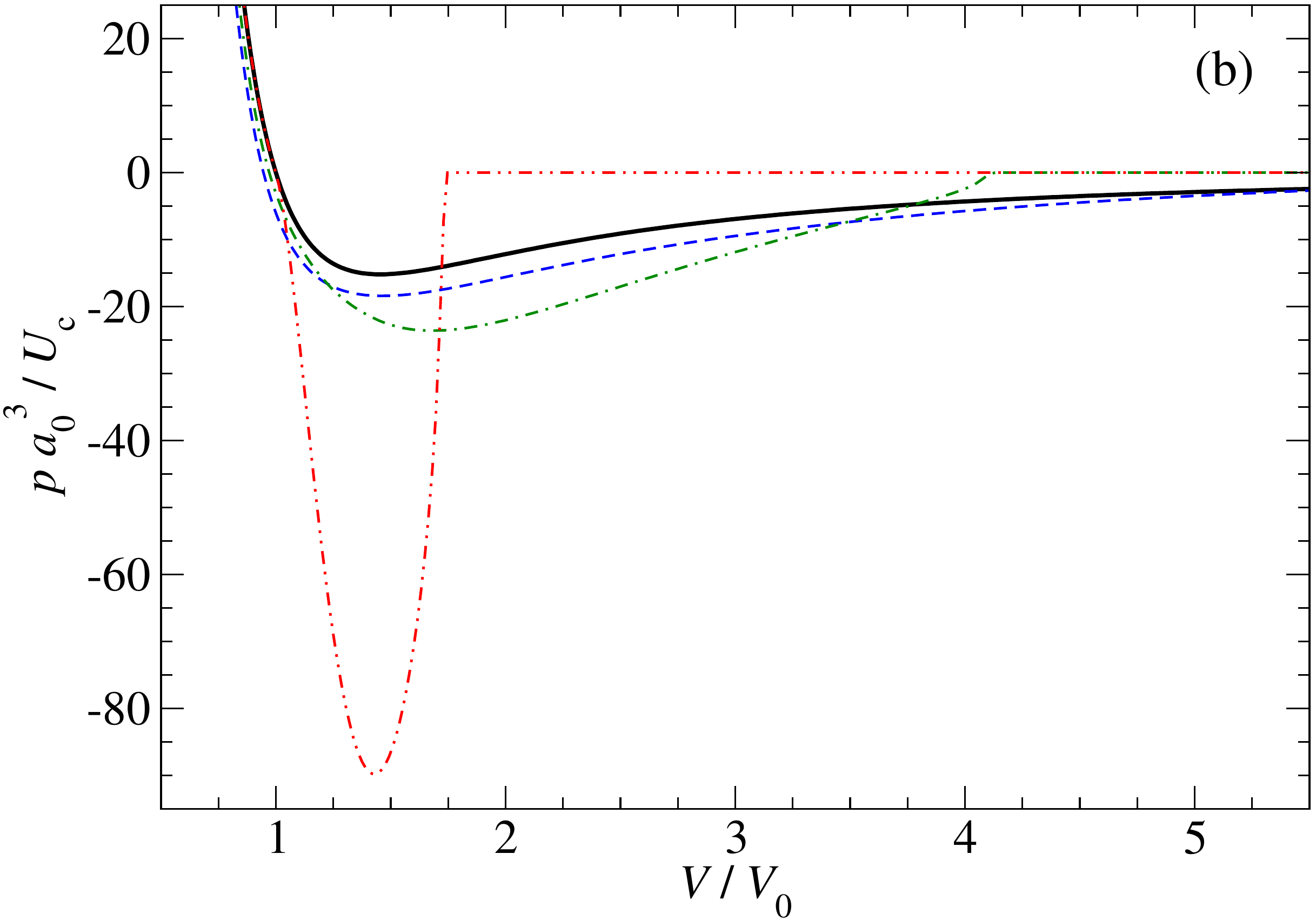}
    \caption{(a) Energy per atom as a function of distance and (b) pressure as a function of volume  for a Cu fcc crystal described by a Ducastelle potential~\cite{Jalkanen2015MSMSE}.
    The MEAM cutoff function given by Eq.~\eqref{eq:meam_cut} was applied to the repulsive potential and the density.
    $U_{\rm{c}}$ and $a_0$ are the cohesive energy and the nearest-neighbor distance for an fcc crystal described by a Ducastelle potential, respectively.
    }
    \label{fig:gupta-screening}
\end{figure}

Smoothing functions are not only applied to pair potentials but also in many-body potentials, sometimes with the goal to imitate screening.
In a crystalline structure, interactions with atoms in the nearest-neighbor shell should be unmodified but those with the next-nearest neighbors should be screened -- often those are approximated to be fully screened.
Figure~\ref{fig:gupta-screening} illustrates the effect of cutting the interaction in a Ducastelle potential at various distances.
When cutting with a distance-based criterion, energies (Fig.~\ref{fig:gupta-screening}a) are forced to drop to zero somewhere between first and second neighbor shell of the equilibrium crystal (since we do not want interaction with second neighbors at equilibrium - they are screened).
This leads to a force or pressure overshoot during homogeneous volumetric dissociation of the crystal, illustrated in Fig.~\ref{fig:gupta-screening}b.
Attempting to break such a crystal with an external stress overestimates the critical stress for fracture by a factor of four.
For this reason, some works have adopted cutting forces rather than energies for modeling fracture~\cite{shenderova_atomistic_2000}, despite the fact that this leads to a nonconservative force-field.
The proper solution to this issue is the use of screening functions, discussed in the next section.
They yield the true nearest-neighbor potential (shown by the black line in Fig.~\ref{fig:gupta-screening}) and hence smooth dissociation of the crystal.

\subsection{Screening}

%\subsubsection{Screening functions}

In order to reflect screening through screening rather than through cutting,
Baskes \textit{et al.}~\cite{baskes_modified_1992,baskes_atomistic_1994} proposed to replace cutoff functions with  environment-dependent screening functions, $S_{ij}$, which themselves are products of three-body functions $0<S_{ijk}<1$,
\begin{equation}
    S_{ij} = \prod_{k\not=i,j} S_{ijk}.
\end{equation}
The $S_{ijk}$ are constructed such that $S_{ijk}=0$ when atom $k$ screens the interaction between $i$ and $j$ completely, while  $S_{ijk}=1$ when atom $k$ does not screen the $ij$ interaction at all. 
Thus, the screening function acts like a topological cutoff function, $0 \le S_{ij} \le 1$.
Baskes proposed different functional forms for $S_{ijk}$.
The refined expression is~\cite{baskes_atomistic_1994}
\begin{equation}
    S_{ijk} = \begin{cases}
    1 & \text{if}\quad C_{ijk}\leq C_\text{min} \\
    \exp\left\{-\left[(C_\text{max}-C_{ijk})/(C_{ijk}-C_\text{min})\right]^2\right\} & \text{if}\quad C_\text{min} < C_{ijk} < C_\text{max} \\
    0 & \text{if}\quad C_{ijk}\geq C_\text{max}
    \end{cases},
\end{equation}
which drops from unity to zero between $C_\text{max}$ and $C_\text{min}$. The quantity $C_{ijk}$ characterizes the geometry of the ellipse
that passes through the three atoms $i$-$j$-$k$. It is given by
\begin{equation}
    C_{ijk} = \frac{2\left( X_{ik} + X_{jk} \right) - \left(X_{ik} - X_{jk}\right)^2 - 1}{1-\left(X_{ik} - X_{jk}\right)^2},
\end{equation}
the square of the ratio of the lengths of the two half-axes of that ellipse. Here $X_{ik}=r_{ik}/r_{ij}$ is the distance between atoms $i$-$k$ normalized by the length of the central $i$-$j$ bond.

\begin{figure}
    \centering
    \includegraphics[width=0.85\textwidth]{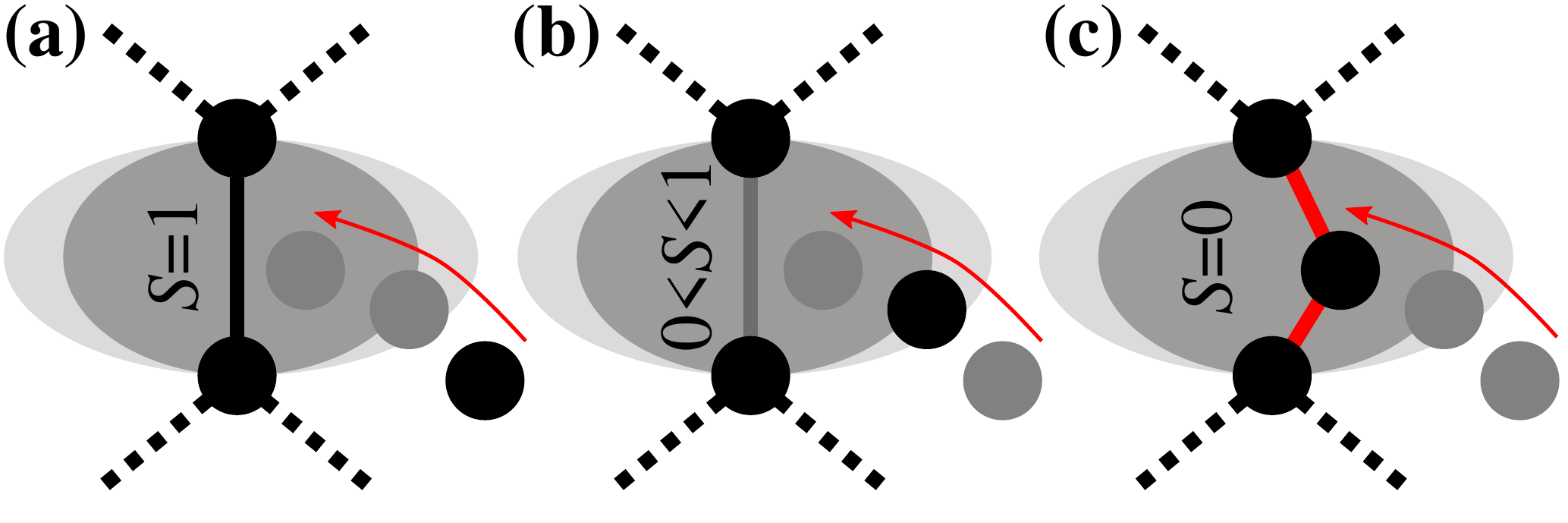}
    \caption{Illustration of the screening concept. The black atom moves along the hypothetical trajectory indicated by the shaded atoms and the red arrow. (a) Initially, the thick, black bond is unscreened, but as (b) the atom enters the region of influence (light gray ellipses) the bond weakens. (c) The atom has moved into the close vicinity of the bond (dark gray region), effectively disabling it while creating two new bonds (red lines). The Baskes screening functions~\cite{baskes_atomistic_1994} are defined the aspect ratio of the light gray region ($C_\text{max}$) and the dark gray ($C_\text{min}$), defining a measure of bond screen that is independent of the absolute lengths of the bonds in the system.}
    \label{fig:screening-functions}
\end{figure}

Figure~\ref{fig:screening-functions} illustrates this concept.
The bond is unscreened ($S=1$) if there is no third atom in its vicinity (Fig.~\ref{fig:screening-functions}a).
As a third atom moves into the shaded ellipsis (Fig.~\ref{fig:screening-functions}b -- the geometry of the ellipsis is described by the values of $C_\text{min}$ and $C_\text{max}$) the bond becomes screened ($S=0$, Fig.~\ref{fig:screening-functions}c).
It becomes clear, that if we simply rescale all coordinates (as in the homogeneous, volumetric dissociation of a crystal), $S$ remains constant for each bond and an energy would follow the true nearest neighbor curve, see the black line in Fig.~\ref{fig:gupta-screening}a.
The screening function $S$ is hence a scale-invariant formulation of a ``cutoff'' procedure.

The Baskes screening functions~\cite{baskes_atomistic_1994} were applied to empirical bond-order potentials independently by Pastewka~\etal~\cite{pastewka_describing_2008,pastewka_bond_2012,pastewka_screened_2013} and Kumagai~\etal~\cite{kumagai_development_2009}. Both groups emphasized that the screened potentials significantly improved the properties of amorphous carbon modeled with REBO2 or Tersoff-type potentials. In addition, the screening functions served to overcome the issue with dissociation of a bond under external stress discussed in Refs.~\cite{mattoni_atomistic_2007,pastewka_describing_2008,pastewka_bond_2012,pastewka_screened_2013}. This enabled modeling of fracture in crystalline and amorphous carbon systems~\cite{khosrownejad_quantitative_2021}. Perriot~\etal~\cite{perriot_screened_2013} presented slightly different screening concept requiring the REBO potential to be refitted.

Screening functions can be rationalized as originating from nonorthogonality in a tight-binding framework.
This nonorthogonality leads to an environment-dependence of the bond-integrals, when the nonorthogonal tight-binding is ``coarse-grained'' to an equivalent orthogonal tight-binding model.
Nguyen-Manh, Pettifor and Vitek~\cite{nguyen-manh_analytic_2000,nguyen-manh_environmentally_2003} showed, that the theory of the bond-order expansion, briefly touched upon in Sect.~\ref{sec:beyond-second-moments}, can be used to derive screening functions from a nonorthogonal tight-binding model.
This first-principles construction lends additional support to Baskes' screening functions and other screening approaches, such as empirical environment-dependence introduced in the context of orthogonal tight-binding shortly after Baskes work~\cite{tang_environment-dependent_1996,tang_erratum:_1996,Haas1998-hc,wang_environment-dependent_1999}.

\section{Summary and perspectives}

\begin{figure}
    \centering
    \includegraphics[width=\textwidth]{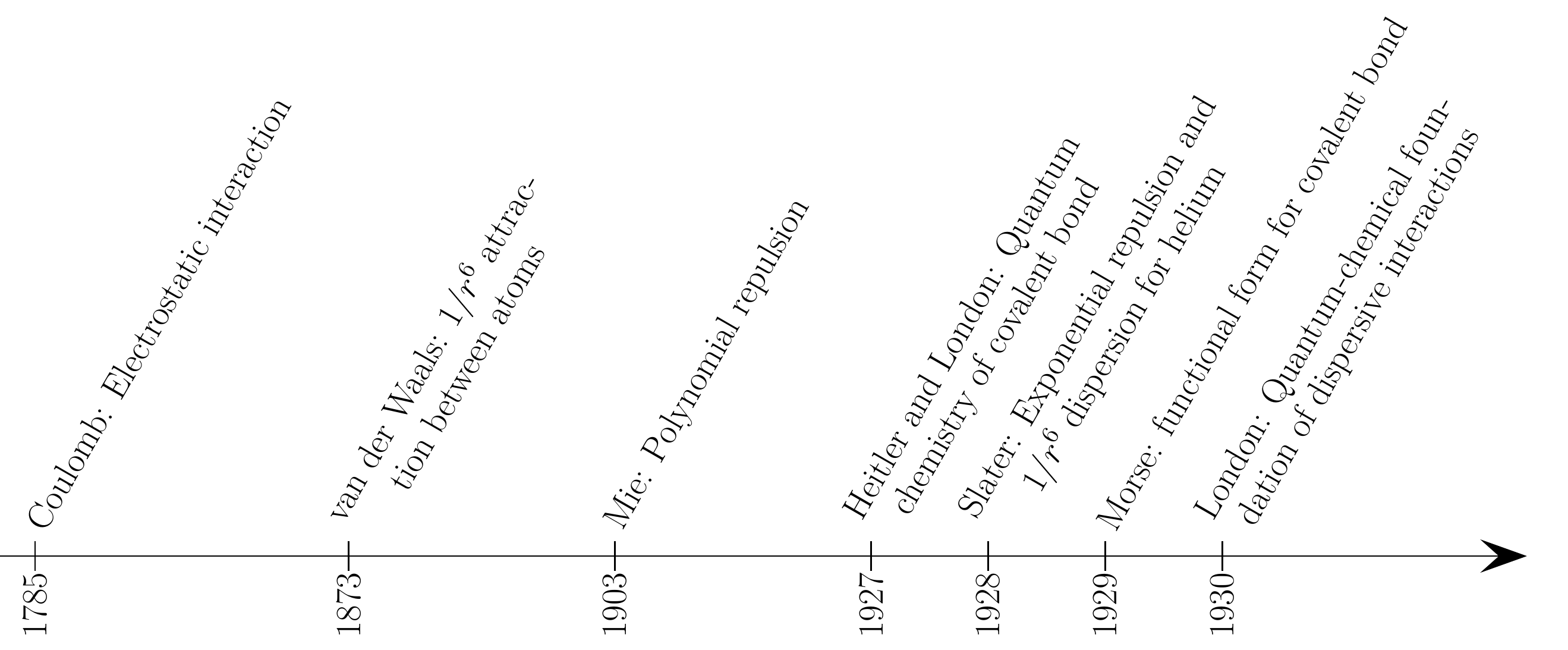}
    \includegraphics[width=\textwidth]{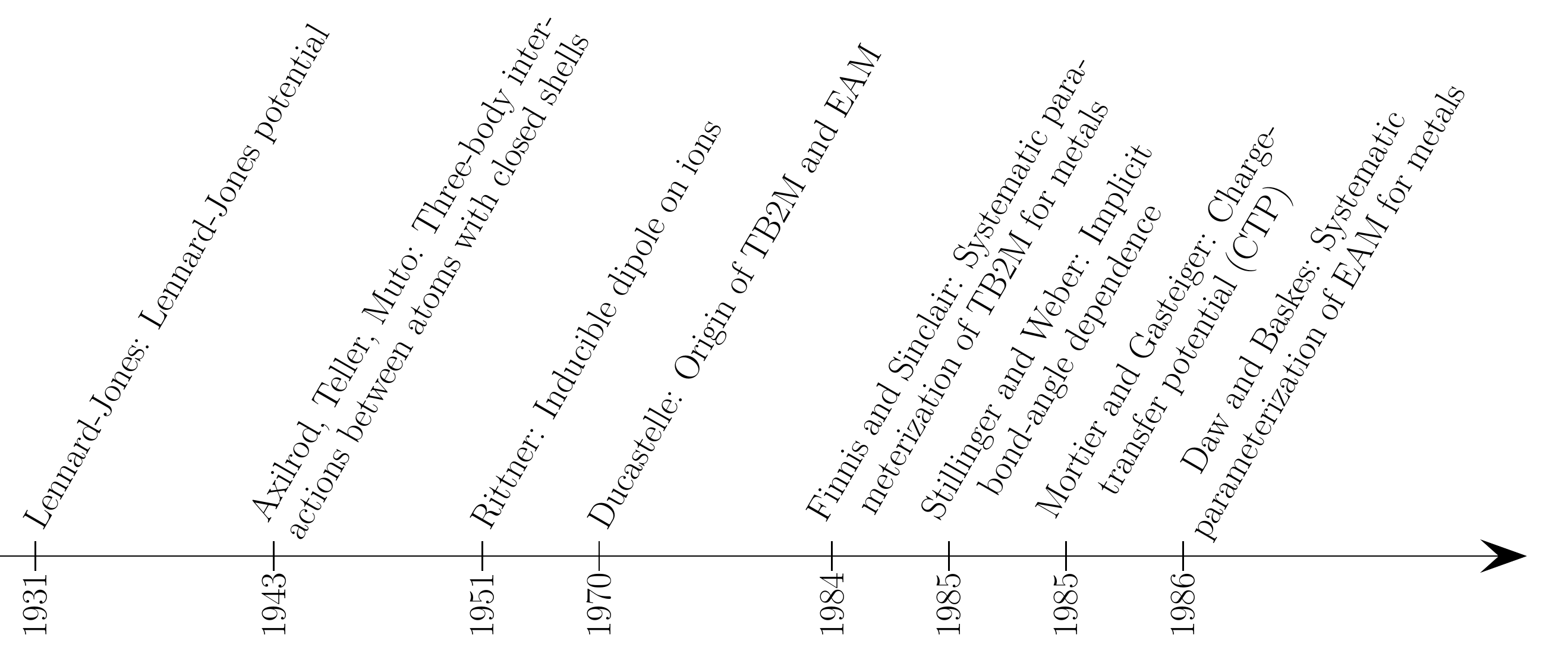}
    \includegraphics[width=\textwidth]{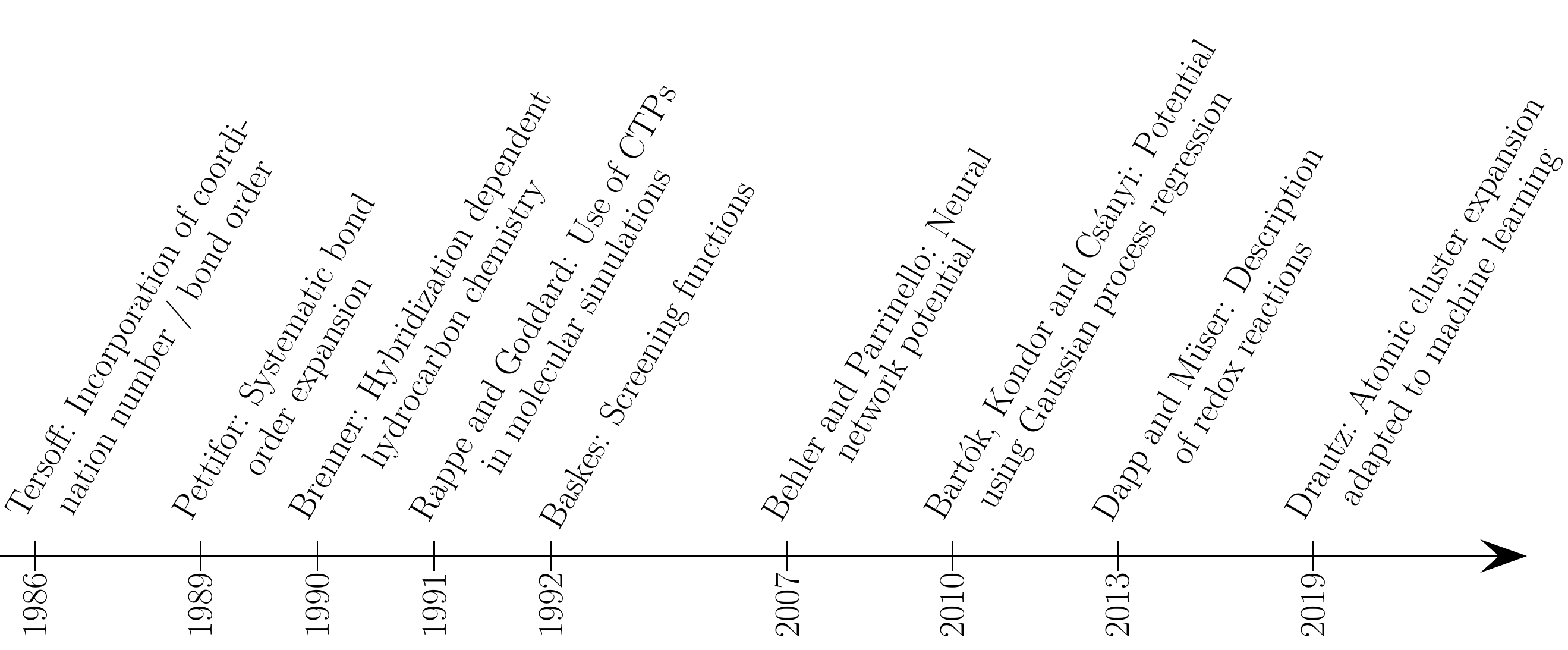}
    \caption{Selected highlights of the development of interaction potentials.}
    \label{fig:Zeitstrahl}
\end{figure}

Eugene Wigner allegedly said: ``\emph{It is nice to know that the computer understands the problem. But I would like to understand it too.}''
One main motivation for writing this review was to assist people with similar ambitions as Wigner.
To this end, we summarized our understanding of what properties in condensed-matter systems can be induced by the functional form of the potentials used for their description. 
In our endeavor, we felt compelled to create much own data and new figures with the purpose to create insight and to convey trends and differences between potential classes rather than to produce numbers for a specific system.
When doing so, we did our best to embed anything written into a historical context, which is summed up in Fig.~\ref{fig:Zeitstrahl}.
Yet many times, we could not find appropriate references or quotes, which we are certain do exist.
As one of numerous examples, we found many papers computing shear and bulk moduli of metals or vacancy-defect and cohesive energies, but always missed the argument why their respective ratios are correlated and how they relate to the ratio of melting and boiling temperature.
We expect our discussion to have satisfied Wigner's desire for understanding the correlation between these ratios. 
Despite certainly having missed well known studies, we did find some old works, which may have been underappreciated, such as Slater's paper on the interaction between helium atoms~\cite{Slater1928PR}.
As mentioned earlier, Slater derived the exponential repulsion between atoms with closed valence shell, which promoted Born and Mayer~\cite{Born1932ZFP}  as well as Buckingham~\cite{Buckingham1938PRSL} to use this or slightly modified forms for the repulsion in the potentials now carrying their names. 
However, Slater also derived the dispersive coefficient for helium to within 15\% accuracy, two years before London~\cite{London1930ZFP} generalized the results to other closed-valence shells.

Despite the length of this article, we could only scratch the surface of the large field of interatomic potentials. 
Many central aspects were not touched upon.
Most importantly, we barely discussed how to adjust parameters, in particular the pros and cons to fit to experimental or to \textit{in-silico} data. 
There are good reasons to follow the main-stream opinion that the quality of a potential increases the less empirical the data on which the potential is parameterized. 
Computer-generated reference data is much more versatile than that provided from experiments.
Forces on individual atoms can be used for characteristic bonding situations or rare but important configurations like a transition state occurring during a chemical reaction or a collective phase transformation~\cite{Ercolessi1994-dr}.
Moreover, \textit{in-silico} data does not contain quantum effects, which frequently need to be accounted for when comparing computer-generated data to experiments. 
Describing how to do that properly would have required us to outline how to approximately subtract the nuclear quantum effects from experimental data or how to incorporate quantum fluctuations into the simulations, e.g., through path-integral techniques, or, by encoding their effect into effective temperature-dependent, many-body potentials, which would have been beyond the scope of this review.

Thus, there is scarcely any argument to gauge parameters on experiments, if there was not the small but important detail that experimental data is by and large more accurate than density-functional theory, which cannot be deemed exact, as long as the exact functional has not been identified. 
It could be argued that we base one theory or potentials with uncontrolled approximations on another one with uncontrolled approximations, which has trouble to predict two dislike molecules or clusters separated by a large distance to each acquire an integer charge~\cite{Bally1997JPCA,Zhang1998JCP}. 
When dismissing empiricism as fundamentally problematic, one may also keep in mind that one of the greatest theoretical achievements in chemistry, arguably in all of science, was the construction of the periodic table by Mendeleev.
% https://en.wikipedia.org/wiki/Mendeleev%27s_predicted_elements
%
He even predicted the existence of unknown elements including some of their physical and chemical properties with an accuracy that people using potentials or even DFT might have a hard time to match if they did not know what they had to predict, or, rather postdict.
Moreover, the amount of data that Mendeleev could build on was noticeably less than what is required in machine learning.

The potentials discussed in this review pertain mostly to situations, in which bonds can be clearly classified as dispersive, metallic, covalent, or ionic.
For situations, where this simple categorization cannot be made, different potential classes are combined in a mix-and-match fashion into compound potentials.
Prominent examples are the adaptive intermolecular REBO (AIREBO)~\cite{Stuart2000-jc} (combining Brenner's potential with nonbonded Lennard-Jones interaction), the Streitz-Mintmire potential~\cite{Streitz1994-zw} (combining EAM with charge transfer), the charge-optimized many-body potential (COMB)~\cite{Yu2007-es,shan_second-generation_2010,Martinez2017-is} (combining Tersoff's potential with charge transfer),
a merger of REBO with split-charge equilibration~\cite{Mikulski2009JCP,knippenberg_bond-order_2012}, as well as early combinations of Keating-type with charge-transfer potentials potentials~\cite{Martin1970-fy,Vashishta1990-yc}.
Of course, the widely-used ReaxFF potential~\cite{Senftle2016-nn,VanDuin2001-iv}, which merges a bond-order approach (different in nature than the approaches discussed in this review) with non-bonded interactions and charge transfer, must also be mentioned. 

While compound approaches can be extremely powerful, many of them simply add different energy terms.
This can be problematic even for seemingly simple alloys or intermetallics formed by elements of large electronegativity difference.
Put simply, negatively charged atoms grow in size while positively charged atoms shrink.
This symmetry breaking between negative and positive charge is not reflected when simply adding charge equilibration to a (post) Ducastelle potential.
Yet, it is supposedly responsible for why the negatively charged atoms in intermetallics have the tendency to close pack while positive atoms occupy interstitial positions, as it happens, for example, for Al$_2$Au, also called the purple plague: Au atoms form an fcc lattice while Al atoms assume interstitial positions. 
Although promising steps toward true compound potentials have been taken~\cite{Bhattarai2019PRB}, e.g., by augmenting or reducing the valence density of a neutral atom with a term proportional to its partial charge, systematically merged potential remain a dream.

Novel paths that are taken with machine learning potentials seem extremely promising.
However, a puzzling question is why machine-learned potentials outperform parameterized potentials.
The claim that they are parameter free or free of functional constraints is not entirely justified. 
Many of the local descriptors are suspiciously close to what is used in potentials, as indeed they are often ``physics-inspired''~\cite{Musil2021-lm}.
However, the big advantage of MLPs is that they do not make strong assumptions like pair-wise additive repulsion, which might be one of the most important sources of error in classical interatomic potentials.

A show-stopping problem central to all potentials is the curse of dimensionality.
Fitting multi-species (or alloy) potential requires a number of pair-parameters that scales asymptotically as $N_\textrm{s}^2$ with the number of atomic species $N_\textrm{s}$.
The scaling becomes even less favorable if we need specific parameters for triplets ($\propto N_\textrm{s}^3$), quadruplets ($\propto N_\textrm{s}^4$) and so on, quickly becoming intractable for a large number of species.
The compression of chemical fingerprints has recently been proposed to circumvent the curse of dimensionality for MLPs~\cite{Darby2021-xq}.
Using explicit functional forms, it can be possible to circumnavigate the curse of dimensionality with combining rules.
However, they are only available for few interactions types and may be plagued with poor transferability.

As a final note, we would like to point out that despite the fact that (with the exception of bare Coulomb interaction) all potentials discussed here are local, chemistry can be quite non-local.
By non-locality we do not mean the range of the bare interaction, such as the range of the bond-integrals in a tight-binding formulation.
We mean the non-locality intrinsic in the diagonalization of the quantum mechanical Hamiltonian.
In hydrocarbon chemistry, the non-locality manifests itself for example in bond conjugation and in metals through an algebraic decay of the density matrix~\cite{Goedecker1999-jm}, while in group 15--17 in the periodic table, it is reflected in the Peierls deformation causing elemental crystals to reduce from the simple cubic to less symmetric structures~\cite{Gaspard1998PMB}.
As another example, carbon chains -- also called carbynes -- can exist in a polyynic form of alternating single and triple bonds or a cumulenic form of repeating double bonds~\cite{Smith1982-nr,Heimann1983-md,Whittaker1985-wc}.
Which form is chosen depends on whether the chain is odd or even numbered and how it is terminated.
This crucially affects how they interact with their environment, for example with oxygen~\cite{Moras2011-zd,Moras2011-yo}.
Such non-local effects even manifest in bulk materials: Force-locality tests on amorphous carbon by Deringer and Csányi showed that chemistry in low-density, graphite-like amorphous carbon is much longer ranged than in denser more diamond-like carbon~\cite{deringer_machine_2017}.
Approaches for incorporating true quantum non-locality into potentials currently do not appear to exist.
Modeling it appears to require new classes of potential, e.g., the ability of an EAM or MEAM potential to make atoms adjust their donating charge density in response to the environment in a fashion that allows for multistability.

We hope that this review was successful in highlighting the incredible achievements throughout the last century in understanding the bonding of matter, and molding these insights into simple analytical expressions.
The wide availability of high-accuracy electronic structure calculations and advances in statistical modeling have moved the field into exciting new directions.
We would also like to add that the wide availability of present-day interatomic potentials in the form of open-source software, ideally embedded in a standard database~\cite{Tadmor2011JOM} or a standard code~\cite{plimpton_fast_1995,Thompson2022-qg}, is accelerating quick adoption of potentials into practice --- not to mention the savings in students' lifetimes, by not having to dissect which of the $50$ parameters just manually copied from printed publication XYZ is missing a $0$ in print.
(Yes, we are thinking about our own PhD theses.)
Of course, significant challenges remain, both for traditional fixed-form as well as machine-learned interatomic potentials, of which we believe the curse of dimensionality and the coupling of electron transfer and Coulomb interaction to the electronic bond as most crucial.

\section*{Acknowledgement(s)}

We thank G\'abor Cs\'anyi, Volker L. Deringer, Christian Elsässer, Peter Gumbsch, Judith Harrison, James Kermode, Pekka Koskinen, Gianpietro Moras, Michael Moseler, Matous Mrovec, Toon Verstraelen and Michael Walter for many enlightening discussions over the years. We are further indebted to James Kermode for comments on the machine learning section of the manuscript as well as Joshua Wei{\ss}enfels and Jan Grie{\ss}er for proofreading and commenting on the full manuscript. We used \textsc{gpaw}~\cite{Enkovaara2010-bp} for all DFT calculations shown here that are not obviously taken from third sources.

\section*{Disclosure statement}
No potential conflict of interest was reported by the author(s).

\section*{Nomenclature/Notation}

\subsection*{Abbreviations}

\begin{tabular}{ll}
ACE & atomic cluster expansion \\
ATM & Axilrod-Teller-Muto interaction potential \\
DFT & density-functional theory \\
EAM & embedded-atom method \\
EOS & equation of state \\
GAP & Gaussian approximation potential \\
LJ & Lennard-Jones  \\
MEAM & modified EAM \\ 
%MD & molecular dynamics \\
ML & machine learning \\
MLP & machine-learned potential \\
QEq & charge equilibration \\
REBO & reactive empirical bond-order potential \\
REBO2 & second generation REBO \\
SQE & split-charge equilibration \\
SW  & Stillinger-Weber \\
TB, TB$n$M & tight-binding, TB $n$-th order moment expansion \\
\end{tabular}

\subsection*{Symbols}

\begin{tabular}{ll}
$\alpha$, $\alpha'$ & polarizability in SI and atomic units \\
$\alpha_\text{M}$ & Madelung constant \\
$\delta_{\alpha\beta}$ & Kronecker delta \\
$\epsilon$ & Lennard-Jones energy parameter\\
$\varepsilon_0$ & vacuum permittivity \\
$\varepsilon_\textrm{r}$ & dielectric constant \\
$\varepsilon_{\alpha\beta}$ & element of the Eulerian strain tensor \\
$\eta_{\alpha\beta}$ & element of the Lagrangian strain tensor \\
$\rho$ & charge density, number density \\
$\sigma$ & length scale parameter \\
$\sigma_{\alpha\beta}$ & element of the Cauchy stress tensor
\end{tabular}

\begin{tabular}{ll}
$A$ & electron affinity \\
$B_n$ & $n$-th order virial coefficient \\
$C_{\alpha\beta\gamma\delta}$ & element of elastic tensor \\
$C_{ij}$ & element of elastic tensor in Voigt notation \\
$C_n$ & dispersion coefficient of order $n$ \\
%H & Hartree (unit of energy) \\
$\hat{H}$ & Hamilton operator \\
$H_{i\alpha j\beta}$ & Hamiltonian integral between orbital $\alpha$ on atom $i$ and orbital $\beta$ on atom $j$ \\
$I$  & ionization energy \\
$B$ & bulk modulus \\
%$G$ & symmetrized shear modulus \\
$N$ & particle number \\
$Q_i$ & charge of atom $i$ \\
$S_{i j}$ & square of the distance between atoms $i$ and $j$ \\
$U$ & interaction energy \\
$U_0$ & dimer/molecular binding energy \\
$U_\text{pa}$ & potential energy per atom \\
$U_\text{pa}^\text{eq}$, $U_\text{coh}$ & equilibrium potential energy per atom (cohesive energy) \\
$U_\text{pb}$ & potential energy per bond \\
$U_\text{pb}^\text{eq}$ & equilibrium potential energy per bond \\
$U_1$, $U_1^{(i)}$ & single-body interaction energy \\
$U_2$, $U_3$ & pair and triplet interaction energy \\
$V$ & volume \\
$Z_0$ & coordination number \\
$Z_s$ & number of atoms in  $s$'th nearest-neighbor shell
\end{tabular}

\begin{tabular}{ll}
$a_n$ & distance between an atom with a $(n+1)$-nearest neighbor \\
$a_0^\textrm{eq}$ & equilibrium bond length \\
$a_\textrm{B}$ & Bohr radius \\
e & elementary charge \\
$f_\text{c}$ & cutoff function \\
$k_\textrm{B} T$ & thermal energy \\
$m$ & mass \\
$n(\varepsilon)$ & density of states \\
$n_{i\alpha}(\varepsilon)$ & local density of states of orbital $i\alpha$ \\
$\hat{p}_i$ & momentum operator \\
$\mathbf{p}_i$, $p_i$ & dipole moment of species $i$ and its magnitude \\
$p$ & pressure \\
$\v{q}_i$ & descriptor of the environment of atom $i$ \\
$q_{ij}$ & bond charge donated from atom $i$ to atom $j$\\
$r$, $r_{ij}$ & (pair) distance \\
$r_0$ & equilibrium distance in a diatomic molecule \\
$r_\textrm{c}$ & cutoff radius \\
$\nu_\textrm{s}^{\alpha\beta}$, $\nu_\textrm{s}^{\alpha\beta\gamma\delta}$ & second- and fourth-rank shell tensor for the $s$'th nearest-neighbor shell\\
$v_\text{pa}$ & volume per atom
\end{tabular}

\printbibliography %ARXIV uncomment

\end{document}